\documentclass[12pt]{article}
\usepackage{jheppub}

\usepackage{amsmath,epsfig}
\usepackage{amssymb,amsfonts}
\usepackage{latexsym}
\usepackage[latin1]{inputenc}
\usepackage{overpic}
\usepackage{subeqnarray}
\usepackage{array}
\usepackage{xcolor}

\usepackage{graphicx}
\usepackage{subfigure}
\usepackage{longtable}
\usepackage{mathdots}
\usepackage{physics}
\usepackage{enumitem}
\usepackage{oubraces}
\usepackage{cancel}

\relax
\renewcommand{\theequation}{\arabic{section}.\arabic{equation}}
\def\be{\begin{equation}}
\def\ee{\end{equation}}
\def\bea{\begin{eqnarray}}
\def\eea{\end{eqnarray}}
\def\bs{\begin{subequations}}
\def\es{\end{subequations}}
 
\newcommand\fverb{\setbox\pippobox=\hbox\bgroup\verb}
\newcommand\fverbdo{\egroup\medskip\noindent%
                        \fbox{\unhbox\pippobox}\ }
\newcommand\fverbit{\egroup\item[\fbox{\unhbox\pippobox}]}
\newcommand{\ha}{{1 \over 2}}

\newcommand{\bear}{\begin{eqnarray}}
\newcommand{\eear}{\end{eqnarray}}
\newcommand{\de}{\partial}
\newcommand{\bsea}{\begin{subeqnarray}}
\newcommand{\esea}{\end{subeqnarray}}
\newbox\pippobox

\let\oldsqrt\sqrt
\def\sqrt{\mathpalette\DHLhksqrt}
\def\DHLhksqrt#1#2{%
\setbox0=\hbox{$#1\oldsqrt{#2\,}$}\dimen0=\ht0
\advance\dimen0-0.2\ht0
\setbox2=\hbox{\vrule height\ht0 depth -\dimen0}%
{\box0\lower0.4pt\box2}}

\newcommand{\eql}[2]
{ \begin{equation} \label{#1}
 #2
\end{equation}}

\def\d{\delta}
\def\g{\gamma}

\def\lab{\label}
\def\6{\partial}
\def\f{\varphi}
\def\a{\alpha}

\def\le{\left}
\def\ri{\right}

\def\cO{{\cal O}}

\def\m{\mu}
\def\n{\nu}

\def\sp{\;\;\;,\;\;\;}

\def\sq
\def\a{\alpha}
\def\b{\beta}
\def\l{\lambda}

\def\hri#1#2{\href{http://arxiv.org/abs/#1}{[ArXiv:#1]#2}}
\def\hre#1#2{\href{http://arxiv.org/abs/#1/#2}{[ArXiv:#1/#2]}}

\def\d{\delta}

\def\D{\Delta}
\def\Dp{\Delta_+}
\def\Dm{\Delta_-}

\def\dd{{\rm d}}
\def\MM{{\cal M}}

\def\nd{{\text{and}}}

\def\<{{\langle}}
\def\>{{\rangle}}

\title{Exotic holographic RG flows at finite temperature}

\author[1]{Umut G\"ursoy}
\author[2,3]{, Elias Kiritsis}
\author[3]{, Francesco Nitti}
\author[3]{, Leandro Silva Pimenta}

\affiliation[1]{\href{http://web.science.uu.nl/itf/}{Institute for Theoretical Physics and Center for Extreme Matter and Emergent Phenomena}, Utrecht University, Leuvenlaan 4, 3584 CE Utrecht, The Netherlands}
\affiliation[2]{\href{http://hep.physics.uoc.gr}{Crete Center for Theoretical Physics},
Department of Physics,\\
University of Crete, 71003 Heraklion, Greece}
\affiliation[3]{\href{http://www.apc.univ-paris7.fr}{APC, AstroParticule et Cosmologie}, Universit\'e Paris Diderot, CNRS/IN2P3, CEA/IRFU, \\
Observatoire de Paris, Sorbonne Paris Cit\'e,\\
 10, rue Alice Domon et L\'eonie Duquet, 75205 Paris
Cedex 13, France}

\preprint{CCTP-2018-01,~ITCP-IPP 2018/25}

\abstract{Black hole solutions and their thermodynamics are studied in Einstein-scalar theories. The associated zero-temperature solutions are non-trivial holographic RG flows. These include solutions which skip intermediate extrema of the bulk scalar potential or feature an inversion of the direction of the flow of the coupling (bounces). At finite temperature, a complex set of branches of black hole solutions is found. In some cases, first order phase transitions are found between the black-hole branches. In other cases, black hole solutions are found to exist even for boundary conditions which {\em did not} allow a zero-temperature vacuum flow. Finite-temperature solutions driven solely by the vacuum expectation value of a perturbing operator (zero source) are found and studied.  Such solutions exist generically (i.e. with no special tuning of the potential) in theories in which the vacuum flows feature bounces. It is found that they exhibit conformal thermodynamics. In special theories with a moduli space of vacua (at zero temperature), it is found that finite temperature destroys the existence of the moduli space.}

\begin{document}
\maketitle

\section{Introduction and summary}

The gauge/gravity duality \cite{maldacena,witten,GKP} gives a geometric representation of  the
renormalisation group (RG): the RG flow in a $d$-dimensional field
theory is realised as the radial flow  of $d$-dimensional
hyper-surfaces  which foliate a solution of a  higher-dimensional
(super)gravity theory \cite{Boonstra:1998mp,Girardello:1998pd,Balasubramanian:1999jd,Freedman,deboer,Bianchi:2001de,deHaro:2000vlm,papaskenderis}. When the ($d+1$)-dimensional solution is
 driven by  scalar fields, and when the $d$-dimensional radial slices are flat, holographic RG flows with a regular interior geometry connect
two extrema of the scalar potential $V(\phi)$, where the geometry
approaches $AdS_{d+1}$.  Usually, a maximum
of $V$ maps  to the field theory UV, while a minimum corresponds to
the IR (although in special cases there may be exceptions to this rule, \cite{multibranch}).

In  this context, the holographic RG flow often captures
non-perturbative features of the dual field theory: phenomena
 such as  IR fixed points, confinement, and the condensation of
scalar operators, which are usually inaccessible in
perturbation theory around a UV conformal fixed point,
can be realised   in  holographic RG
flow solutions of simple 2-derivative gravitational theories.

Besides reproducing holographic versions of the
field-theoretical  non-perturbative features mentioned above, holographic RG flows
have been shown to display some unusual, or {\em exotic},
phenomena. Recently, a systematic  exploration of different  classes of these exotic
features has been initiated in \cite{multibranch} for single scalar field
models. This  work focused
on the analysis of asymptotically $AdS$ holographic RG-flow solutions
in $d+1$-dimensional Einstein gravity coupled to a single  scalar
field, dual to renormalisation group flows of a $d$-dimensional CFT
deformed away from the UV by a single operator. This analysis was  extended in
\cite{multifield} to multi-field models.

Depending on the details of the bulk scalar
potential $V(\phi)$, some of the following exotic feature may arise:
\begin{enumerate}
\item {\bf Multiple vacuum flows.}
If $V(\phi)$ displays several maxima and  minima, there may be several
regular flows connecting the same UV maximum to different IR
minima. This corresponds to the existence, in the same UV-deformed
CFT, of multiple vacua which are distinguished (in the UV) by the value of the scalar
condensate $\< O\>$
 and which reach different IR fixed points. 
   Only one of them (the one with lowest free energy) is the true vacuum.

\item {\bf Bouncing RG flows.} These correspond to solutions where  the flow of the scalar field is non-monotonic. For
  example, after starting
  off   away from the UV fixed point in the positive direction,  the
  scalar field reaches a maximum ({\em bounce}), start decreasing again, and  eventually
  reaches an IR fixed point situated on the opposite side of the
  initial UV
  fixed point. At the turning point, the holographic $\beta$-function has a square-root branch singularity.

\item {\bf Irrelevant vev flows.}
Usually, holographic RG flows with a regular interior correspond to
solutions which asymptote in the UV to a maximum of the scalar
potential. However in special cases, if the scalar potential is
appropriately tuned, regular solutions may exist which connect two
minima of the bulk potential (one in the UV, one in the IR). Although
they do not exist generically, these flows are interesting because
they correspond to a deformation of the theory by the vev of an
irrelevant operator (for which turning on a source is not an
option).  These flows  have a continuous moduli space, parametrised by
 the vacuum expectation value (vev) of the condensate $\< O\>$,  with degenerate free energy
(also degenerate with the fixed-point $AdS$ solutions where $\<
 O\>=0$). For $\< O\> \neq 0$, these solutions break conformal invariance
 spontaneously and the presence of a moduli space indicates that
 there is a massless dilaton in the spectrum of excitations, \cite{dilaton1}.
\end{enumerate}

The discussion above refers to vacuum (i.e. Poincar\'e invariant) solutions. Probing these
features by turning on additional sources, or by considering non-vacuum
states, is important in order to test the robustness and consistency of these
solutions, most of which arise  in bottom-up holographic models. A
first step in this direction was taken in \cite{curvedRG}, where
solutions were considered in which the dual field theory lives on a
maximally symmetric curved manifold: in this case,  the relevant gravity dual solutions are
 holographic RG flows in which each  slice transverse to the
 (holographic) radial coordinate is a maximally symmetric curved
 space-time.

Turning on non-zero curvature
introduces a new source on the boundary CFT, and correspondingly a new
scale beyond the (dimensionful) coupling of the relevant CFT
deformation. As  shown in \cite{curvedRG}, this leads to several
interesting new effects. For  theories admitting  multiple
vacuum RG flows (point 1 above) a curvature-driven quantum phase
transition  is found. Also, certain CFT deformations which do not correspond to
 any regular vacuum solution become allowed if a sufficiently
large curvature is turned-on.

A different way to probe these models is by turning on  finite
temperature and this is the subject of the present
work. Specifically,  we
consider black-hole solutions of the bulk Einstein-scalar
theories, like those discussed in \cite{multibranch}, which allow
``exotic'' vacuum  RG-flow solutions\footnote{A specific example of the black hole solution obtained by finite T extensions of these vacuum RG flows were first studied in \cite{Gursoy:2016ggq,Janik:2016btb}.}.

In the same way as space-time curvature,  a finite temperature
introduces a new scale in the system. Indeed, many of the results we
describe here (space of solutions, phase diagrams) closely resemble
those found in \cite{curvedRG}, if we simply replace curvature by
temperature as a control parameter.

However there is a very
important  difference: while
a non-zero curvature is a modification of the theory itself, i.e. it
introduces new sources (in this case non-trivial components of the
metric, which turns on a relevant  deformation proportional to the stress
tensor), going to  finite temperature  corresponds to considering
different {\em states} of the same theory whose vacuum state is a
Poincar\'e-invariant solution. 

Indeed, one of the main motivation{\bf s} for considering states beyond the
vacuum is to probe whether the theories displaying exotic
behaviour{\bf s} discussed above (and in particular the ``bouncing'' geometries)   are ``good'' holographic theories. From the
analysis in \cite{multibranch}, no pathology emerged
neither in the vacuum (e.g. singular nature of some curvature
invariant), nor in the spectrum of small fluctuations
around the vacuum, which was shown in general to be perturbatively
stable. We note however there may be dynamical instabilities found in the corresponding finite temperature blackhole extensions of these solutions, as was first observed in \cite{Gursoy:2016ggq,Janik:2016btb}.

By going to finite temperature, we will probe different states of the  theory in a way which is
complementary to turning on small excitations above the vacuum. In
this way, we will test the consistency from the  thermodynamic
standpoint. As we will see, this will enable us to detect, in some
cases, certain pathologies which indicate that some of the holographic
models under investigation  may not be self-consistent after all.

A particularly interesting question concerns the fate of the
irrelevant operator flows discussed in point 3 above. As we will see,
at $T\neq 0$
no regular black hole solutions with
the same UV asymptotics at the minimum of the potential can be found.
This does not come as a surprise as we expect that turning on finite temperature in a theory with a moduli space, lifts generically the moduli space and the moduli acquire a non-trivial potential.

However, surprisingly, we find that finite temperature allows new
irrelevant vev flows from a minimum to exist for {\em generic}
(i.e. non-finely tuned) potentials. These solutions display interesting
properties such as a conformal thermodynamics and temperature-driven
operator condensation and they may also have
interesting hydrodynamic properties.

\subsection{Summary of results}

In this work we consider $(d+1)$-dimensional Einstein gravity
coupled to a scalar field $\phi$. The action contains  a scalar potential $V(\phi)$ with
 several AdS extrema, which  play the role of UV and IR fixed points
for asymptotically AdS RG-flow  solutions. These are dual to RG flows driven by deforming  a UV CFT by a
relevant operator $O$, of dimension $\Delta<d$. Its coupling $j$ is
encoded in the leading term of the UV asymptotics of the scalar field
about the fixed point value $\phi_{UV}$,
\be\label{intro0}
\phi(u) \simeq \phi_{UV} + j \, \ell^{(d-\Delta)} \exp\left[(d-\Delta)u/\ell
\right], \qquad u \to -\infty,
\ee
where $\ell^2 = -d(d-1)/V(\phi_{UV})$ is the  squared UV AdS length.

We consider the theory
at finite temperature by looking at static, planar black hole solutions, of the form
\be \label{intro1}
ds^2  = {du^2 \over f(u)} + e^{A(u)} \left[- f(u) dt^2 + \delta_{ij}
  dx^i dx^j\right], \quad  \phi = \phi(u)
\ee
where $u$ is the holographic radial coordinate and $(t,x^i)$ are the
space-time coordinates of the dual field theory, whose temperature equals the Hawking temperature of the black hole. 

The main goal of this paper is to  analyse the space of solutions of
the form (\ref{intro1}) and study their thermodynamics in the canonical
ensemble, i.e. at fixed temperature $T$ and fixed UV coupling
$j$.

The main features of the solutions depend only on
the dimensionless parameter ${\cal T} = T
|j|^{-1/(d-\Delta)}$. Requiring the solution to be regular, both
${\cal T}$ and
 the dimensionless vev parameter $C \equiv \< O\>
 |j|^{-\Delta/(d-\Delta)}$ are completely determined by the value the
 scalar field takes at the horizon, where $f(u)=0$.
 Therefore, a useful way to scan the space of
 black hole solutions is by changing the horizon endpoint $\phi_h$ of the flow,
 and studying  the behaviour of the functions ${\cal T} (\phi_h)$ and $C(\phi_h)$.
The free energy is then computed by standard holographic techniques.
We should note that $\phi_h$ parametrises in a faithful way the black-hole solutions, as there is at most a single solution for each value of $\phi_h$.

After a brief general discussion of the thermodynamics of holographic
 RG-flows  in Section \ref{s2}, we study black-hole solutions
 and their thermodynamics in two specific models:
\begin{enumerate}
\item[a)]  The first one
admits two distinct regular  vacuum RG-flows from the same UV theory
to two different IR fixed points  (Section \ref{sec:skipping}).
\item[b)]
Next we turn to a  models for which the vacuum
RG-flow presents a bounce  (Section \ref{sec:bouncing}).
\end{enumerate}
Finally, in section  \ref{sec:vev} we
consider the special black-hole solutions with $j=0$,  which correspond to
black holes   driven only by  the vev of the deforming operator, and
have special properties.

In the rest of this subsection  we briefly summarise our results.

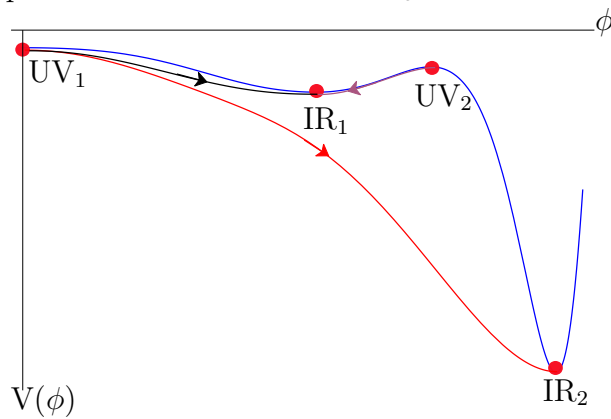
\begin{figure}[h!]
\centering
\begin{overpic}
[width=0.5\textwidth]{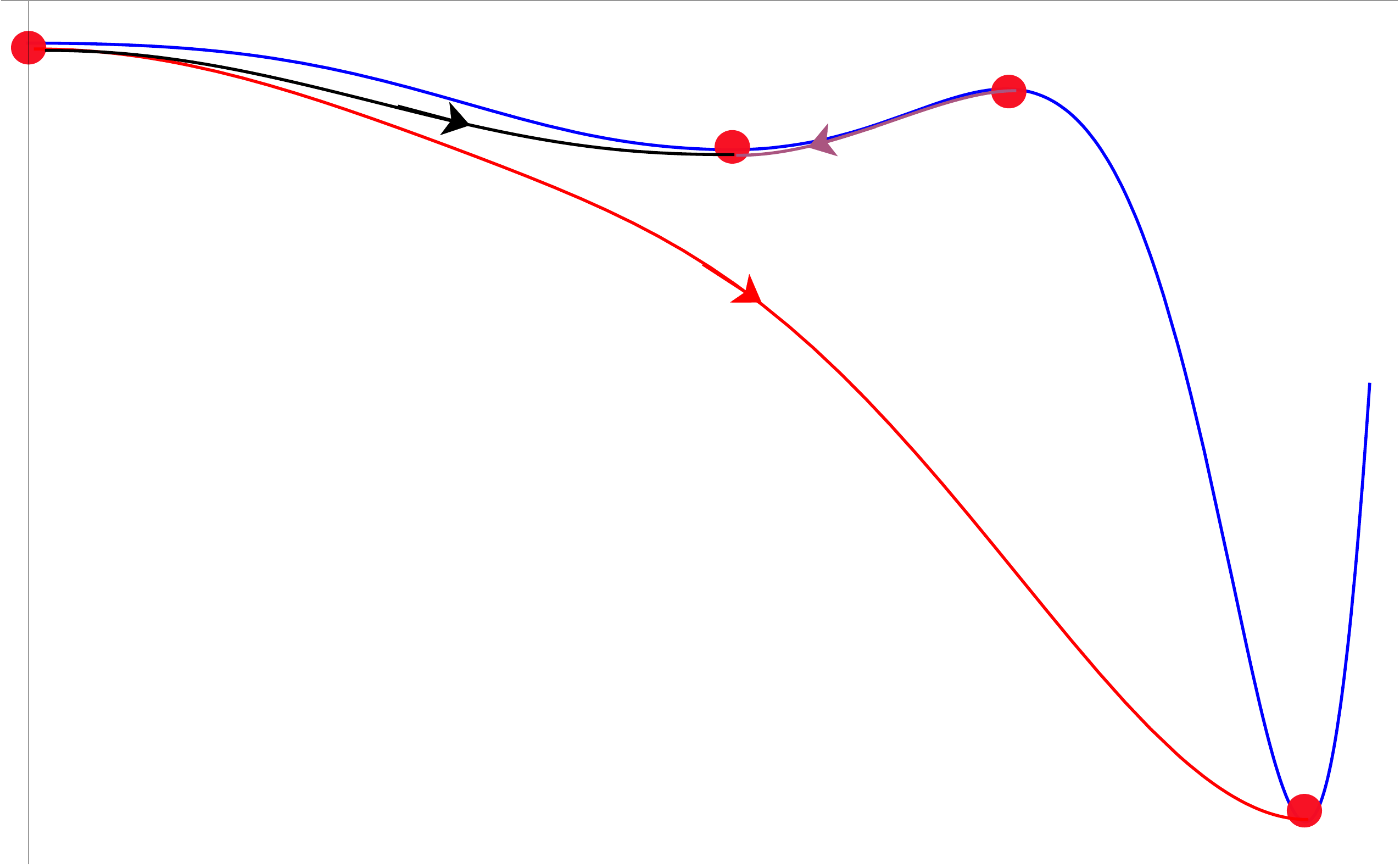}
\put (0,-3) {V($\phi$)}
\put (3,53) {UV$_1$}
\put (50,45) {IR$_1$}
\put (69,49) {UV$_2$}
\put (91,-1.5) {IR$_2$}
\put (100,62) {$\phi$}
\end{overpic}
\caption{Schematic plot of the  potential which allows skipping
  solutions, with a representation of the corresponding  zero-temperature RG flows connecting different fixed points.}
  \label{skipintro}
\end{figure}

\paragraph{First order transitions in multi-vacuum theories.} The
first model we consider has  a potential  shown schematically in  Figure \ref{skipintro}.  For $T=0$, there  are
two regular solutions  connecting the fixed point  UV$_1$ situated at the origin with
each of the two IR fixed point at the two minima. In
\cite{multibranch} it was shown that the favoured solution is the {\em
  skipping} one, i.e. the one that does not stop at IR$_1$
but  reaches the next available fixed point IR$_2$ . At finite, fixed  temperature $T$
we find there are up to three competing black-hole solutions, all with UV
asymptotics at the origin. Two of them skip IR$_1$, but
exist only up to a maximal temperature $T_{max}$. Calculation  of the free energy shows a
first order phase transition at $T_c< T_{max}$ above which the {\em
  non-skipping} solution becomes dominant. This solution  
continuously connect to the zero temperature flow ending at
IR$_1$. Therefore, in this model there is a transition between
skipping and non-skipping behaviour as the temperature is increased
above $T_c$.

\paragraph{Thermal desingularisation  of ill-defined vacua.}
We next consider the same potential shown in figure \ref{skipintro},
but focusing on the black holes which asymptote in the UV to the fixed point
UV$_2$. For $T=0$, there is  only one regular RG flow going from UV$_2$
to IR$_1$, but no solution reaching IR$_2$ from  UV$_2$. From the dual field theory
standpoint, this means that we can
only deform the UV$_2$ CFT for $j<0$, but there is no well-defined
vacuum with $j>0$. This is not uncommon in  perturbative field theory,
i.e. pure YM theory exists only for $g^2>0$ or $\lambda \phi^4$ theory exists only for $\lambda >0$.

At finite temperature the situation is richer:
\begin{enumerate}
\item For $j<0$ we find two black-hole solutions at all temperatures,
  one of which displays a bounce (i.e. an inversion in the flow of the
  scalar field). Calculation of the free energy shows that the
  dominant solution is the non-bouncing one at all temperatures
\item For $j>0$, although there was no regular solution at $T=0$,
    two black-hole solutions appear {\em above a minimal
  temperature $T_{min}$}. This implies that, at high enough temperature,
the theory  with $j>0$ may be well-defined, though the
zero-temperature vacuum did not exist.
\end{enumerate}

\begin{figure}[h!]
\centering
\begin{overpic}
[width=0.5\textwidth]{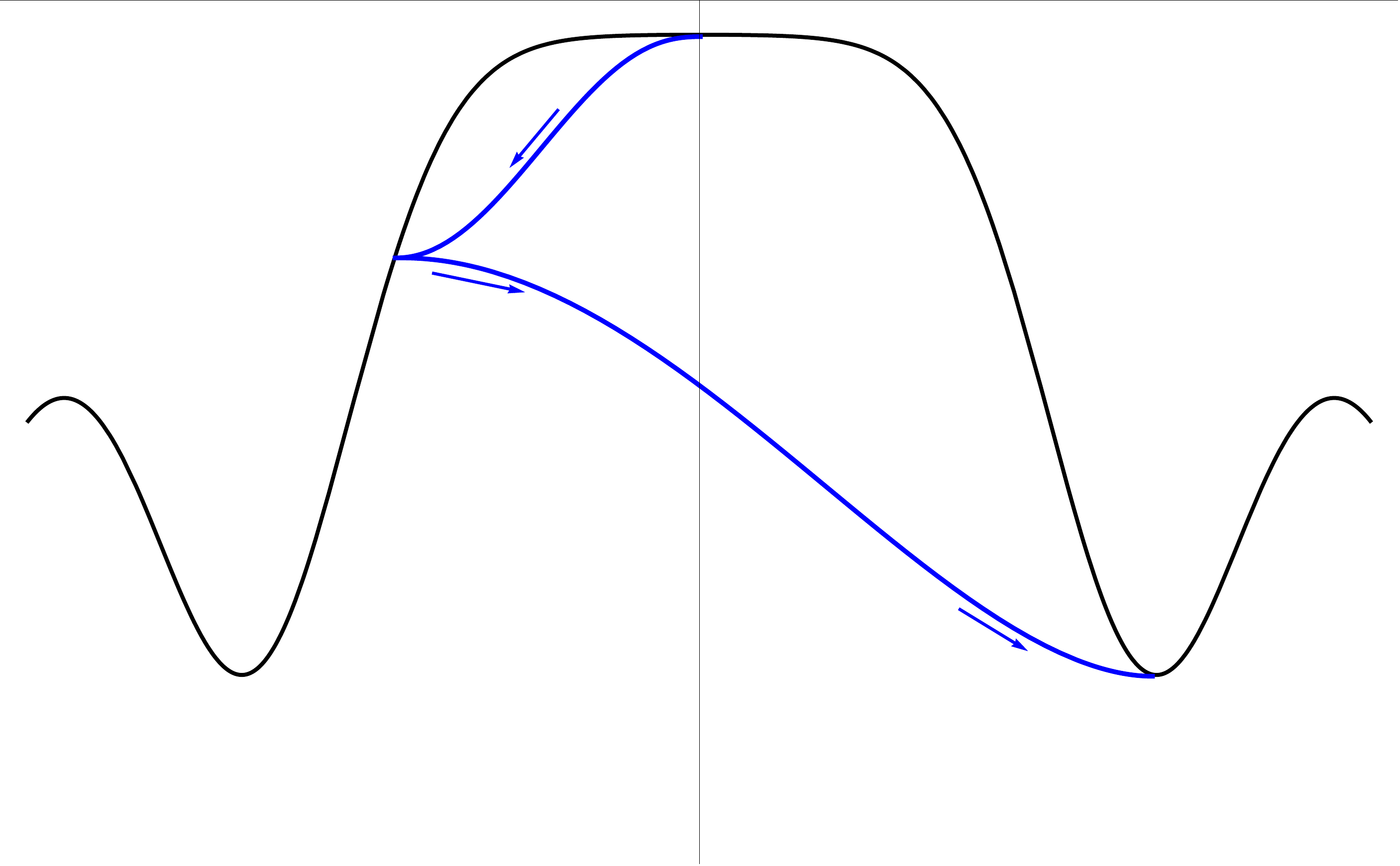}
 \put (48,-4) {V($\phi$)}
 \put (47,62) {UV}
 \put (79,7) {IR}
 \put (100,62) {$\phi$}
\end{overpic}
\caption{Sketch  of a bouncing RG-flow at $T=0$, with $j<0$.}
  \label{bounce-intro}
\end{figure}
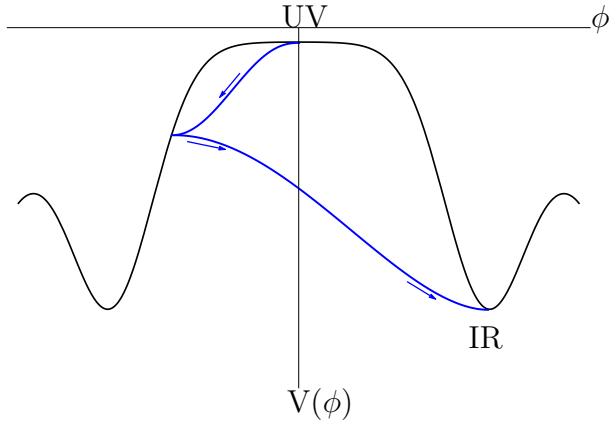

\paragraph{Theories with bouncing
  vacua and their thermodynamic (in)consistency.}
The second example we analyse in detail is a potential for which, at
zero temperature, the RG-flow solution {\em bounces}:  the scalar
field starts decreasing away from the UV,  reaches a minimum, then it
increases again  past the starting point to reach an IR fixed
point on the opposite side (see figure \ref{bounce-intro}).

For the potential considered in  Section \ref{sec:bouncing}, the flow
represented in figure \ref{bounce-intro} is the only regular solution
for $T=0$ and $j<0$ (for $j>0$ we have its mirror solution with $\phi
\to -\phi$, as the potential we are considering is an even function of
$\phi$).
At finite temperature, for fixed $j$, we find up to {\em five}
different black-hole branches. One of them connects to the vacuum
$T=0$ solution, one connects with the AdS-Schwarzschild black hole at
the origin as $T\to +\infty$. At low temperature, all solutions
exhibit a bounce as in figure \ref{bounce-intro}, while above a
certain  temperature, new solutions appear which do not bounce, but
have horizon (for $j<0$) on the negative side of the UV fixed point.

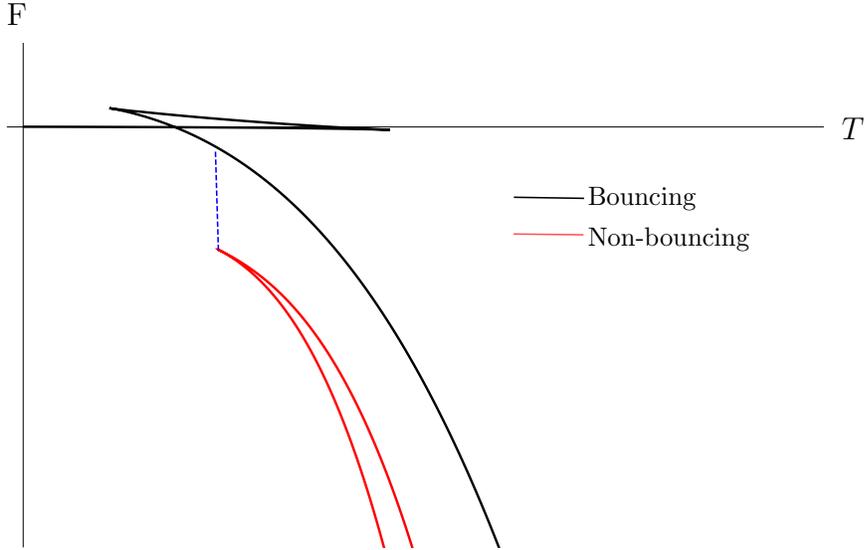
\begin{figure}[h!]
\centering
\begin{overpic}
[width=0.7\textwidth]{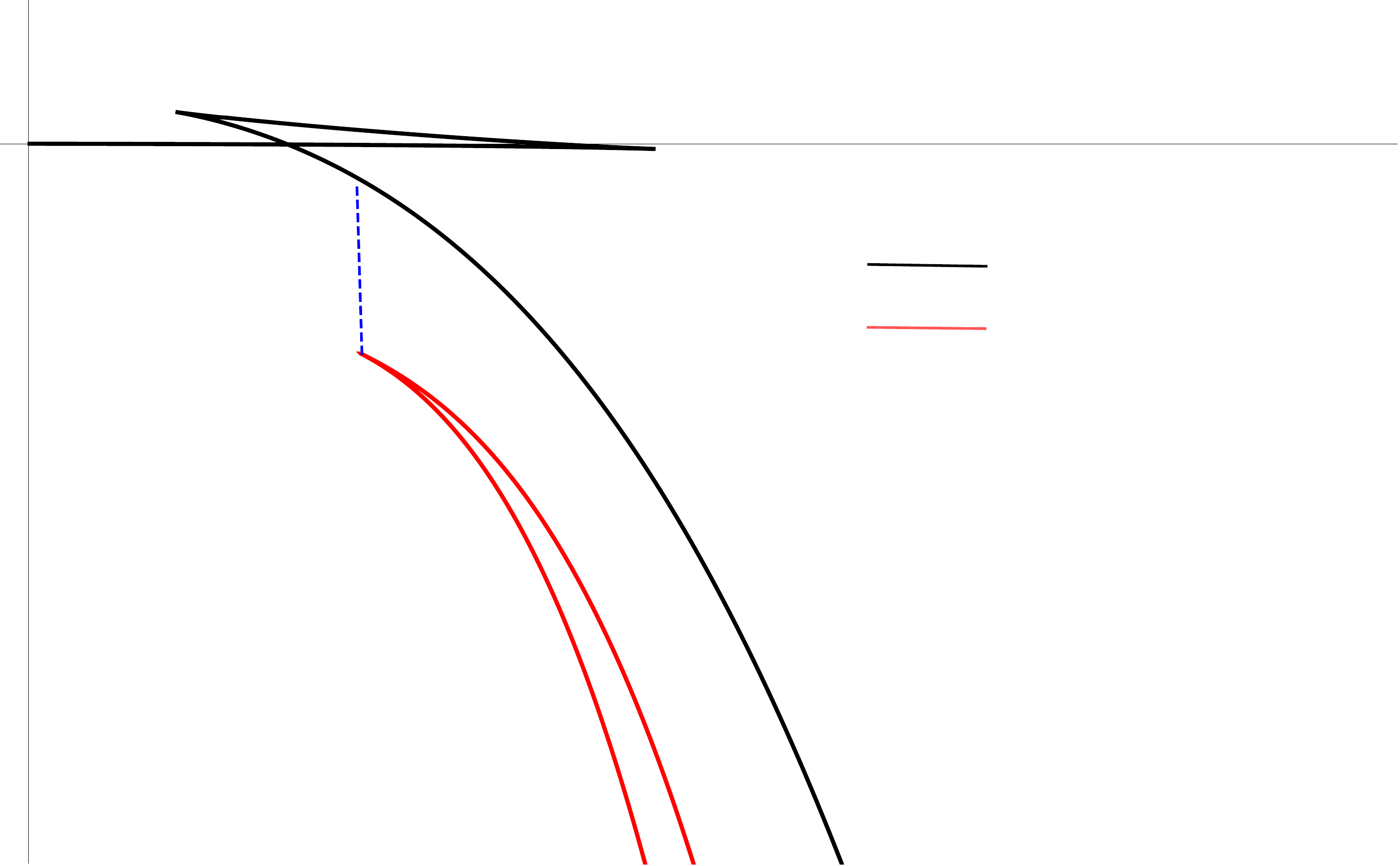}
 \put (0,64) {{\cal F}}
 \put (71,42) {{\footnotesize Bouncing}}
 \put (71,37) {{\footnotesize Non-bouncing}}
 \put (102,50) {$T$}
\end{overpic}
\caption{Phase diagram of $j<0$ black holes in the model with bouncing
vacuum solution. Bouncing black holes dominate at low temperature,
then the  free energy $F$ jumps by a finite amount (dashed blue line)
on the non-bouncing branch.}
  \label{jump-intro}
\end{figure}

The computation of the free energy reveals a puzzling situation. While
at high temperature the dominant solution is, as one may have
suspected, the large black hole whose horizon is closest to the UV
fixed point, the transition to this solution is discontinuous: the free
energy shows a jump  from the bouncing to the non-bouncing solution as
soon as the latter one appears (see figure \ref{jump-intro}) . This situation does not allow for a
consistent Maxwell construction of the phase diagram, and it may
indicate  that this is not a good
holographic theory, albeit the vacuum was found in \cite{multibranch} to be perturbatively
stable {\em and} the dominant branch at finite $T$ is also
thermodynamically stable (the specific heat is positive). Other options are also possible, as we discuss in section \ref{sec:disc},   and at this stage we cannot determine with certainty the reason behind the unusual behaviour.\\

\paragraph{Vev-driven black holes.}
When the black-hole horizon $\phi_h$
approaches a UV or an IR fixed point, this corresponds to a high-temperature or low-temperature limit, respectively. However, in the
space of solutions,  new infinite-temperature and zero-temperature limits
 appear which
are not connect to any fixed point solution and for which the
horizon is far from  the extrema of $V$. These limits signal the
existence of {\em vev-driven black holes} for which, at fixed $T\neq 0$, the
source of the deforming operator is set to zero.

These solutions arise in two different situations:
\begin{enumerate}
\item At the interface between bouncing and non-bouncing black holes branches
  asymptoting to the same UV maximum of $V(\phi)$, since at this
  interface the source $j$ changes sign;
\item When the bounce in a solution coincides with  a minimum of the potential.
\end{enumerate}
In the latter case, the $j=0$ solution can be shown to asymptote in
the UV to a {\em minimum} of the potential rather than a
maximum. The corresponding deformation is driven by an {\em
  irrelevant} operator, unlike the case  in which the UV
fixed point is a maximum of $V(\phi)$. Vev-driven black holes are
isolated in the space of solutions in the following sense: they
correspond to special values of $\phi_h$ at which the dimensionless
temperature ${\cal T}(\phi_h) \to 0,+\infty$.

Vev-driven flows have a very simple thermodynamics, which turns out to
be exactly conformal, with the free energy given, for all $T>0$, by
$$
{\cal F}  = -\sigma T^d,
$$
where $\sigma$ is a temperature-independent coefficient.
In addition, the vev of the operator dual to $\phi$ is completely
determined by the temperature,
$$
 \< O \>  = c T^{\Delta},
$$
where $c$ is another constant.

Finally, we analysed a  particular case (with a specially tuned potential) where
a regular vev-driven flow solution  does exist for $T=0$, and was
constructed in   \cite{multibranch}. The corresponding vacuum flow connects
two minima of $V$ and it provides a zero-temperature example of a
regular flow driven by the vev of an irrelevant operator. We find  that in this case,
after turning on temperature, one cannot find any regular  black-hole
solution in which the scalar has a non-trivial flow.

\subsection{Discussion and outlook} \label{sec:disc}

The results described  in the previous section show that the space of black-hole solutions,
built around holographic RG flows may have an extremely rich
structure and may display unexpected phenomena.

The  structure of the different branches  of solutions at
finite temperature  closely parallel similar structures that were found
in \cite{curvedRG} when the dual field theory is defined on a positively curved
space-time. Furthermore, as we will discuss below, both curvature and
temperature  destroy moduli spaces which can be found at zero
temperature for specially tuned potentials.

This similarity is not surprising, as in some sense the theory responds
in the same qualitative way to the introduction of an additional dimensionful
parameter, be it curvature or temperature.
As we have  mentioned however, in the case of temperature  we
are dealing with different states in the same theory, whereas
curvature introduces a change in  the definition of the theory itself.
Therefore, the finite temperature analysis can tell us something about
the consistency of the theory itself,  if we
 require that, given a consistent QFT,  it should always be possible to
 couple it  consistently to a thermal bath.

This consistency criterion puts strong constraints on  theories where the vacuum state is
a  bouncing RG-flow. These solutions  were shown to be regular and free of
instabilities \cite{multibranch}.  At finite temperature however, the phase
diagram shows a jump in the free energy by a finite amount (figure
\ref{jump-intro}),  which
should not be allowed in a consistent Maxwell construction. The
interpretation is open to discussion: one possibility is that this
kind of potentials lie in a holographic ``swampland'' and result from
an inconsistent truncation of a more complete theory, e.g. a
multi-field model in which the dynamics of the extra scalar cannot be
neglected. Another possibility is that these models may be consistent
but there are other phases, which we have neglected, and which make
the phase diagram well-behaved. Since we have exhausted all spatially
homogeneous solutions, one option  is that these new phases
break rotational or translational invariance (e.g. they may be striped phases). An instability to
one such phase may be signalled by unstable quasi-normal modes, as was
indeed found in models with bounces \cite{Gursoy:2016ggq}.
We leave these questions for future work.

We have also observed the opposite phenomenon: certain deformations of
a CFT, which are not allowed at zero-temperature (because they do not
lead to  regular solutions),  lead  instead to consistent solutions
above a certain $T$. This {\em thermal desingularisation} occurs in
our examples around asymmetric extrema, where only one sign of the
source leads to  a consistent RG flow. The field theory interpretation
is that, presumably, the ``wrong sign'' deformation is
inconsistent because of some infrared instability. This is eliminated
at a sufficiently large temperature, which effectively acts as an IR
cutoff. In fact, the same feature was observed if instead of a thermal
state we put the theory on a sphere with sufficiently small radius
\cite{curvedRG}. It would be interesting to better understand the details of
this mechanism from the field theory point of view, and/or in a
top-down model.

A particularly interesting class of special solutions, which can be
seen as separating various different branches of black holes, are the
vev-driven  black holes with $j=0$. These are special solutions with a
fixed value of the horizon parameter $\phi_h$. They exist at any non-zero
temperature and exhibit conformal thermodynamics.   Interestingly,
 at zero temperature, purely vev-driven holographic
RG-flows are generically singular bulk solutions. In other words, for $j=0$, the only regular
solution is AdS space, with constant scalar field and $\< O\> =0$, except if $V(\phi)$
has some tuned parameters. In contrast, for $T\neq 0$, existence of regular
vev-driven black holes does not require tuning the potential.
 This is another example of what we
referred above by ``thermal desingularisation,'' whereby a  regular
black-hole solution has no regular vacuum counterpart with the same
value of the sources.

Since they have $j=0$, vev-driven black holes satisfy the same
  UV boundary conditions as the AdS-Schwarzschild black holes with
  constant scalar field (fixed at an extremum of the potential) and
  the same temperature: these solutions are therefore in thermodynamic
  competition with each other. In all cases we have considered, it is
  the Schwarzschild black hole which dominates the canonical ensemble
  at all $T\neq 0$. It is an open (and interesting) question whether this
  is generic in Einstein-scalar theories, or whether there may be
  cases in which the vev-driven black hole is the dominant
solution. Because of the relation $\<O\>\propto T^{\Delta}$ in these solutions, these would
provide a holographic  example of temperature-driven condensation of a
scalar operator\footnote{Examples of this kind exist for  charged
  black holes and are the basis for holographic superconductors,  where
condensation  of the scalar operator can be understood as due to an IR
instability of the AdS-Schwarzschild solution \cite{super}.}.

For those models where a regular vacuum
vev-driven flow {\em does} exist (as in the non-generic potentials
considered in \cite{multibranch}, which allow for minimum-to-minimum
holographic RG flow solutions), the situation is quite different from
the one described above. At $T=0$, there is a
one-parameter family of solutions, parametrised by the arbitrary value
of $\< O\>$, all flowing between the same  two minima  of the
potential. All these solutions are  degenerate in free energy,
therefore forming a moduli space and admit a massless dilaton excitation (the
Goldstone mode of spontaneously broken conformal invariance).
Going to $T\neq 0$, the entire moduli space disappears: in the example
we have studied, there are no black-hole solutions with a
non-trivial flow of the scalar field,
which reach the same UV as the vacuum solutions.  This  means that
finite temperature destroys the moduli space, and leaves the
AdS-Schwarzschild black hole, with constant $\phi$ and $\<O\>=0$, as
the only solution\footnote{This has an analogy in weakly-coupled field
theories with a moduli space, where at finite temperature an
effective potential of the form  $m^2(T)\varphi^2$ may be generated, with $\varphi$ a scalar representing the appropriate scalar operator.
This leaves as the only minimum the one at  $\<\varphi\>=0$.}.
It is an open question  whether this behaviour is generic, or specific to the model we considered:  in general, one cannot exclude that another
branch of regular vev-driven solutions could appear, with conformal
thermodynamics and $\< O\> \propto T^\Delta$, which would still lift the
moduli space but have a non-trivial flow.

The existence of regular $j=0$ black holes seems tied to
the presence of bounces, because the transition across $j=0$ occurs
between bouncing and non-bouncing solutions. It is unclear at present
under which conditions, given a generic extremum of $V$, regular vev-driven
black holes will or will not exist. For example, in the model with the
potential shown in figure \ref{skipintro}, purely vev-driven black
holes exist which asymptote the points UV$_2$ and IR$_1$, but not
UV$_1$. It would be interesting to understand what features of the scalar
potential determine the existence or non-existence of these
solutions.

More generally, it is an interesting but highly non-trivial question  to understand
which features of the potential determine whether the vacuum solution
will bounce, or skip a fixed point.  Although some qualitative
criteria can be roughly guessed by the experience with different cases
(for example, a ``steeper'' potential is more likely to admit bouncing
vacuum flows) it would be very interesting to obtain some quantitative
criteria similar to those existing for other phenomena
(e.g. confinement).

\subsection*{Note Added}
During completion of this work, we became aware that a study very
similar to ours was  being performed independently by Y. Bea and D.
Mateos  \cite{David}. The results of that work are in
agreement with those presented here.

\section{Holographic RG flows at finite temperature\label{s2}}

In this section we consider the finite-temperature generalisation of
the exotic RG flows found in \cite{multibranch}.
In that paper,  solutions of
$(d+1)$-dimensional  Einstein-Scalar gravity were considered,  which
corresponded to  holographic RG flows of the dual field theory {\em in
  the vacuum}, i.e. those solution had full $d$-dimensional Poincar\'e
invariance.  In the following  subsections we review the
finite-temperature generalisations of such solutions, which contain a
black hole in the interior and are only symmetric under the $d-1$ Euclidean
group plus time translations and we discuss the corresponding
thermodynamics in terms of the free energy and of the thermal
effective potential.

\subsection{Black-hole solutions}

We consider the two-derivative action of gravity coupled to a single
scalar field, with a generic potential $V(\phi)$.
\begin{subequations}\label{b0}
\begin{align}
& S = - M_P^{d-1}\int_{\MM} \dd^{d+1}x \sqrt{-g} \left[
R
-\frac{1}{2}g^{ab}\partial_a \phi\partial_b\phi -V(\phi)
\right]+ S_{GHY} + S_{ct}
~,\\
& S_{GHY}=-2M_P^{d-1} \int_{\partial \MM} \dd^dx \sqrt{\gamma} K\, .
\end{align}
\end{subequations}

The sign of the action is the one appropriate for Euclidean
signature. $S_{GHY}$ is the  Gibbons-Hawking-York boundary
term. $S_{ct}$ is an extra boundary term which is needed for
holographic renormalisation and will be specified later.

We write the  Euclidean black-hole solutions in the form:
\be
\label{b1}
\dd s^2=\frac{\dd u^2}{f(u)}+e^{2A(u)}\le(f(u)\dd t^2+\dd x^i \dd x^i\ri)~, ~~
\phi=\phi(u), \quad i = 1\ldots d-1,
\ee
where $f(u)$ is a monotonically decreasing function taking values
between zero and one
, and $t\sim t+\beta$ where $\beta$ is the inverse
temperature.

Einstein's equations read
\begin{subequations}\label{b3}
\begin{align}
&
	f \ddot\phi +\le( {\dot f} +d~ f \dot A
	\ri)\dot \phi
	-{d V\over d\phi}=0
\label{b3a}
\\
&
(d-1)\dot f\dot A+\le(d(d-1)\dot A^2-\ha \dot\phi^2\ri)f+ V(\phi)=0
\label{b3b}
\\
&\ddot f+d \dot A \dot f=0
\label{b3c}
\\
& 2(d-1)\ddot A +\dot \phi^2=0
\label{b3d}
\end{align}
\end{subequations}
The equations of motion are invariant under the following transformations:
\begin{subequations}\label{inv}
\begin{align}
&A\to\tilde  A= A-\bar A \label{inv1}
\\
& u\to \tilde u=u+v \label{inv2}
\\
&\le( u, f( u)\ri)\to \Big(\tilde u,\tilde f(\tilde u)\Big)= \Big(\l
 u,\l^2  f\le( u\ri)\Big) \label{inv3}
\end{align}
\end{subequations}
where $\bar A$, $v$ and $\l$ are constants.

The metric (\ref{b1}) describes planar black holes, whose horizon is located at $u_h$, i.e. $f(u_h) = 0$. Temperature
and entropy density (per unit $d-1$-volume $V_{d-1}$) are given, respectively, by:
\be \label{b1-1}
T = { e^{A(u_h)} \over 4\pi} |\dot{f}(u_h)|, \qquad s = 4\pi M_p^{d-1}
e^{(d-1)A(u_h)}.
\ee
Euclidean  time is compactified on a circle of length $\beta = 1/T$.
The zero-temperature case  (no black hole) corresponds to taking
$f(u)=1$. The $AdS$ black-hole solution corresponds to taking constant
$\phi=\phi_0$ such that $V'(\phi_0) = 0$. Defining $\ell \equiv \sqrt{-d(d-1)/V(\phi_0)}$,  this soluton  is
\be
\label{b2}
A(u)=-{u\over \ell} ~, ~~
f(u)=1 - e^{d(u- u_h)/\ell}, ~~ \phi(u)= \phi_0 ~ .
\ee
and
\be \label{Tsconf}
T_{conf} =   {d\over 4\pi \ell}
e^{-u_h/\ell}, \qquad  s_{conf} = 4\pi M_p^{d-1} e^{-(d-1)u_h/\ell} .
\ee

In the general case, equation  \eqref{b3c} can be integrated once to obtain
\be
\dot f(u) e^{dA(u)}= - D , 
\label{a7}\ee
where $D$ is a non-negative integration constant. We can relate it to the black-hole
temperature and entropy by evaluating equation (\ref{a7}) at the
horizon $u=u_h$ and using equation (\ref{b1-1}), leading to
\be
\label{a10}
D= {T s \over M_p^{d-1}}.
\ee

We restrict to  solutions which reach an asymptotically AdS region
(UV) where the scalar field approaches  a maximum $\phi_{UV}$ (which
without loss of generality we take to be at $\phi=0$) of the potential, i.e.
\eql{V}{
V(\phi) \simeq -{d(d-1)\over \ell^2} +{m^2 \over 2} \phi^2 + \ldots
}
In this region the solution  takes the asymptotic form as $u  \to
-\infty$,
\begin{subequations}\label{max0}
\begin{align}
&\phi= \phi_-\,e^{\D_- u/\ell}+\dots+ \phi_+\,
e^{\D_+ u/\ell}+\dots,  \qquad \Delta_\pm =
{d\over 2} \pm{1\over 2} \sqrt{d^2 + 4 m^2 \ell^2},  \label{max0a}
\\
&A(u)=-{u\over \ell}+ \dots \label{max0b}
\\
&f(u)=1 - {\ell D \over d}e^{du/ \ell}~ + \ldots\label{max0c}
\end{align}
\end{subequations}
 The parameters  $\phi_-$  and
 $\phi_+$ are related to  the
 UV coupling $j$ and to the vev of the
dual operator $O$ (whose  dimension\footnote{We will only discuss ``standard
  quantisation'', where the dimension of the deforming operator
  $\Delta \geq d/2$.} is $\Delta=\Delta_+$) by
\be \label{vev}
  \phi_- = j \, \ell^{\Delta_-}, \quad  \phi_+ = { \langle O
    \rangle   \ell^{\Delta_+} \over
  (M_p \ell)^{d-1}  (d-2\Delta_-) }.
\ee
 Finally, the constant $D$ is related to the
 temperature and entropy by equation (\ref{a10}). Generically
 $\phi_-\neq 0$ and
 the flows are driven by a deformation of the CFT by adding a source
 to a  relevant
 operator. For special
 solutions with $\phi_- = 0$  the asymptotic expansion starts at
 sub-leading order with $\phi_+$.  These flows  are driven by a vev of the
dual operator, and as we will see they play a special limiting role  in the
space of solutions.

\subsection{Dimensionless thermodynamic parameters}\label{ssec:dim}

It is useful to classify black-hole solutions in terms of a
dimensionless and diff-invariant quantity. One useful choice is  the horizon value of the scalar field,
\be
\phi_h \equiv \phi(u_h).
\ee
As was shown in \cite{gkmn,Gursoy:2016ggq} and explained in detail in appendix
\ref{app:first}, $\phi_h$ determines the
dimensionless quantity
\be\label{FE8}
{\cal T} \equiv {T \over |j|^{1/\Delta_-}},
\ee
Thus, for each $\phi_h$  there is a one-parameter family of black-hole
solutions with fixed ${\cal T}$, and fixing either the temperature or
the UV source $j$ selects a single solution in this family. In
other words, we can build a map
\be\label{FE9}
(\phi_h, j) \rightarrow T (\phi_h, j) = j^{1\over \Delta_-}
{\cal T}(\phi_h).
\ee

In what follows, we will use ${\cal T}$ itself,  rather than $\phi_h$, as an
independent parameter\footnote{Notice that the map ${\cal T}(\phi_h)$
  is singled-valued, but not necessarily invertible. Therefore, we can
  use ${\cal T}$ as an independent parameter piecewise, i.e. there may
be more than one black-hole branch with the same values of ${\cal T}$
but different $\phi_h$. This is usually the case when using
temperature to parametrise
asymptotically AdS black holes.}, since the former is directly related to the
boundary quantities $T$ and $j$.
Similarly, any physical quantity measured in units of $j$
(entropy, vev, etc) only depends on ${\cal T}$ (or equivalently on
$\phi_h$). This is the case in particular for the rescaled entropy density
\be\label{FE10}
 {\cal S}( {\cal T}) \equiv {s\over |j|^{(d-1)/\Delta_-}}.
\ee

\subsection{First order formalism}

To classify black-hole solutions we will often resort to a first order
formalism. This is a finite-temperature extension, developed in
detail in appendices \ref{app:first} and \ref{AppXY}, of the
standard first order formulation of holographic RG flows, \cite{deboer,hami}.

Black-hole solutions can be  classified in terms of  a {\em
  superpotential}, a  function $W(\phi)$ which determines the scale
factor and  scalar field profile by the equations
\be\label{f1}
\dot{A}(u) = -{1\over 2(d-1)} W(\phi(u)),  \quad \dot{\phi}(u) = {dW
    \over d\phi}(\phi(u)).
\ee
The superpotential $W(\phi)$ and the blackness function $f(u)$
(more precisely, the function $f(\phi)$ defined such that $f(u) = f(\phi(u))$ ) are  determined
together by solving a coupled non-linear third-order system of
equations in the independent variable $\phi$ , equations
(\ref{a8}-\ref{a9-ii}). Because this system couples $W$ to $f$,
$W(\phi)$  depends on temperature.

Below we list the most important properties of the superpotential
$W(\phi)$  which we will be useful for our analysis
\begin{enumerate}
\item
Imposing regularity at the horizon, the functions $W(\phi)$ and $f(\phi)$
 are completely specified by assigning a single parameter, i.e. the
 value $\phi_h$  of the scalar field at the horizon, or equivalently (at least piecewise) the dimensionless
temperature parameter ${\cal T}$.
\item
The superpotential is monotonically increasing along the
flow, as  a function of the holographic coordinate $u$.  However it
can be multi-valued as a function of $\phi$  if the
scalar field profile $\phi(u)$ is non-monotonic \cite{multibranch}.
\item  For source-driven flows, i.e. solutions with $j \neq 0$ in
  equation (\ref{max0a}),  the superpotential (denoted in this case by
  $W^-(\phi)$) takes the form of
  a universal (i.e. ${\cal T}-$independent) analytic expansion around the boundary value $\phi=0$, plus  a one-parameter, ${\cal T}-$dependent, non-analytic
  contribution controlled  by an integration constant $C({\cal T})$
  (or equivalently, $C(\phi_h)$),
\be \label{f2}
W^-(\phi)  = {2(d-1) \over \ell} + {\Delta_-\over 2\ell} \phi^2 +
{\cal O}\left(\phi^4\right) + {C({\cal T})\over \ell} |\phi|^{d/\Delta_-} \Big[1 + O(\phi)\Big] +\ldots
\ee
where the ellipsis refers to terms of higher non-analytic order
\cite{KN,Bourdier} but with no new free parameter. The
value of the source $j$ enters the full solution as the integration
constant of the flow equation (\ref{f1}) for $\phi$.  The
quantity $C({\cal T})$ determines the sub-leading term $\phi_+$ in the scalar
field expansion, and consequently the vev of the dual operator by
equation (\ref{vev}), by
\be
\phi_+  =
{d \over \Delta_-}
 {  C({\cal T}) \over (d-2\Delta_-)} |j|^{\Delta_+/\Delta_-} sign(j) \,. \label{FE6-2}
\ee
\item
For vev-driven flows ($j=0$) the superpotential (denoted in this
case by
$W^+(\phi)$  consists of a
purely analytic expansion in $\phi$, with no additional ${\cal
  T}$-dependent deformation parameters,
\be \label{f3}
W^+(\phi) =  {2(d-1) \over \ell} + {\Delta_+\over 2\ell} \phi^2 +
{\cal O}\left(\phi^4\right)
\ee
In this case the integration constant of the first order equation for
$\phi(u)$ is $\phi_+$, which is a free parameter for these solutions.
 \end{enumerate}

Alternatively, as explained in detail in appendix \ref{AppXY} one can define the {\em scalar variables} (that transform as scalars under a diffeomorphism of $u$):
\begin{equation}\label{XY0}
  X(\phi)\equiv \frac{1}{\sqrt{2d(d-1)}}\frac{\dot \phi}{\dot A(u)}, \qquad  Y(\phi)\equiv
  \frac{1}{d}\frac{\dot g(u) }{ \dot A(u)}\, .
\end{equation}
where the function $g$ is defined as $g = \log{f}$. Then Einstein's equations can be reduced to two coupled first order equations for $X$ and $Y$ as detailed in appendix \ref{AppXY}. The functions $X$ and $Y$ contain all physically relevant information on the system both in the vacuum and at finite temperature. For example the free energy can be read off directly from the boundary asymptotics of the functions $X$ and $Y$. One can think of the boundary values of $Y$ and $X$ as the enthalpy $s\ T$ and a combination of energy with the enthalpy, respectively. The first order formalisms in terms of the superpotential and the scalar variables are completely equivalent, e.g. $X$ is the logarithmic derivative of $W$.

\subsection{The free energy}

The free energy associated to the solution is given by the Euclidean
renormalised on-shell action,
\be
{\beta \cal F}  = S_{on-shell}^{(ren)}
\ee

Here we will focus on source-driven flows  leaving  the
special case of vev-driven flows for a later section.
An explicit calculation, which is performed with two independent
methods in Appendices \ref{app:onshell} and \ref{AppXY:free}, leads to the expression
\be\label{free}
{{\cal F} \over  V_{d-1}} = - {Ts \over d} - (M_p\ell)^{(d-1)}  C ({\cal
    T}) |j|^{d/\Delta_-},
\ee
where $T$ is the black-hole temperature, $s$  the BH entropy density and
$C({\cal T})$ is the parameter controlling  the
sub-leading asymptotics of  the superpotential near the boundary, see
equation (\ref{f2}).

In the canonical ensemble we are using, the boundary data which are
kept fixed are $T$ and $j$,  and ${\cal F} ={\cal F} (T,j)$.
Black-hole thermodynamics implies that the black-hole entropy density
$s$ and
energy density $\epsilon$ are
\be\label{FE12}
s = -{1\over V_{d-1}}{\de {\cal F} \over \de T} , \qquad \epsilon =
{{\cal F}  \over V_{d-1}} + Ts
\ee
The dual operator vev is the conjugate variable to $j$, 
\be\label{EP1}
\langle O \rangle (T,j) = -{1\over V_{d-1}}{\de {\cal F}(T,j) \over
  \de j }
\ee
and  it can be shown (see appendix
\ref{app:vev}) that the right-hand side of the equation above agrees
exactly with the definition of $\langle O \rangle$ from the
holographic  dictionary, equation  (\ref{FE6-2}). The differential
form of the first law (written in terms of the pressure $p={\cal
  F}/V_{d-1}$)  is then:
\be
dp = -s d T - \langle O \rangle d j.
\ee

Taking the conformal  limit $j\to 0$ in (\ref{free}) we recover conformal thermodynamics, ${\cal F} = V_{d-1} Ts/d$, which implies $s \propto T^{d-1}$ as found by integrating the differential relation  (\ref{FE12}).  Finally, at $T=0$ we recover the known result
\cite{Papadimitriou:2007sj,KN}:
\be\label{freeT0}
 {\cal F}_{T=0}  =  -  \left(M_p\ell\right)^{d-1}V_{d-1} \, C_0 |j|^{d/\Delta_-}
 = -{\Delta_- \over d} V_{d-1}\,  \langle O \rangle_{T=0} \,j.
\ee
where $C_0$ is the value of  $C({\cal T})$ in the zero-temperature vacuum.

It is convenient to rewrite the free energy in terms of the
dimensionless variables introduced in section (\ref{ssec:dim}) ,
\be\label{FE11}
{{\cal F}(T,j)  \over V_{d-1}} = T^{d}\left(-{\sigma ({\cal T})\over d} + \gamma({\cal T})
\right),
\ee
where we have defined the dimensionless quantities
\be\label{FE11-b}
\sigma({\cal T}) \equiv  {s\over T^{d-1}} = {{\cal S}({\cal T}) \over {\cal
    T}^{(d-1)}}, \qquad \gamma({\cal T}) \equiv (M_p\ell)^{d-1} {C({\cal
    T})\over {\cal T}^{d}} .
\ee

Apart from the overall $T^d$ scaling,  all the
non-trivial dependence  on $T$ and $j$   in the free energy only appears through the  combination ${\cal T}$ defined in equation (\ref{FE8}).

\subsection{The thermal effective potential}  \label{ssec:veff}

From the free energy, we can define the finite temperature effective
potential by a Legendre transform. First, we trade $j$ for its
conjugate variable, i.e. the dual operator vev $O$ (in this section we
omit the brackets for simplicity of notation):
inverting the relation (\ref{EP1}) to obtain $j(T,O)$,  the effective potential is then defined as
\be\label{EP2}
V_{eff}(T,O) = {\cal F}  + O j ,\qquad   d V_{eff} =  V_{d-1} \bigg[ -s dT +j d O\bigg],
\ee
and it satisfied the relations:
\be\label{EP3}
{1\over V_{d-1}}{\de V_{eff}\over \de O} = j, \qquad {1\over
  V_{d-1}} {\de V_{eff} \over  \de T} = -s .
\ee
As for the temperature,  we can introduce a dimensionless vev parameter,
\be\label{EP4}
{\cal O}  = {O \over T^{\Delta_+}} .
\ee
Starting from equation (\ref{FE11}) it is then easy to show that one can
write (\ref{EP2}) in the form
\be\label{EP5}
V_{eff}(T, O)  = T^d {\cal V}({\cal O}).
\ee
where ${\cal V}({\cal O})$ is the Legendre transform of ${\cal F}/T^d$ with
respect to  ${\cal T}^{-\Delta_-} $ (which  is indeed the dual
variable to ${\cal O}$).

Equation (\ref{EP5}) is useful because it allows the treatment of theories
 in which the source $j=0$, i.e. the case of pure vev flows:
 these are the values of ${\cal O}$ which extremise the function
 ${\cal V}$. We will see examples of these flows in the following
 sections, and we will discuss them in detail in section
 \ref{sec:vev}.

\section{Thermal phase transitions in multi-vacuum theories}
\label{sec:skipping}

In the previous section we have  developed a general expression for the
free energy of any black-hole solution, in terms of the UV source,  temperature,
entropy and vev of the operator dual to
$\phi$. 

We are now ready  to  study the phase diagram of black-hole
solutions in situations where the zero-temperature  RG flow displays
exotic features.

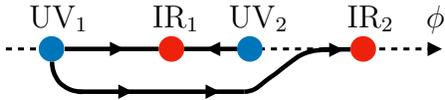
\begin{figure}[h!]
 \centering
\begin{overpic}
[width=0.45\textwidth]{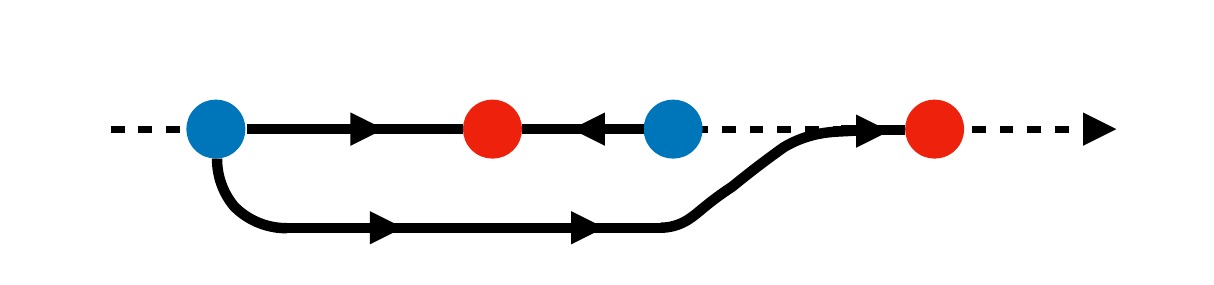}
\put (89,17) {$\phi$} \put (13.5, 17) {UV${}_1$} \put (37, 17) {IR${}_1$} \put (51.5, 17) {UV${}_2$} \put (74, 17) {IR${}_2$}
\end{overpic}
\caption{Schematic structure of a field theory which presents multiple
  RG-flows: In particular, there are two flows starting at the fixed
  point UV$_1$, one going to the closer IR fixed point IR$_1$, the
  second skipping IR${}_1$ and ending at  IR${}_2$. On the other hand there is no flow from $UV_2$ to  $IR_2$.}
\label{fig:skipschematic}
\end{figure}

In this section we concentrate on situations where {\em skipping
  flows} are present: in the presence of several maxima and minima of
the scalar potential, these are flows which skip an intermediate
potential IR fixed point and end at a  fixed point further away in
field space, as schematically represented in figure
\ref{fig:skipschematic}.

Vacua of the dual  field theory correspond to IR-regular flows.
In the  models at hand there may be multiple distinct IR-regular  RG
flows with the same UV boundary conditions. These are interpreted, under  the holographic map, as different vacua  of
the  same theory, with different $\beta$-functions and different IR
endpoints. At zero temperature, the true vacuum is the one with the lowest free energy. In \cite{multibranch} it was shown that this is the flow where the
parameter $C(T=0)$ is the largest, i.e. the one with the largest vev
at fixed source $j$ (cfr. equation \ref{freeT0}). This guarantees that the relevant solution has the lowest free energy.

\subsection{Skipping RG flows at zero-temperature}

\begin{figure}[t]
\centering
\begin{overpic}
[width=0.5\textwidth]{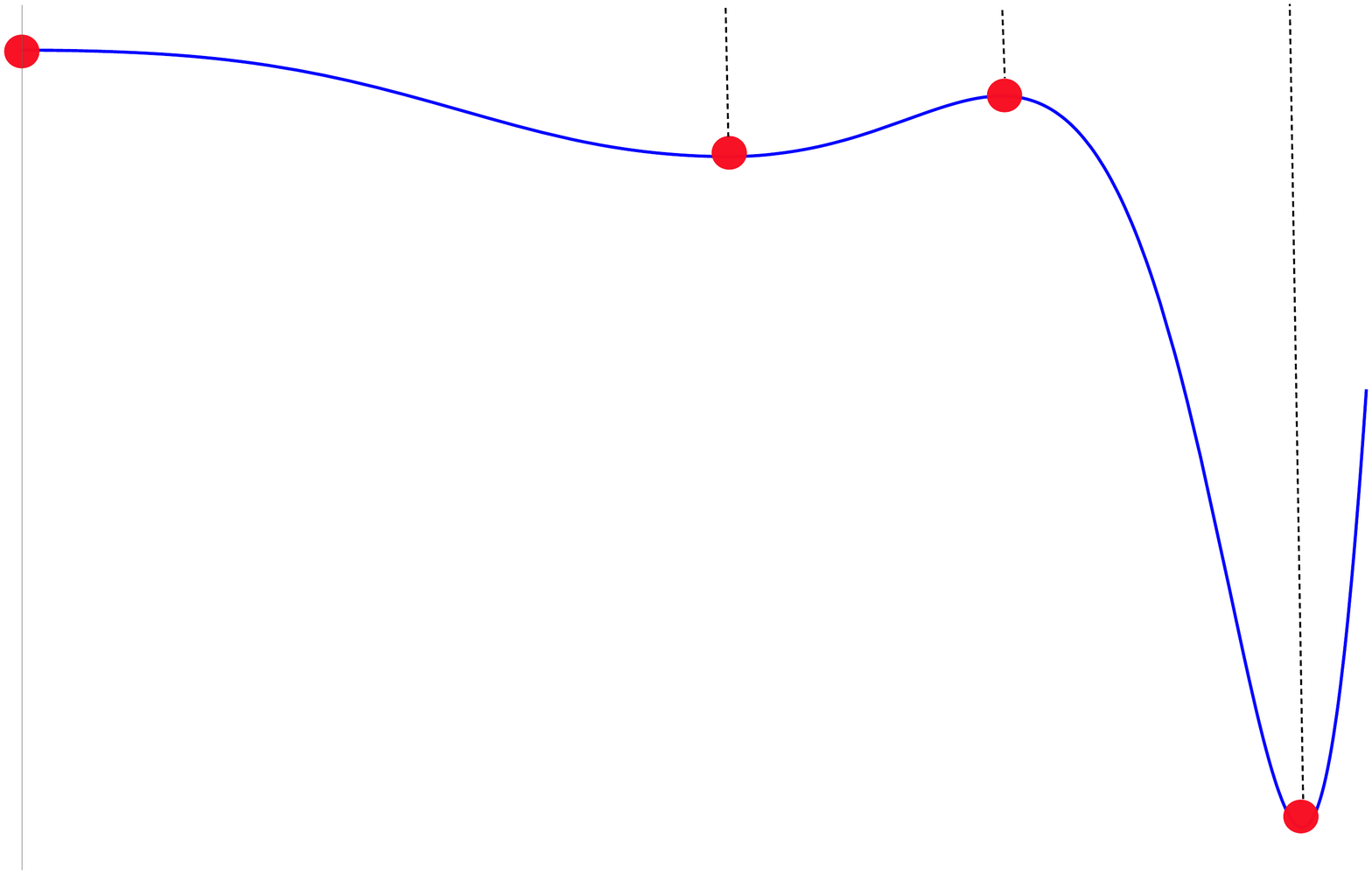}
\put (0,0) {V($\phi$)}
\put (3,53) {UV$_1$}
\put (3,65) {0}
\put (50,45) {IR$_1$}
\put (50,65) {$\phi_0$}
\put (69,49) {UV$_2$}
\put (69,65) {$\phi_1$}
\put (91,-1.5) {IR$_2$}
\put (91,65) {$\phi_2$}
\put (100,65) {$\phi$}
\end{overpic}
\caption{Plot of the degree-12 potential which allows skipping
  solutions. This potential has several extrema which are denoted as
  UV$_1$, UV$_2$, IR$_1$ and IR$_2$. }
  \label{skipnew}
\end{figure}

As an example of the behaviour described above,  in \cite{multibranch}
the following 12th-order  potential was considered,
\eql{V12}{
		V(\phi)=-{d(d-1)\over \ell^2} + \int_0^\phi V'(x) dx.
}
where
\eql{V12'}{
	  \ell^2 V'(\phi):=-\phi\le(\phi^2-\phi^2_0\ri)\le(\phi^2-\phi^2_1\ri)\le(\phi^2-\phi^2_2\ri)\le(\phi^2-\phi^2_3\ri)\le(\phi^2-{\Delta(\Delta-d)\over\phi^2_0~\phi^2_1~\phi^2_2~\phi^2_3}\ri),
}
The potential has  extrema at the points
$0<\phi_0<\phi_1<\phi_2<\phi_3$. We make  the specific choice:
\begin{align}
	&d = 4\quad \phi_0 = 1.0837\quad \phi_1= 1.1316 \nonumber\\
	&\Delta = 2.8  \quad \phi_2 =1.9200 \quad\phi_3 = 2.1500 \, .\label{ParamSkip}
\end{align}
With these choices, the operator dimensions at the various fixed
points are given by:
\be\label{deltas}
\Delta_{UV_1} = 2.8, \quad   \Delta_{UV_2} = 3.1, \quad
\Delta_{IR_1} = 4.5, \quad  \Delta_{IR_2} = 11.6 .
\ee
The  potential is shown in figure \ref{skipnew}, where the
correspondence between the values $\phi_i$ and  the UV and IR fixed
points is made manifest. The explicit expression of $V(\phi)$ can be found in appendix A of \cite{multibranch}.

\begin{figure}[h!]
\centering
\begin{overpic}
[scale=.45,angle=90]{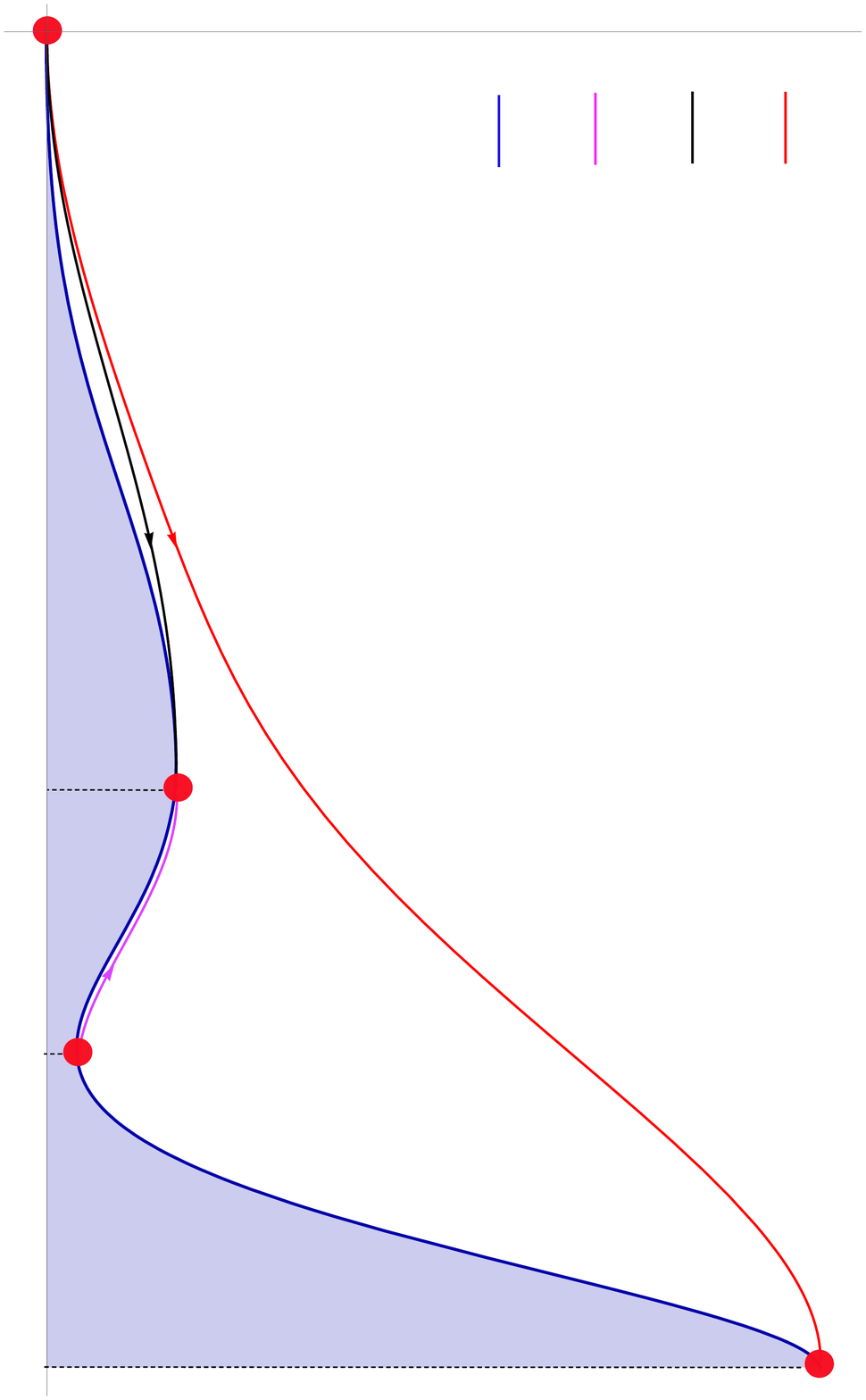}
\put(100,8){$\phi$}
\put(1,65){$W(\phi)$}
\put(2,11){UV$_1$}
\put(3,6){0}
\put(55,19){IR$_1$}
\put(55,6){$\phi_0$}
\put(72,13){UV$_2$}
\put(73,6){$\phi_1$}
\put(95,64){IR$_2$}
\put(95,6){$\phi_2$}
\put(13,59){$W_{12}(\phi)$}
\put(13,52.5){$W_{11}(\phi)$}
\put(13,46.5){$W_{21}(\phi)$}
\put(13,39.5){$B(\phi)=\sqrt{-3V(\phi)}$}
\end{overpic}
\caption{The vacuum RG flow solutions arising from the potential
  \eqref{V12}. All flows interpolate between one of the  UV fixed points and
  one of the IR fixed points. The plotted lines are the
  superpotentials corresponding to each flow (see Appendix \ref{app:first}).  The arrows represent the direction of
  the flow from the UV to the IR. The blue area is the forbidden
  region below  the  curve $B(\phi) = \sqrt{-4(d-1)\, V(\phi)/d}$ (where
  we set $d=4$), which bounds from below  any solution to the
  superpotential equation at $T=0$ (as explained in  Appendix \ref{app:bound}).
  }\label{skipflatnew}
\end{figure}

At $T=0$ there are several IR-regular RG-flow solutions,  displayed
schematically in figure \ref{skipflatnew}, which shows the
zero-temperature superpotential of each flow\footnote{Figure
  \ref{skipflatnew} displays the {\em super}potentials and the critical curve $B\sim
\sqrt{-V}$ (which bounds the space of solutions, see
\cite{multibranch}). Therefore  what is presented as a maximum
(minimum) in that figure actually corresponds to a minimum (maximum)
of the potential in figure \ref{skipnew}.}.
\begin{itemize}
\item {\bf UV$_1$ $\to$ IR$_1$.} This solution is the standard holographic
  RG flow which connects a maximum  of the potential (UV) to the
  nearest minimum (IR).
\item {\bf UV$_1$  $\to$ IR$_2$.} This solution on the other hand skips the
  first minimum and ends at the next available IR fixed point at
  $IR_2$. This kind of solution is not found in  {\em generic} potentials
  admitting several extrema: for it to exist the extremum $IR_1$ has
  to be sufficiently shallow.

\item {\bf UV$_2$ $\to$ IR$_1$. } This solution corresponds to a
  standard  flow with  a {\em negative} source from the second maximum
  of $V(\phi)$, reaching the closest available IR fixed point.  Notice that there is no solution connecting UV$_2$ to IR$_2$. The
reason is that there already is a regular flow arriving at IR$_2$ (the
one from UV$_1$: since flows reaching (from a given direction) a
minimum of the potential are isolated, this prevents  other flows to
reach the same IR.
\end{itemize}

The two solutions leaving UV$_1$, correspond to two vacua of the same
UV theory. The one with the lowest free energy (\ref{freeT0})  is the skipping one
UV$_1$$\to$ IR$_2$, since it has the largest vev parameter $C_0$, as can
be  seen immediately from the fact that the corresponding
superpotential  ($W_{12}$ in figure \ref{skipflatnew}) increases faster close to
the origin.

\subsection{Finite temperature solutions}

We now move to finite $T$ by considering black-hole solutions, of the
form (\ref{b1})  in the
model  with the same potential in figure \ref{skipnew}. These black holes
are  uniquely characterised by a dimensionless number: the value $
\phi_h$  of the scalar field at
the horizon. If we keep the value of the UV  source $j$ fixed,
we expect  $\phi_h$  to determine  all other quantities, (temperature,
entropy, free energy, etc.) Therefore we are interested in
constructing all solutions with $\phi_h$  ranging from zero (UV$_1$) to
$\phi_2$ (IR$_2$). Indeed, we will see that for every value of
$\phi_h$ there exists at most one black-hole solution with all
other UV data fixed.

It has to be stressed that, in order to be
considered as different states in the same dual QFT, two solutions must
connect to the same UV fixed point, with the same value of the source
parameter $j$.

As we will see below, depending on the value reached at the horizon by
the scalar field, integrating the solution ``backwards''  away from
the horizon may lead  either to UV$_1$ or to UV$_2$. These represent two
disconnected classes of solutions, since they have different boundary
conditions at the UV boundary. From the dual field theory standpoint,
they represent thermal states in different (deformed) CFTs. For this
reason, we analyse them separately.

\subsubsection*{Flows from UV$_1$}

\begin{figure}[h!]
\centering
\begin{overpic}
[scale=.65]{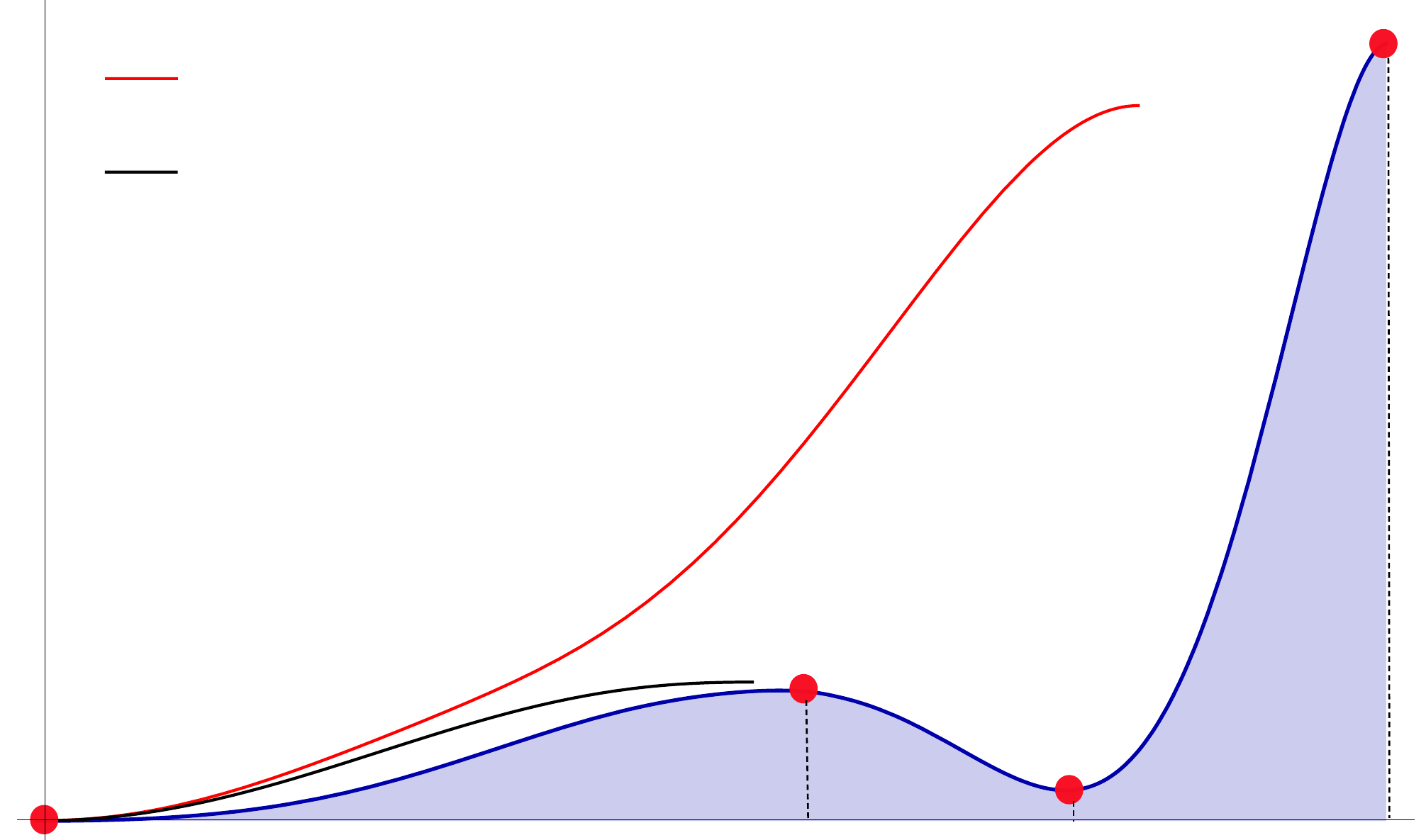}
\put(100,0){$\phi$}
\put(1,60){$W(\phi)$}
\put(2,4){UV$_1$}
\put(3,-1){0}
\put(55,12){IR$_1$}
\put(55,-1){$\phi_0$}
\put(72,6){UV$_2$}
\put(74,-1){$\phi_1$}
\put(95,58){IR$_2$}
\put(96,-1){$\phi_2$}
\put(13,53){$W_{skip}(\phi)$}
\put(13,46.5){$W_{Non-skip}(\phi)$}
\end{overpic}
\caption{Finite temperature solutions arising from the potential
  \protect\eqref{V12} and connecting to $UV_1$.} 
\label{skipflat-finiteT}
\end{figure}

We first consider solutions connecting to the UV$_1$ fixed point at
$\phi=0$. The two zero-temperature vacuum flows are the black and red curves
in figure \ref{skipflatnew}.  Turning on temperature, the situation is
represented in figure \ref{skipflat-finiteT}, where the endpoint of
each flow is now at the horizon, where $\phi=\phi_h$.

\begin{figure}[h!]
\centering
\begin{overpic}
[width=10cm]{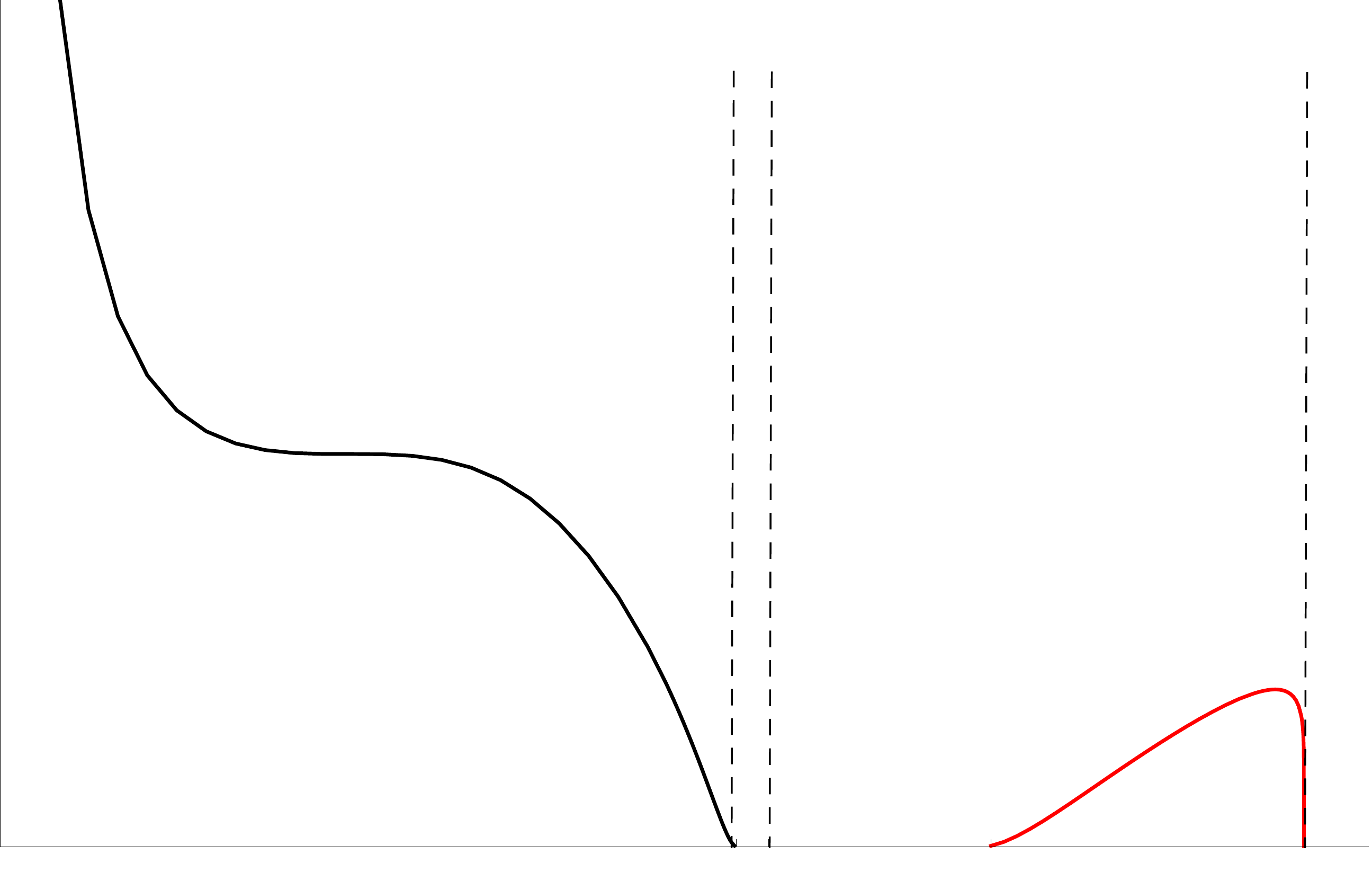}
\put(100,1){$\phi_h$}
\put(-2,65){${\cal T}$}
\put(-3,55){{\scriptsize UV$_1$}}
\put(0,-1){0}
\put(49,55){{\scriptsize IR$_1$}}
\put(51,-1){$\phi_0$}
\put(57,55){{\scriptsize UV$_2$}}
\put(56,-1){$\phi_1$}
\put(70,-1){$\phi_*$}
\put(94,55){{\scriptsize IR$_2$}}
\put(94,-1){$\phi_2$}
\put(90, 15){{\scriptsize ${\cal T}_{max}$}}
\end{overpic} \\
(a) \\
\vspace{2cm}
\begin{overpic}
[width=10cm]{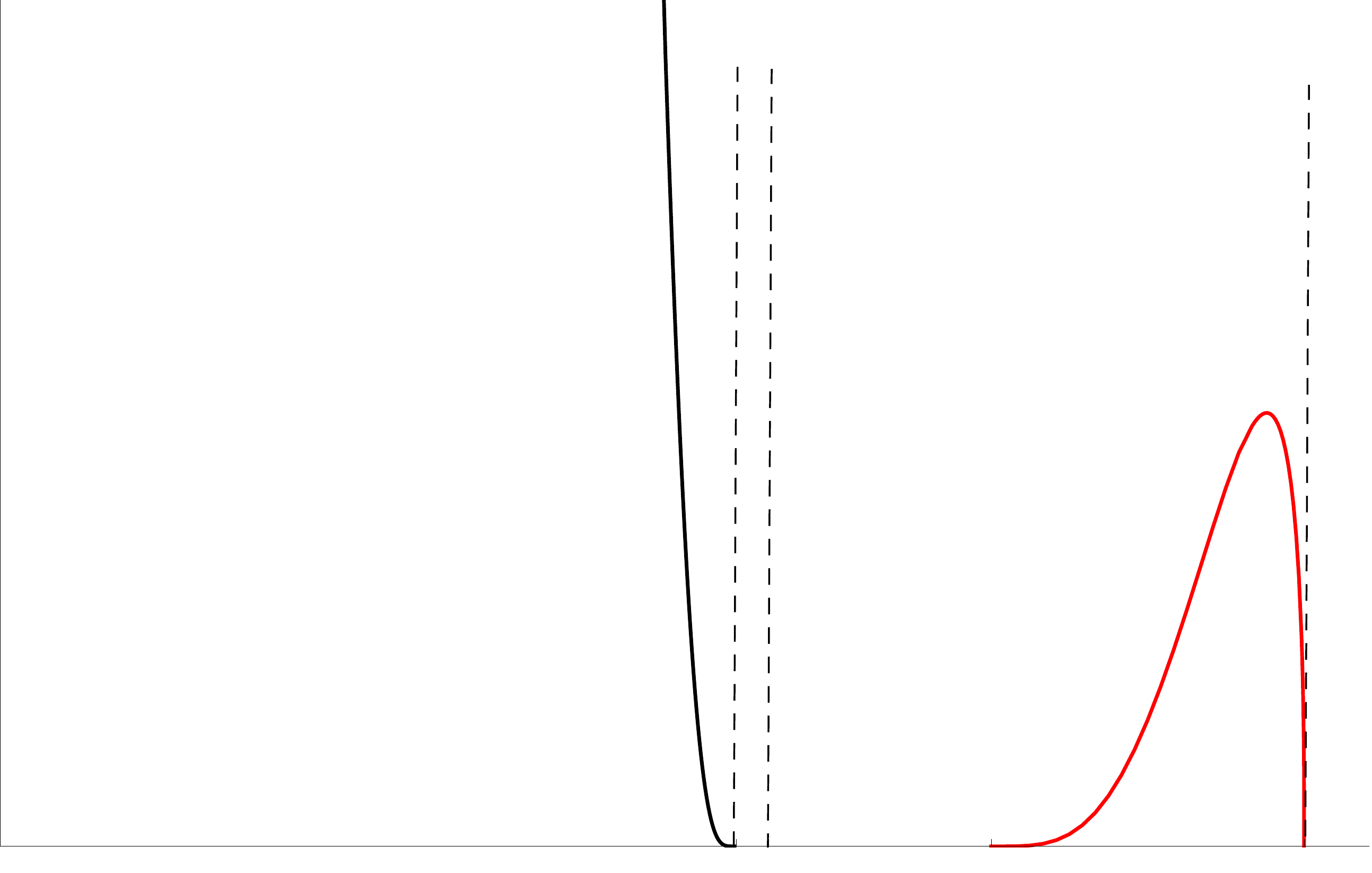}
\put(100,1){$\phi_h$}
\put(-2,65){${\cal S}$}
\put(-3,55){{\scriptsize UV$_1$}}
\put(0,-1){0}
\put(49,55){{\scriptsize IR$_1$}}
\put(51,-1){$\phi_0$}
\put(57,55){{\scriptsize UV$_2$}}
\put(56,-1){$\phi_1$}
\put(70,-1){$\phi_*$}
\put(94,55){{\scriptsize IR$_2$}}
\put(94,-1){$\phi_2$}
\put(90, 35){{\scriptsize ${\cal S}_{max}$}}
\end{overpic}\\
(b)\\
\caption{The dimensionless temperature (a) and the entropy density  (b) as a
  function of the scalar field horizon value $\phi_h$  for black holes connecting to UV$_1$. }\label{UV1}
\end{figure}

\begin{figure}[h!]
\centering
 \begin{overpic}
[width=10cm]{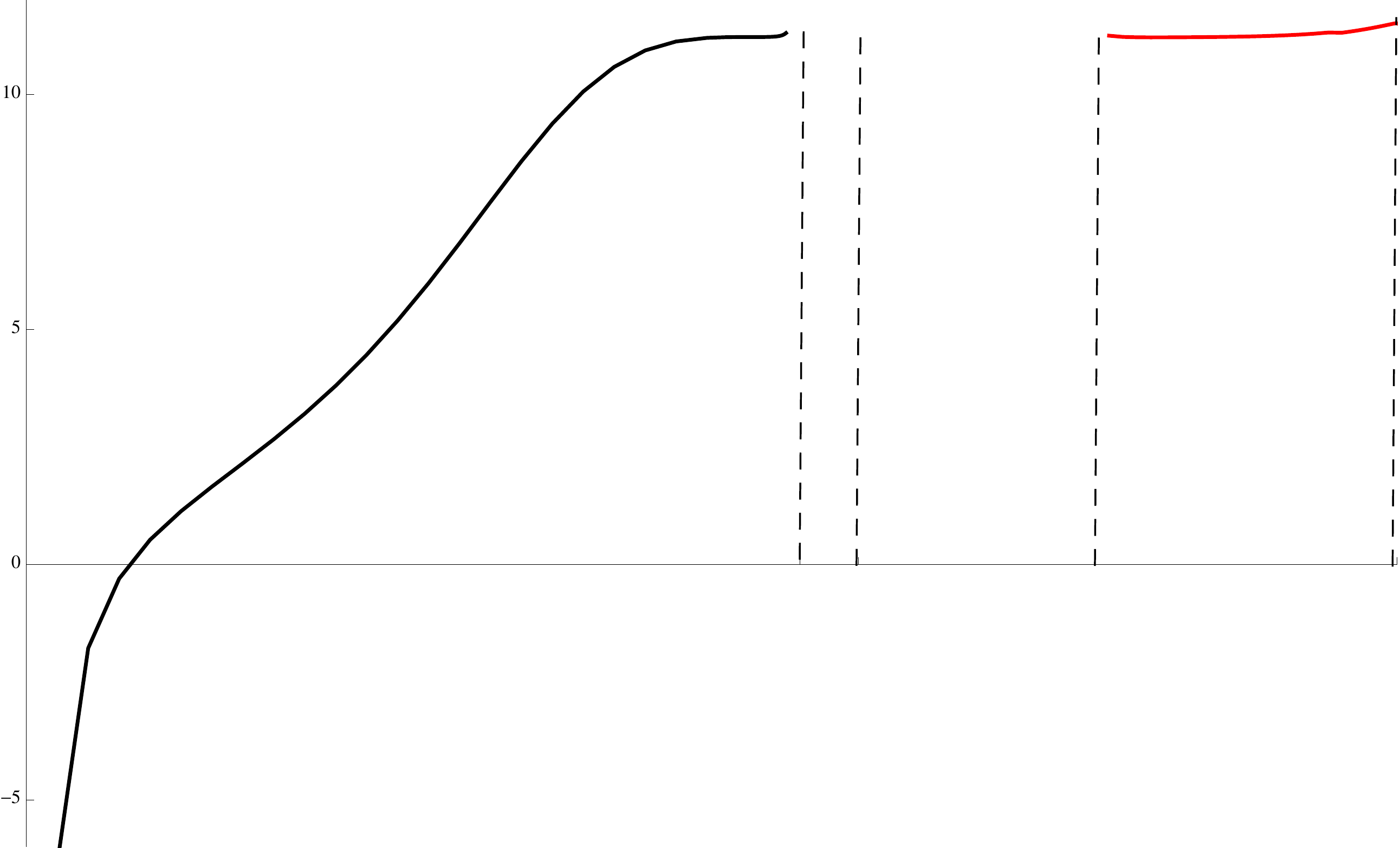}
\put(101,20){$\phi_h$}
\put(0,62){$C$}
\put(3,55){{\scriptsize UV$_1$}}
\put(54,60){{\scriptsize IR$_1$}}
\put(55,17){$\phi_0$}
\put(60,60){{\scriptsize UV$_2$}}
\put(60,17){$\phi_1$}
\put(77,17){$\phi_*$}
\put(98,60){{\scriptsize IR$_2$}}
\put(98,17){$\phi_2$}
\end{overpic}
\caption{The (dimensionless) vev parameter $C$  as a function of the scalar field horizon
  value $\phi_h$  for black holes connecting to UV$_1$. As the horizon
  approaches the UV fixed point at $\phi=0$,  the parameter $C \to
  -\infty$.}\label{CUV1}
\end{figure}
There are now up to {\em three} branches of solutions at fixed
${\cal T}$, whose (dimensionless) temperature and entropy density as a
function of the horizon value $\phi_h$ are  is represented in figure \ref{UV1}. The corresponding  vev parameter $C(\phi_h)$ is shown in figure \ref{CUV1}.
As one can observe, there is  a range  of horizon values, $\phi_h$,  (between $\phi_0$
and a critical point which we denote by $\phi_*$)  for which no
solution exists which continuously connects to UV$_1$.

\begin{enumerate}
\item {\bf Solutions with $\phi_h < \phi_0$.} These are black holes
  {continuously connected to the}  non-skipping vacuum flow from UV$_1$ $\to$ IR$_1$. As the
  temperature is increased, the horizon moves closer and closer to the
  UV fixed point of the vacuum solution. This is the standard behaviour
   at finite temperature for the simplest RG flows, connecting two consecutive
   extrema of the scalar potential.

\item{\bf Solutions with $\phi_*< \phi_h < \phi_2$}.
 These solutions all
  skip IR$_1$ and flow to the region between UV$_2$ and IR$_2$. As one
  can see from figure \ref{UV1}, these solutions  have a maximal
  temperature ${\cal T}_{max}$ and entropy ${\cal S}_{max}$.
 For each ${\cal T} < {\cal T}_{max}$ there are two solutions. Of the two,
  the one with the larger $\phi_h$ is the deformation of the
  zero-temperature skipping flow UV$_1$ $\to$ IR$_2$, for which the
  temperature increases as the horizon moves away from the IR fixed
  point. The second solution is a new branch, which has no
  zero-temperature analogue, and for which the temperature increases
  as the horizon moves {\em towards} the IR fixed point IR$_2$. At the
 critical value corresponding to ${\cal T}_{max}$, the two solutions
 merge. Both branches extend to arbitrarily low temperature, but only
 one of them (the one with higher $\phi_h$) actually  connects to a
 horizonless zero-$T$ solution. The fate of the other branch as ${\cal
   T}\to
 0$, $\phi\to\phi_{*}^+$, will be discussed separately at the end of this section.
\end{enumerate}

\begin{figure}[h!]
\centering
\begin{overpic}[width=14cm]{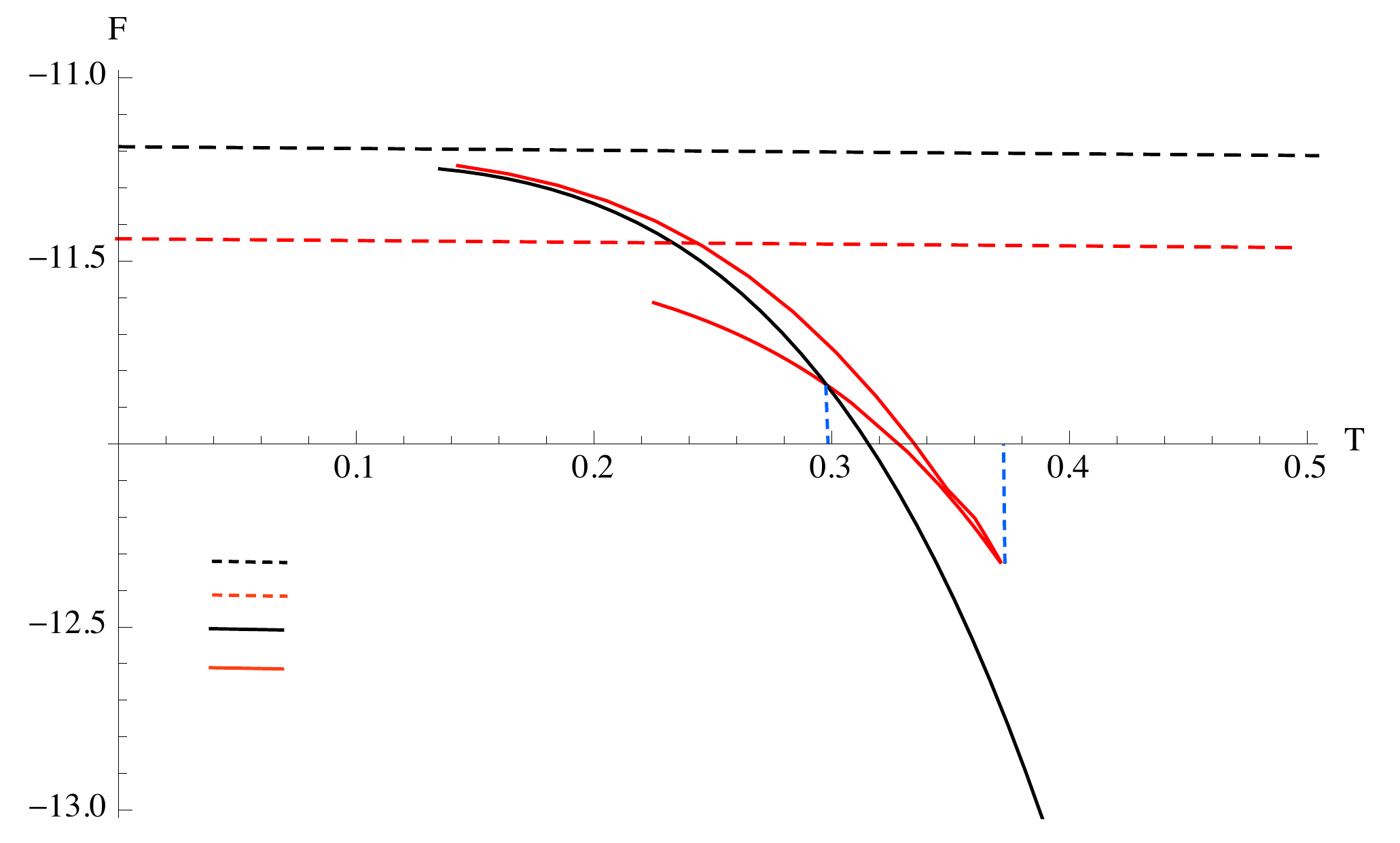}
 \put(70,30){\scriptsize $T_{max}$}
 \put(58,24){\scriptsize $T_c$}
 \put(21,19.5){{\scriptsize $T=0$, non-skipping}}
\put(21,17){{\scriptsize $T=0$, skipping}}
\put(21,14.5){{\scriptsize non-skipping BH}}
\put(21,12){{\scriptsize skipping BH}}
 \end{overpic}
\caption{The figure shows the free energy density density  as a function of temperature
  for the solutions connecting to UV$_1$. The units were fixed  such
  that $j=1$ (so that ${\cal T}=T$).  The black curve corresponds to
   the non-skipping branch, and it extends from $T=0$ to infinity. The
   red curves are the two skipping branches. The dashed horizontal
   lines represent the free energy of the zero-temperature skipping
   (red) and non-skipping (black) solutions.  Of
 the two skipping black-hole solutions, the one with lower free energy  connects to the true
 vacuum skipping flow as $T\to 0$ (The
 curves shown here do not reach all the way to zero temperature due to
 limitations in the numerics). Both skipping branches disappear at
 a maximal temperature, which is numerically found to be $T_{max}
 \simeq 0.37$ in units where $j = 1$. The phase transition
 occurs at the temperature $T_c\simeq 0.3$, where the skipping and
 non-skipping branches cross.}\label{phtr}
\end{figure}
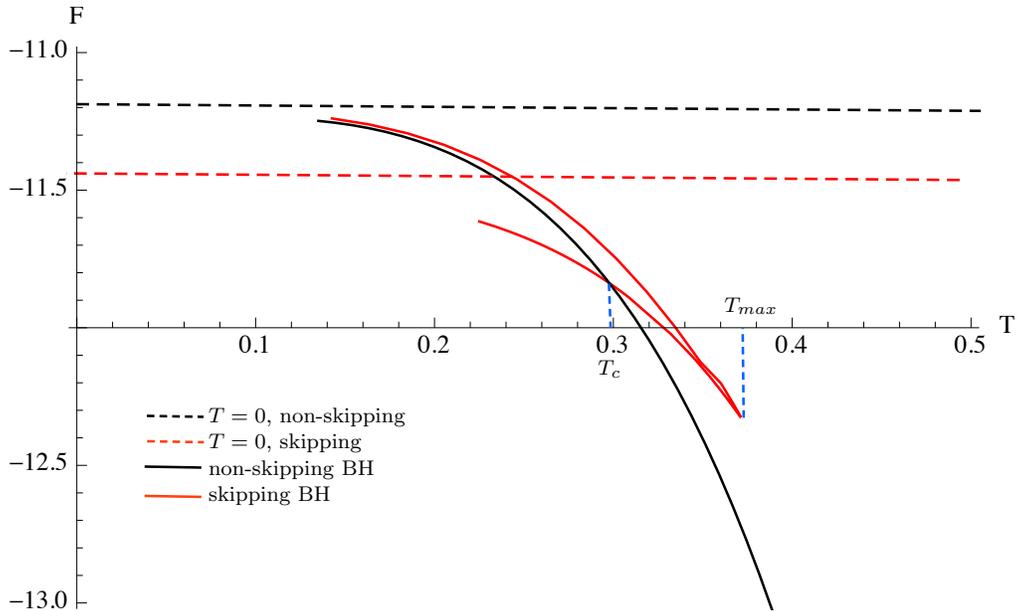

Given the situation described above, it is clear that at ${\cal T}>
{\cal T}_{max}$
there is a unique black hole solution, which belongs to the non-skipping branch.  At zero temperature however, as we discussed at the
beginning of this section, the true ground state is the skipping
branch which reaches IR$_2$.  Therefore, a skipping solution is expected to
continue to be the ground state for small but finite temperature. In
other words, there should be a phase transition between the skipping
and non-skipping black holes at some finite temperature ${\cal T}_c<
{\cal T}_{max}$. This is indeed confirmed by a numerical analysis, and it is
clearly visible in figure \ref{phtr}, where we display the free
energy, as a function of the temperature,
of the three branches of solutions connecting to UV$_1$.

\subsubsection*{Flows from UV$_2$}
Solutions with scalar field  horizon values $\phi_h$ in the range
$\phi_0 < \phi_h < \phi_*$ connect to UV$_2$, rather than UV$_1$. This
explains the empty gap in horizon values  in figures \ref{UV1} and
\ref{CUV1}.  These
solutions belong to a different dual field theory (a deformation of a different
UV CFT) from those flowing from UV$_1$.

\begin{figure}[h!]
\centering
\begin{overpic}
[scale=.65]{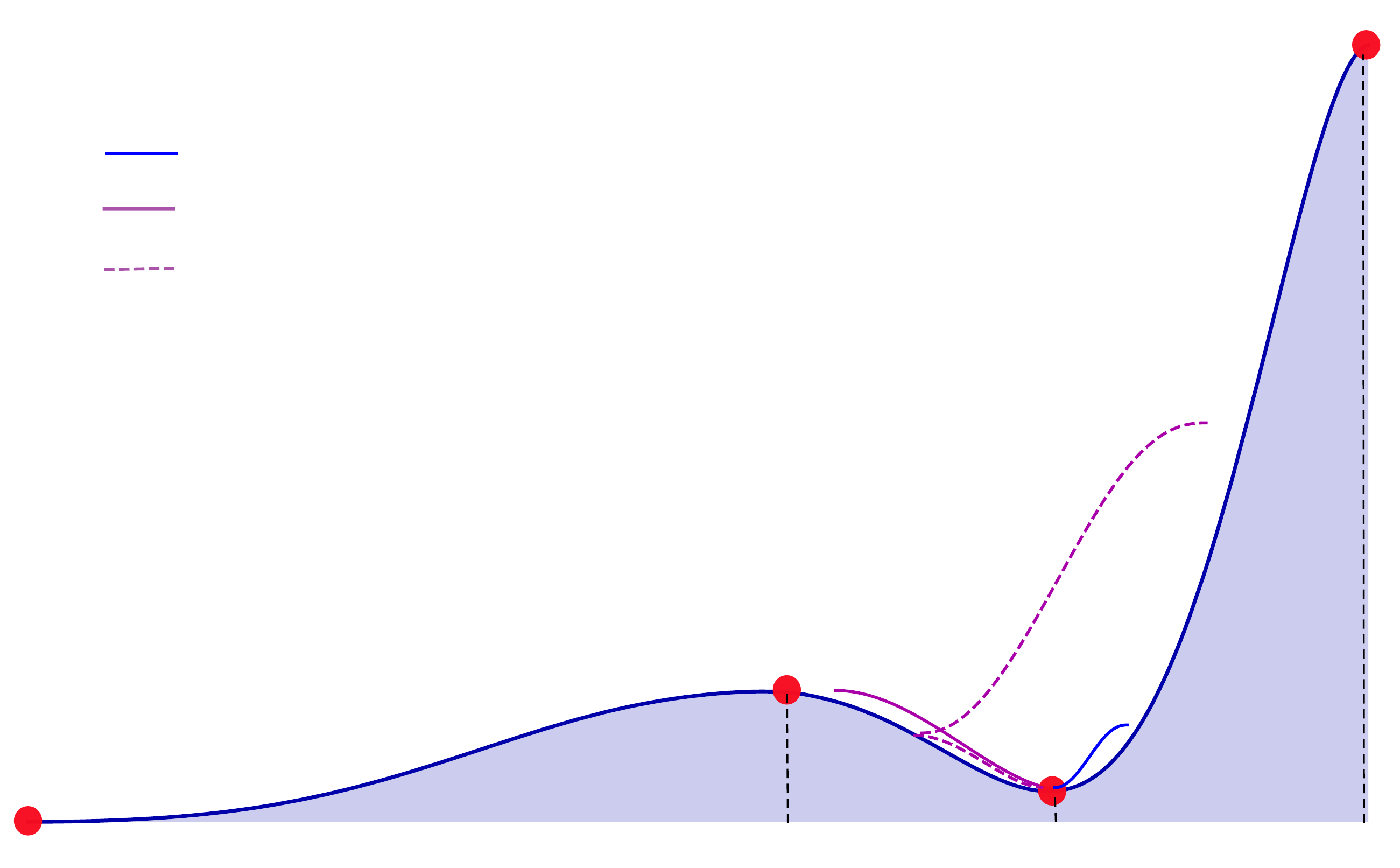}
\put(101,3){$\phi$}
\put(1,65){$W(\phi)$}
\put(2,5){UV$_1$}
\put(2,0){0}
\put(55,14){IR$_1$}
\put(55,1){$\phi_0$}
\put(72,7){UV$_2$}
\put(74,1){$\phi_1$}
\put(95,61){IR$_2$}
\put(96,1){$\phi_2$}
\put(13,50.5){$j>0$}
\put(13,46.5){$j<0$, Non-bouncing}
\put(13,42.5){$j<0$, Bouncing}
\end{overpic}
\caption{Finite temperature solutions arising from the potential
  \protect\eqref{V12} and connecting to $UV_2$. 
  }\label{skipUV2-finiteT}
\end{figure}

The set of flows emerging from UV$_2$ is represented
schematically in figure \ref{skipUV2-finiteT}. These flows can be
divided into three different classes: those with positive source (represented in
blue),  those with negative source which {\em bounce}  (i.e. where the
scalar field inverts its direction along the flow, dashed
purple curve) and do not bounce (solid purple curve). Bouncing
solutions of this kind where discussed at zero-temperature in
\cite{multibranch}, and examples where studied in a model with a different
bulk potential, which will be the subject of  section
\ref{sec:bouncing}.  Here   we see a new feature: although in the
current example with the potential in figure \ref{skipnew}  there
are no bouncing solutions at zero-temperature,  these may appear at
finite temperature. A similar phenomenon was observed in \cite{curvedRG} in
the case of a non-zero boundary curvature.

\begin{figure}[h!]
\centering
 \begin{overpic}
[width=10cm]{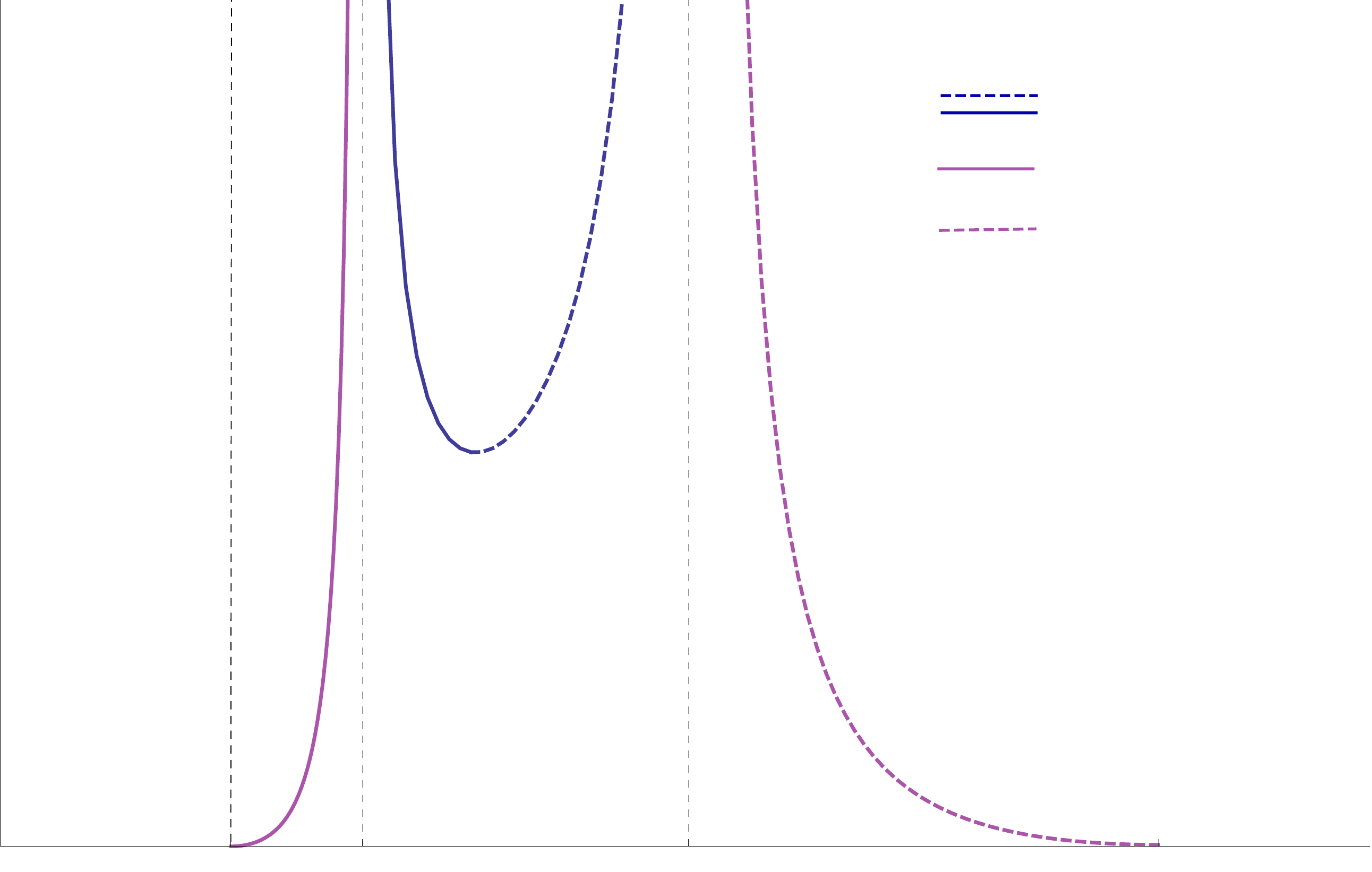}
\put(101,2){$\phi_h$}
\put(0,64){${\cal T}$}
\put(14,65){{\scriptsize IR$_1$}}
\put(15,-1){$\phi_0$}
\put(24,65){{\scriptsize UV$_2$}}
\put(25,-1){$\phi_1$}
\put(49,-1){$\phi_c$}
\put(83,-1){$\phi_*$}
\put(76,55.5){\scriptsize $j>0$}
\put(76,50.5){\scriptsize $j<0$, Non-bouncing}
\put(76,46.5){\scriptsize $j<0$, Bouncing}
\end{overpic}
\\
(a)\\
\vspace{1cm}
\begin{overpic}
[width=10cm]{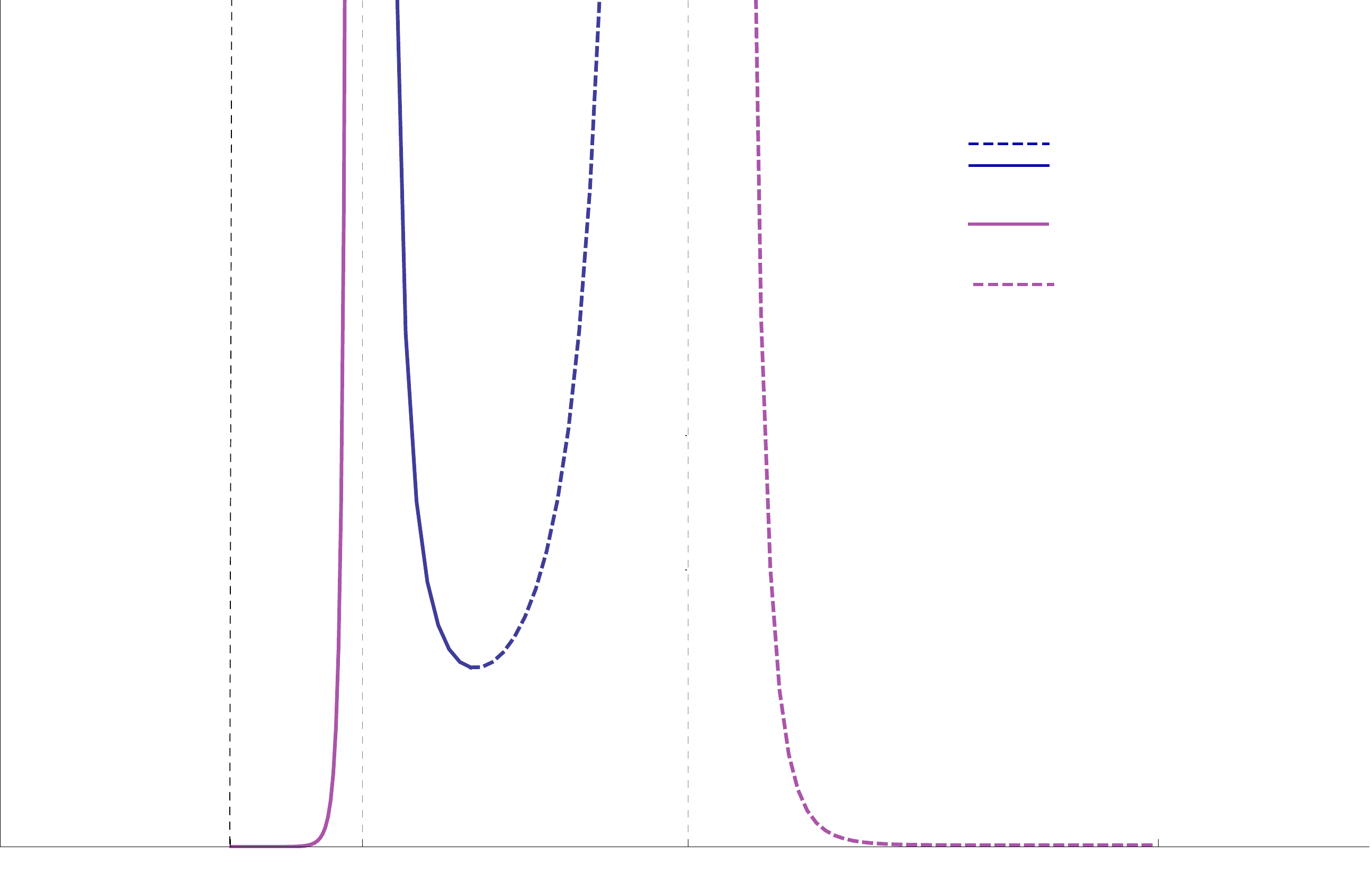}
\put(101,2){$\phi_h$}
\put(0,64){${\cal S}$}
\put(14,65){{\scriptsize IR$_1$}}
\put(15,-1){$\phi_0$}
\put(24,65){{\scriptsize UV$_2$}}
\put(25,-1){$\phi_1$}
\put(49,-1){$\phi_c$}
\put(83,-1){$\phi_*$}
\put(77,51.5){\scriptsize $j>0$}
\put(77,46.5){\scriptsize $j<0$, Non-bouncing}
\put(77,42.5){\scriptsize $j<0$, Bouncing}
\end{overpic}
\\
(b)
\caption{The temperature (a) and entropy density (b) as a function of
  the scalar field horizon  value $\phi_h$  for black holes connecting
  to UV$_2$. The colours and dashes correspond to the flows shown in
  figure \protect\ref{skipUV2-finiteT}.
}\label{UV2}
\end{figure}

\begin{figure}[h!]
\centering
\begin{overpic}[width=10cm]{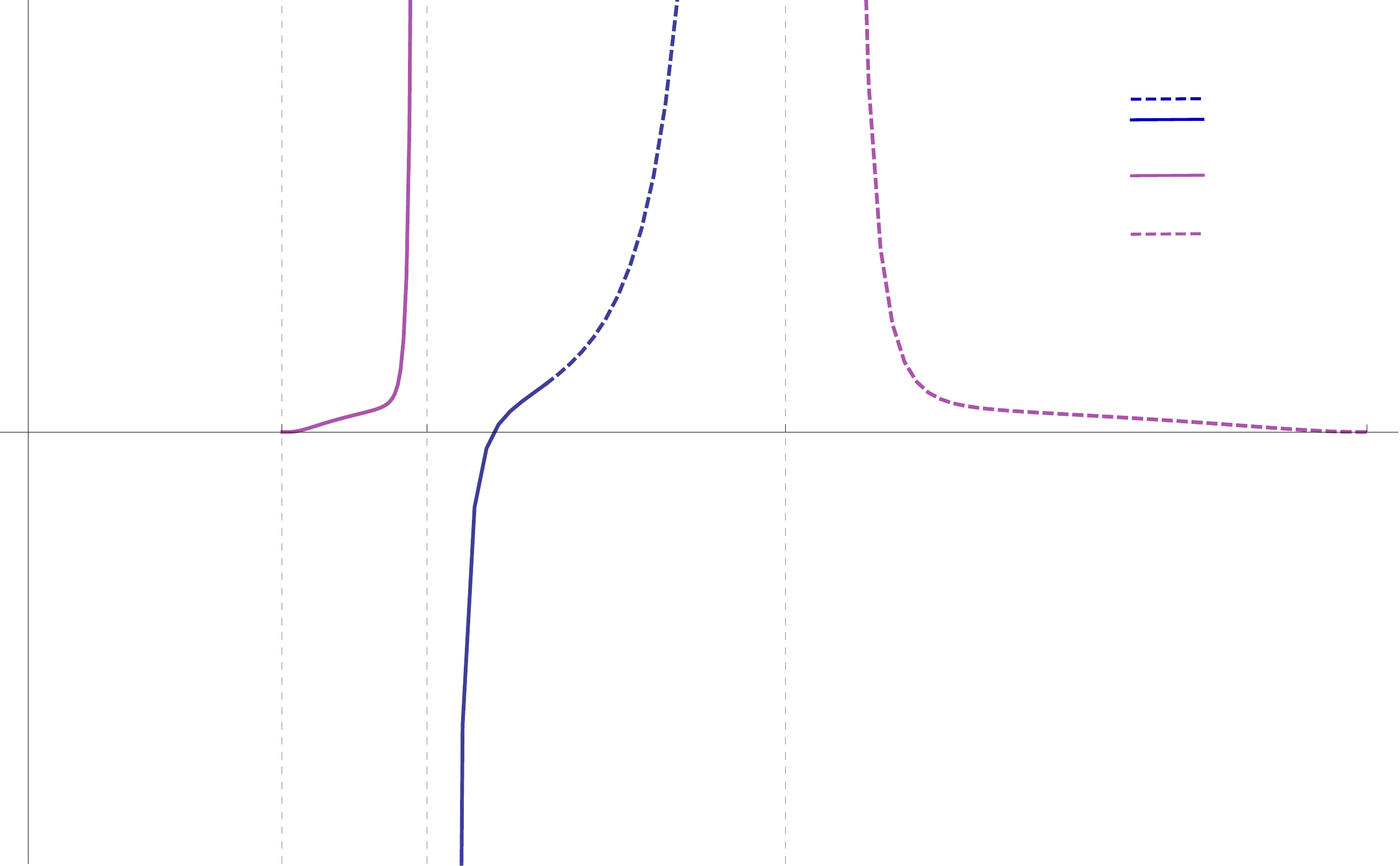}
\put(101,30){$\phi_h$}
\put(0,64){$C$}
\put(18,63){{\scriptsize IR$_1$}}
\put(18,28){$\phi_0$}
\put(26,63){{\scriptsize UV$_2$}}
\put(28,28){$\phi_1$}
\put(53,28){$\phi_c$}
\put(96,28){$\phi_*$}
\put(87,53.5){\scriptsize $j>0$}
\put(87,48.5){\scriptsize $j<0$, Non-bouncing}
\put(87,44.5){\scriptsize $j<0$, Bouncing}
\end{overpic}
\caption{The dimensionless vev parameter $C$  as a function of the scalar field horizon
  value $\phi_h$  for black holes connecting to UV$_2$.}\label{CUV2}
\end{figure}

All these solutions can
be classified
 according to the endpoint value $\phi_h$. Their dimensionless  temperature,
 entropy density, and vev parameter  are represented as a function of $\phi_h$ in  figures \ref{UV2} and \ref{CUV2}  (the complement of figure
\ref{UV1} and \ref{CUV1}). Notice the existence  of a special  point
$\phi_c$,  whose value lies between $\phi_1$ and $\phi_*$, which
separates  positive-source and negative-source black holes.


\vspace{1cm}
\begin{enumerate}
\item {\bf Solutions with $\phi_0 < \phi_h < \phi_1$ (solid purple
    curve in figures \ref{skipUV2-finiteT} and \ref{UV2}).}
These black holes are the finite temperature {continuations} of the vacuum flow
from UV$_2$ to IR$_1$ shown in figure \ref{skipflatnew}. As the
temperature is increased, the scalar field endpoint moves from IR$_1$
to higher values  towards UV$_2$. These flows have a {\em negative}
value of the UV deformation parameter $j$, since $\phi(u)$ decreases along
the flow.

\item {\bf Solutions with  $\phi_1 < \phi_h < \phi_c$ (blue
    curve in figures \ref{skipUV2-finiteT} and \ref{UV2}).}
These solutions flow from UV$_2$ to larger values of $\phi$, and
therefore, although  they originate from the same fixed point, they do
not belong to the same class of deformed CFT as those in the previous
class, since they differ by the sign of the source of the
deformation. Solutions in this class have no zero-temperature analogue
(recall that a regular flow from UV$_2$ with positive source does not
exists at zero temperature), and in fact they only exist above a {\em
  minimum} temperature $T_{min}$, as can be seen from figure
\ref{UV2}. At both ends of this range, the temperature asymptotes to infinity.

\item {\bf Solutions with  $\phi_c < \phi_h < \phi_*$ (dashed purple
    curve in figures \ref{skipUV2-finiteT} and \ref{UV2}).}
These solutions display a bounce in the flow: they start out from UV$_2$ with {\em decreasing}
scalar field, (i.e. they have $j<0$ and therefore
belong to the same UV theories as those  in class 1 above). Before
they reach IR$_1$, the scalar field reaches a minimum, the flow
inverts its direction and starts running towards IR$_2$. The horizon
lies somewhere in between the critical points $\phi_c$ and $\phi_*$,
beyond which we find solutions in the other classes.
\end{enumerate}

Below we discuss several properties of the space of solutions which
asymptote UV$_2$.

\begin{figure}[h!]
\centering
\begin{overpic}
[width=8cm]{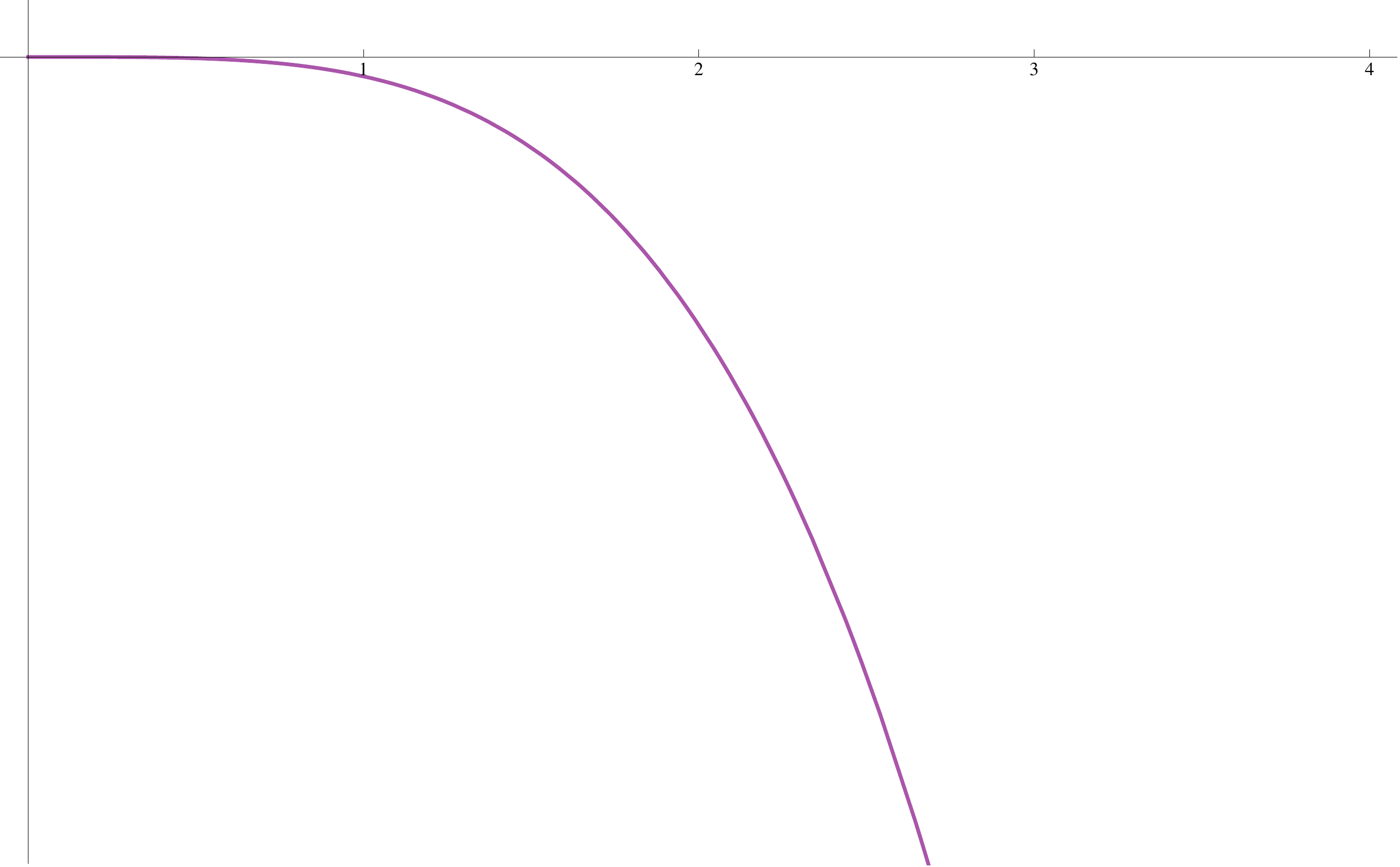}
\put(101,57){${\cal T}$}
\put(-1,56){$0$}
\put(1,60){${\cal F}_{Non-Bouncing} - {\cal F}_{Bouncing}$}
\end{overpic}
\caption{Free energy difference as a function of the (dimensionless)
  temperature  between the two branches of black holes emanating from
  UV$_2$ with negative source (dashed and solid purple curves in
  figures \protect\ref{skipUV2-finiteT} and  \protect\ref{UV2}).
  }\label{UV2-Neg}
\end{figure}

\paragraph{Negative source phase diagram.} From figure \ref{UV2} (a)  we see that, for $j <0$,
there are two black-hole solutions for any given temperature $T>0$. By
computing their free energies numerically, we have shown that the
thermodynamically favoured
black hole (the one with lowest free energy) is,  at all temperatures, the non-bouncing
solution, i.e. the one with horizon between IR$_1$ and UV$_2$, which
continuously connects to the  vacuum flow ending at $\phi_0$ at zero
temperature. This result is shown in figure \ref{UV2-Neg}, where we
plot the free energy difference between the non-bouncing and bouncing
solutions.

\begin{figure}[h!]
\centering
\begin{overpic}
[width=8cm]{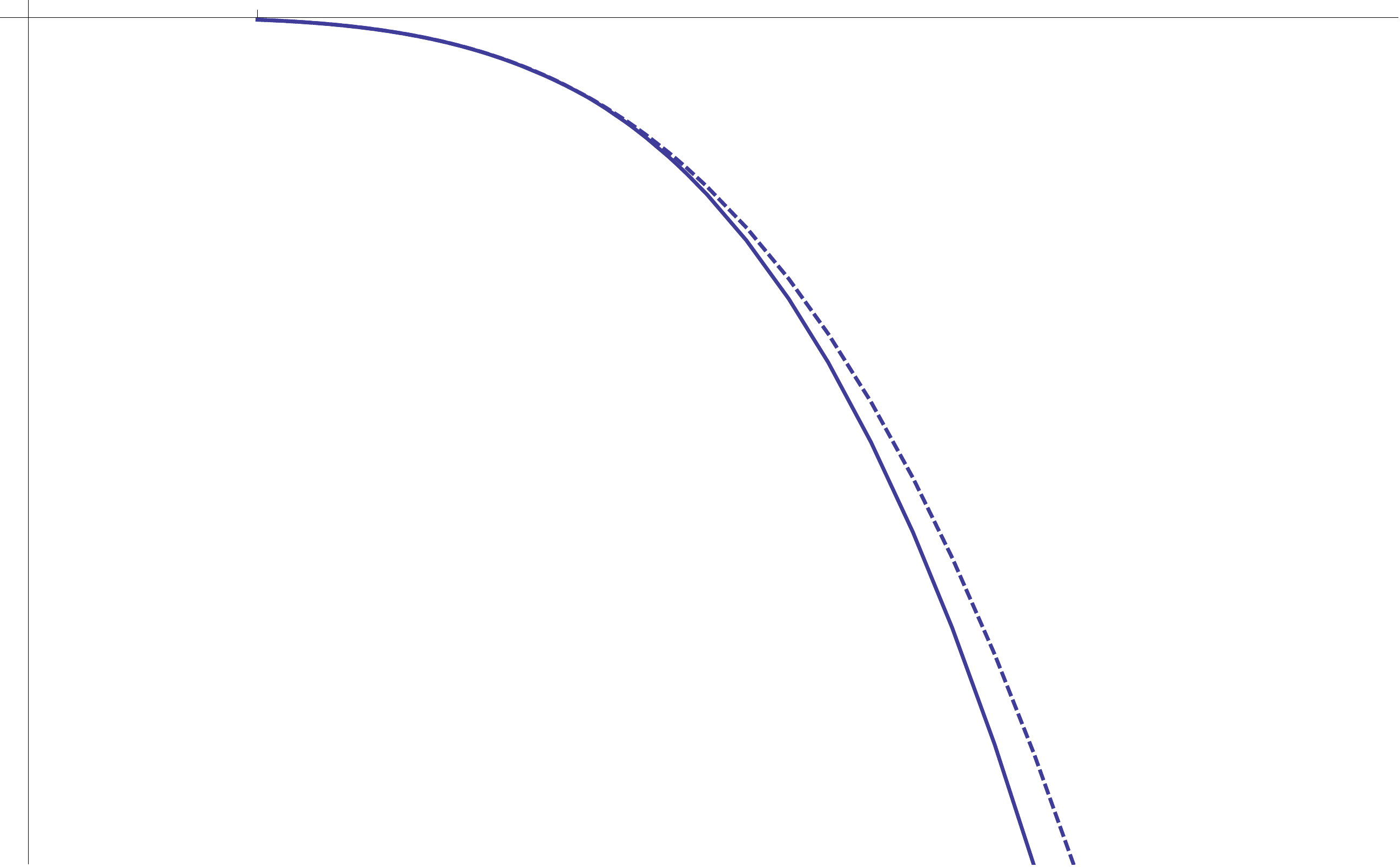}
\put(101,59){${\cal T}$}
\put(-2,59){$0$}
\put(1,62.5){${\cal F}$}
\put(17,62.5){$T_{min}$}
\end{overpic}
\caption{Free energy difference as a function of the (dimensionless)
  temperature  between the two branches of black holes emanating from
  UV$_2$ with negative source (dashed and solid purple curves in
  figures \protect\ref{skipUV2-finiteT} and  \protect\ref{UV2}). 
  }\label{UV2-Pos}
\end{figure}
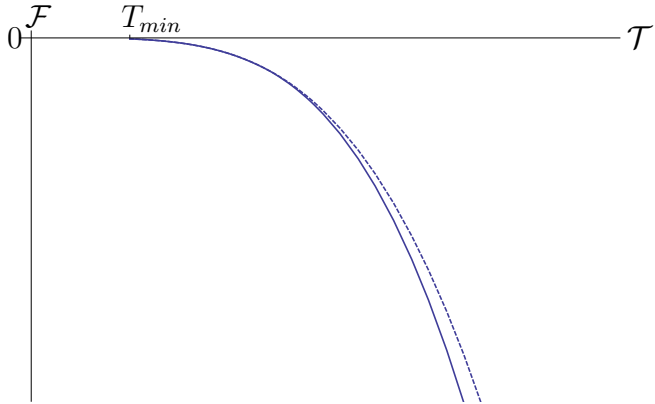

\paragraph{Positive source phase diagram.} We now turn to the black-hole solutions
starting at UV$_2$ with positive source. Recall that no such
regular solutions exist at zero temperature: as one can see  from
figure \ref{skipflatnew} the only regular flow vacuum flow from UV$_2$
is the one ending at IR$_1$, and it has negative source. As was
already noted in \cite{multibranch}, this means
that for $j>0$ the dual field theory is ill defined\footnote{examples of
this behaviour in perturbative field  theories are common, e.g $\lambda
\phi^4$ theory is defined only for $\lambda >0$, and the same can be
said for the sign of the 't Hooft parameter $\lambda = g^2 N$ in
Yang-Mills theory.}.

Interestingly, at finite temperatures larger than a minimal
temperature $T_{min}$, black-hole solutions with $j>0$ start to
exist, as one can see in figure \ref{UV2}. In other words, it is only by turning on
a sufficiently high temperature that we obtain regular solutions with
this sign of the source. A possible interpretation from the field
theory point of view would be the fact that temperature provides an IR
regulator which eliminates some IR pathologies which made the vacuum
theory ill-defined.

At any temperature above $T_{min}$ there are two black-hole
solutions. Computing their free energy, we found that the one which
dominates the ensemble is always the one with smaller $\phi_h$ (the solid blue branch in figure \ref{UV2}). This is shown in figure
\ref{UV2-Pos}. The dominant branch is the one which, as the
temperature rises, approaches the AdS-Schwarzschild black hole with
constant scalar field $\phi= \phi_1$ (see discussion in the following
paragraphs).

\paragraph{Infinite Temperature limits.} There are three situations in
which the dimensionless temperature ${\cal T} \to \infty$: when the horizon
approaches one or the other UV fixed points, and at the point
$\phi_c$. The first case is easy to understand, as it corresponds to
the limit of large AdS-Schwarzschild black holes in which the scalar
is approximately constant. The divergence of the temperature at
$\phi_h = \phi_c$ is more interesting: across this point the UV
source changes sign. From the definition (\ref{FE8}), the divergence in ${\cal T}$ can be interpreted in this case as the limit $j\to 0$ with $T$ finite.   This means that one should be  able to
find a  black-hole solution
with {\em zero} source but finite temperature and with horizon exactly at
$\phi_c$.  These flows are driven by the sub-leading term in the UV
scalar field expansion, corresponding to a  vev of the dual operator
in the absence of a source. They will be discussed in detail in
section \ref{sec:vev}.

\begin{figure}[h!]
\centering
\begin{overpic}[width=9cm,angle=90]{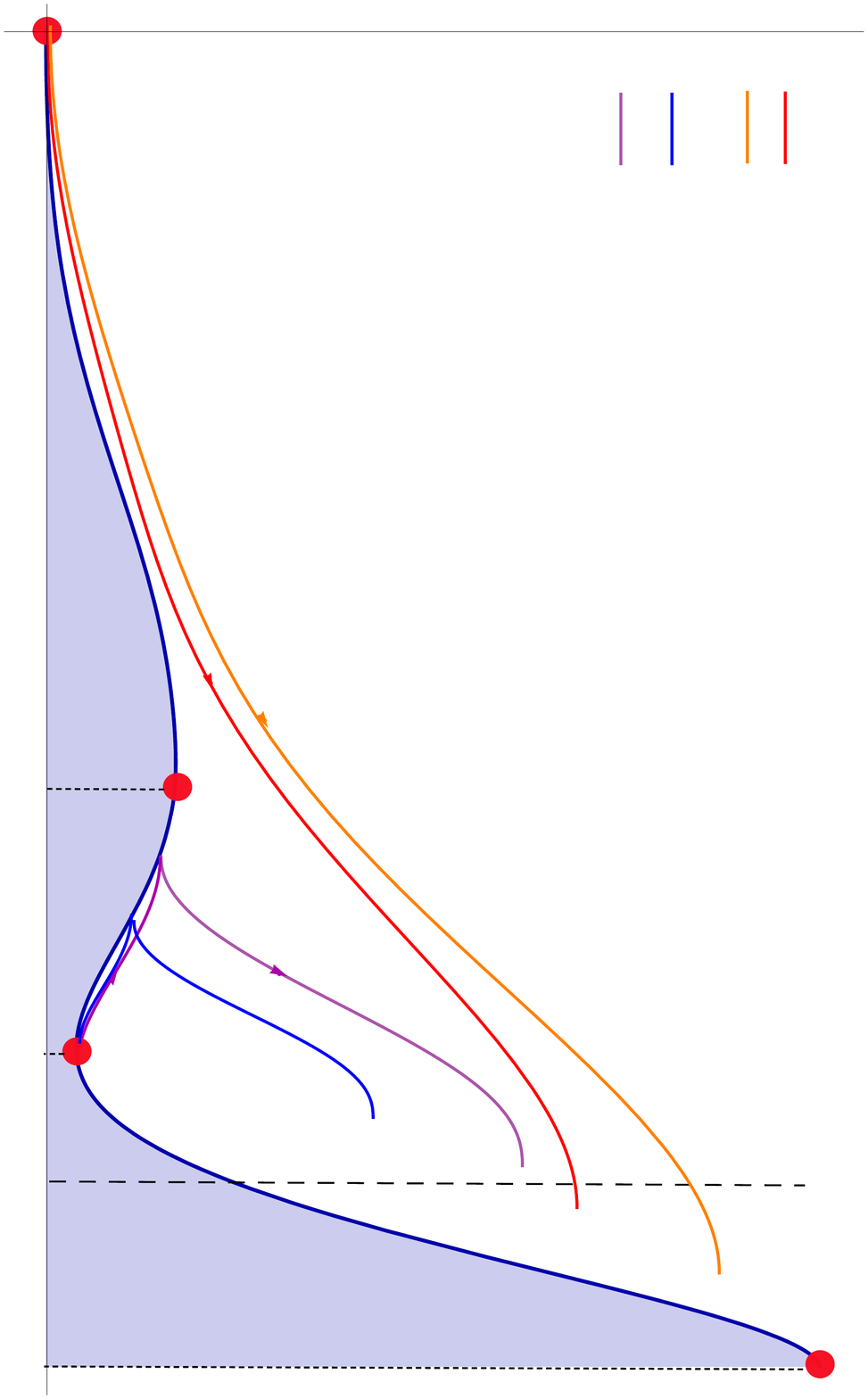}
\put(100,8){$\phi$}
\put(2,5){0}
\put(55,5){$\phi_0$}
\put(74,5){$\phi_1$}
\put(97,5){$\phi_2$}
\put(83,5){$\phi_*$}
\put(1,67){$W(\phi)$}
\put(3,11){UV$_1$}
\put(55,14){IR$_1$}
\put(72,13){UV$_2$}
\put(95,64){IR$_2$}
\put(14,57){{\small Skipping flows from UV1}}
\put(14,48.5){{\small Bouncing flows from UV2}}
\end{overpic}
\caption{The approach to $\phi_h = \phi_*$ from both sides. The curves
show the superpotentials of the corresponding flows. Here we display
flows with different temperatures (longer flows correspond to lower
temperatures).}\label{missing1}
\end{figure}

\begin{figure}[h!]
\centering
\begin{overpic}[width=9cm,angle=90]{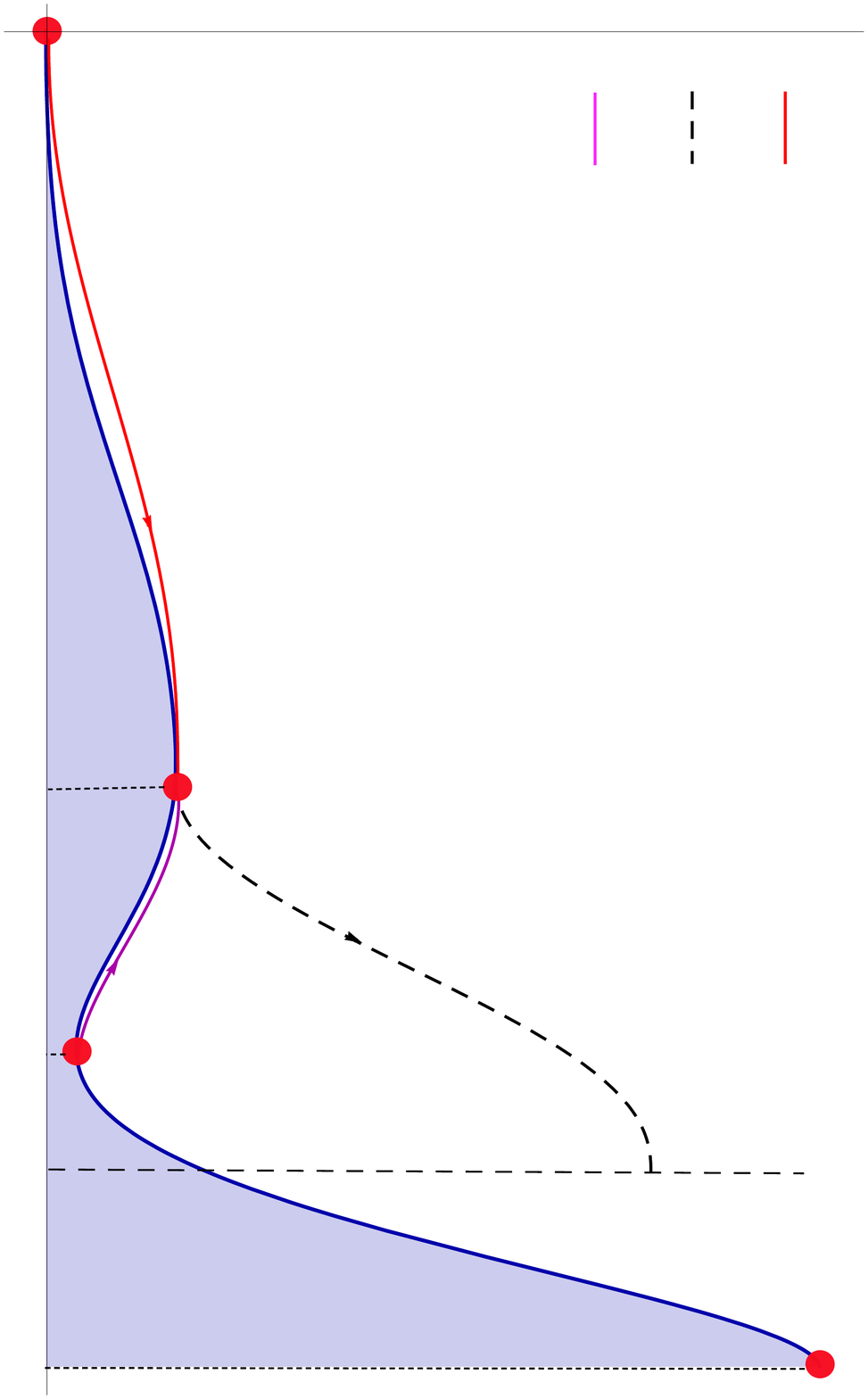}
\put(100,8){$\phi$}
\put(1,65){$W(\phi)$}
\put(2,5){0}
\put(55,5){$\phi_0$}
\put(74,5){$\phi_1$}
\put(97,5){$\phi_2$}
\put(83,5){$\phi_*$}
\put(20,12.5){\scriptsize $W_1$}
\put(62,30){\scriptsize $W_2$}
\put(64,16){\scriptsize $W_1'$}
\put(3,10){UV$_1$}
\put(55,14){IR$_1$}
\put(72,12){UV$_2$}
\put(95,65){IR$_2$}
\put(13,59){Zero-$T$ flow UV$_2$ $\to$ IR$_1$}
\put(13,52.5){Finite-$T$ vev-flow}
\put(13,45.5){Zero-$T$ flow UV$_2$ $\to$ IR$_1$}
\end{overpic}
\caption{In  the limit $\phi_h \to \phi_*$ the solutions
  reduce to the two zero-temperature flows ending at IR$_1$, plus a
  finite-temperature flow starting at IR$_1$ in the UV and having its
  horizon at $\phi=\phi_*$ (dashed black curve). This solution has a
  finite temperature and entropy since it has a regular black-hole horizon.}\label{missing2}
\end{figure}

\paragraph{Zero temperature limits.} The limit ${\cal T}\to 0$  occurs when
$\phi_h$ coincides with one of the IR fixed points (in which case we
recover the two zero-temperature vacuum flows, skipping and
non-skipping. Additionally, ${\cal T}\to 0$ as $\phi_h\to  \phi_*$, which is
somewhere in the middle between $UV_2$ and $IR_2$. This may seem
puzzling as there is no zero-temperature solution which ends at
$\phi_*$, since this does not correspond to an extremum of the
potential. The puzzle can be resolved by  tracking  how the horizon
approaches $\phi_*$ from both sides. This is  represented  in figure
\ref{missing1} in terms of the superpotentials of the various
branches:
\begin{itemize}
\item approaching $\phi_*$ from the left ($\phi_h < \phi_*$), we have  solutions
starting at UV$_2$ with negative source, bouncing before IR$_1$ and
ending close to $\phi_*$. These are represented as the blue and violet flows in
figure \ref{missing1}.  The closer the horizon is to $\phi_*$, the
closer the bounce is to IR$_1$.
\item from the right  ($\phi_h > \phi_*$), solutions start at UV$_1$, skip IR$_1$ and
  end beyond $\phi_*$.  These are represented as the red and orange flows in
figure \ref{missing1}. The closer the horizon is to $\phi_*$, the
  closer the solution approaches IR$_1$ without touching it.
\end{itemize}
 One can see that, {\em in the limit} $\phi_h \to \phi_*$ (represented
 in figure \ref{missing2}),  both
 classes of  solutions (starting from UV$_1$ and UV$_2$)  actually reach IR$_1$
 from both sides,  and coincide with the ones represented in figure
 \ref{skipflatnew}. These flows stops at IR$_1$ as the scale factor goes to zero
  there, and the solution is approaching asymptotically the
  Poincar\'e-AdS horizon. The remaining leftover piece (dashed black
  line in figure \ref{missing2}) starts  from IR$_1$ and arrives to a horizon
  exactly at $\phi_h=\phi_*$.

   However, for this last flow the point  IR$_1$ is
  seen as a {\em UV fixed point}: recall that the superpotential
  always increases along the flow. These black holes therefore are
  thermal states in a different theory, the one for which $IR_1$ is a UV fixed point and  have finite
  temperature {\em as defined in  the UV theory sitting at IR$_1$.}

  To summarise, in the limit
  $\phi \to \phi_*$, each black-hole solution from UV$_1$ and UV$_2$ splits into two
  disconnected solutions: a zero-temperature flow ending at IR$_1$
  (red and purple curves in figure \ref{missing2}),
  and a finite temperature  (vev-driven)
  flow starting in the UV from IR$_1$ and
  having its horizon at $\phi_*$ (dashed black curve in figure
  \ref{missing2}).

Like those ending at $\phi=\phi_c$, these black
  holes are also driven by a vev, and they will be discussed in more
  detail in section 5.

We make one last comment about an interesting feature of the solutions with horizon approaching
$\phi_*$: as can be observed in figure \ref{CUV1}, although the
solutions approaching the ``gap'' from the left and from the right are
disconnected, it appears that the vev parameter $C$ takes the same
value  on each side of the gap, i.e.  $C(\phi_0) = C(\phi_*)$. This
implies continuity of the free energy across the gap, since
temperature and entropy approach zero at both $\phi_0$ and
$\phi_*$. This can indeed be understood analytically. Indeed, consider
one of the solutions starting from $UV_1$ in figure \ref{missing1}. As
$\phi_h \to \phi_*^+$, we can regard it  as composed by a flow $W_1$ which
stops just above $\phi_0$ plus a flow $W_2$ which
starts just above $\phi_0$ and reaches around $\phi_*$.  In the limit
$\phi_h\to  \phi_*$, the flow $W_1$ reduces to the non-skipping
zero-temperature flow (red line in figure \ref{missing2}) while $W_2$
reduces to the flow from $IR_1$ to $\phi_*$ (dashed black line in figure
\ref{missing2}). The free energy can be then written as a sum of the
two contributions,
\be\label{mis1}
{\cal F}(\phi_h\to \phi_*^+) = {\cal F}_1 + {\cal F}_2
\ee
By construction, in the limit we are
considering,   ${\cal F}_1$ approaches the free energy of the
zero-temperature non-skipping solution with endpoint at $\phi_0$,
\be \label{F1}
{\cal F}_1 = {\cal F}(\phi_0).
\ee
As we will now show below, the
contribution ${\cal F}_2$ is vanishingly small as
$\phi_h\to \phi_*$ .   In fact, the flow $W_2$ can be also approached,
in this limit, by the upper branch of the flow $W_1' + W_2$ in  figure
\ref{missing2}, which starts  from UV$_2$, and bounces  very close to IR$_1$: this is the limit of the blue and
purple curves in figure \ref{missing1} when the bounce point
approaches IR$_1$.  We can view the   branch $W_2$  after the bounce as a
flow from a finite UV cut-off scale (with cut-off energy proportional to
$e^{A(u_b)}$)  to the horizon at $\phi_h\simeq \phi_*^-$.
 Its contribution to the total free
energy of the solution is finite, and given by (see appendix
\ref{app:onshell} for details):
\be\label{mis2}
{{\cal F}_2 \over M_p^{d-1}V_{d-1}} = e^{dA(u_b)}W(\phi_b) \sqrt{f(u_b)}
\ee
where $u_b$ is the radial coordinate of the bounce and $\phi_b$ is the
corresponding value of the scalar field. Similarly, the bounce acts as an IR cut-off (with the same
scale $e^{A(u_b)}$) for the lower branch $W_1'$,  connecting to $UV_2$.  Now, as
 $\phi_h\to \phi_*^-$, the bounce point $\phi_b$ approaches the IR fixed
point $\phi_0$, the  cut-off scale  $e^{A(u_b)} \to 0$ and the
contribution (\ref{mis2}) to the free energy vanishes,
\be \label{F2}
{\cal F}_2 \to 0,  \qquad \phi_h \to \phi_*.
\ee
Putting together equations (\ref{mis1}), (\ref{F1}) and (\ref{F2})  we
conclude that
\be
{\cal F} (\phi_*) = {\cal F}(\phi_0).
\ee

\section{Thermodynamics of bouncing RG flows}
\label{sec:bouncing}

In this section we discuss another ``exotic'' kind of behaviour, which
is  unusual  from the perturbative field theory standpoint, but which
can  be found in holographic RG flows
\cite{multibranch,Gursoy:2016ggq}. This consists in solutions where the
flow inverts its direction at some $\phi= \phi_b$ (``bounce''). Although the
corresponding $\beta$-functions are non-analytic at this point, $\beta(\phi)
\sim \sqrt{|\phi - \phi_b|}$, the corresponding gravity solutions are
regular and can be continued past the point $u_b$ where the bounce
takes place. Specifically, around $u=u_b$ the zero-temperature
solutions have the expansion,
\bea\label{bounce1}
&& A(u) = A_b^{(0)} + (u-u_b) A_b^{(1)} + O\Big( (u-u_b)^2 \Big),
\nonumber \\
&& \\
&& \phi(u) = \phi_b + {V'(\phi_b)\over 2} (u-u_b)^2   + O\Big(
(u-u_b)^3 \Big), \nonumber
\eea
where $A_b^{(0)},  A_b^{(1)}$ are constants. From equation
(\ref{bounce1}) it is clear that
bounces can occur at any point in field space such that $V'(\phi_b) \neq
0$, in a such a way that $\phi$ has a maximum at $\phi_b$ if $V'(\phi_b)
<0$, and a minimum if $V'(\phi_b) >0$.

\subsection{Bouncing solutions at zero temperature}

In \cite{multibranch}, several examples were presented of vacuum holographic RG
flow solutions exhibiting bounces. Some of them interpolate between a
UV and an IR fixed point, and go through one or more  bounces
somewhere in the middle. In other examples, the potential  had a
single extremum (a UV maximum at $\phi=0$
and the solution reaches $\phi \to \pm\infty$ after a bounce at a
finite $\phi_b$.  The latter case corresponds to the zero-temperature
solution in a model which was already considered at finite temperature\footnote{The corresponding bouncing RG flow in the vacuum state has been also worked out but not discussed in \cite{Gursoy:2016ggq}.} in \cite{Gursoy:2016ggq}.

\begin{figure}[h!]
\centering
\includegraphics[width=12cm]{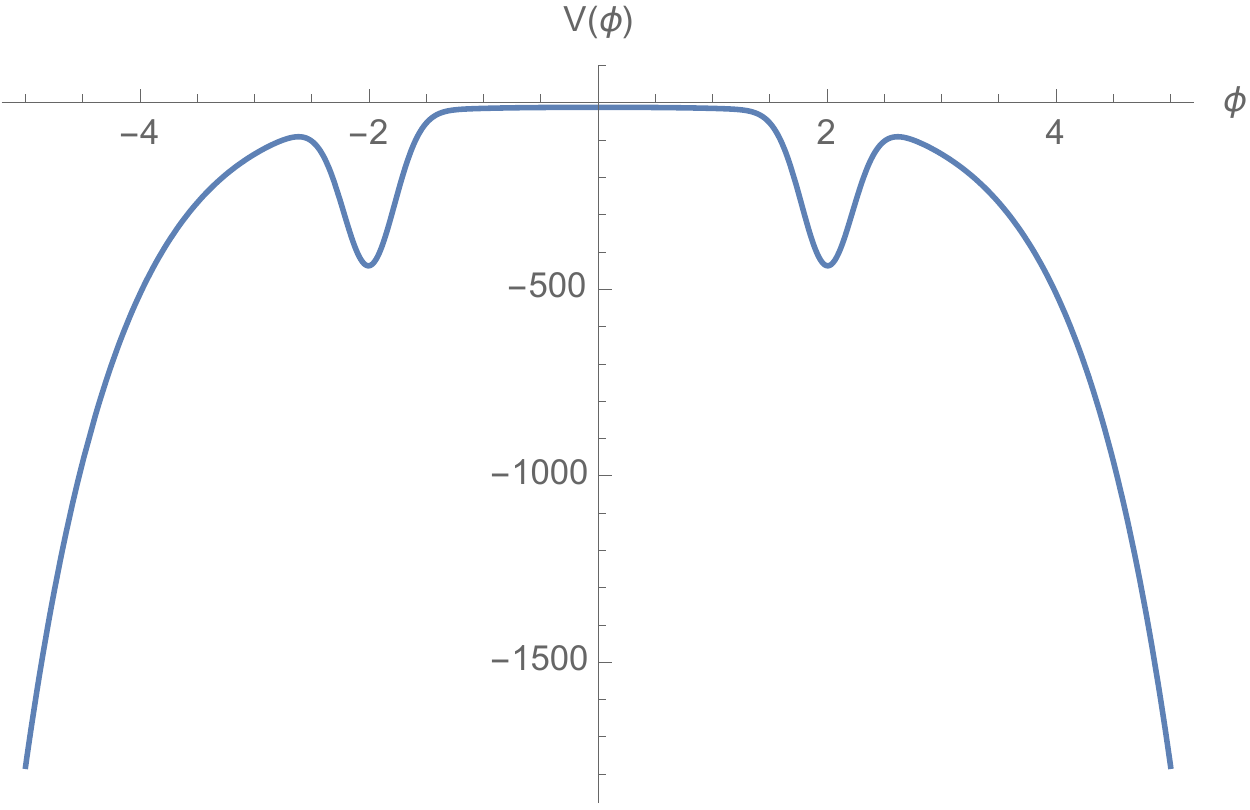}
\caption{Example of a dilaton potential that leads to bouncing solutions. The potential is as in equation (\ref{Vgauss}) with a choice of parameters $d=4$, $\Delta_-=1.7$, $\phi_0=4$, $\Gamma=2/\sqrt{3}$, $c=400$,  $\sigma= 1/20$. The UV fixed point corresponds to $\phi=0$ and the IR fixed point corresponds to $\phi \approx 2.00566$. }\label{FigPot}
\end{figure}

\begin{figure}[h!]
\centering
\includegraphics[width=12cm]{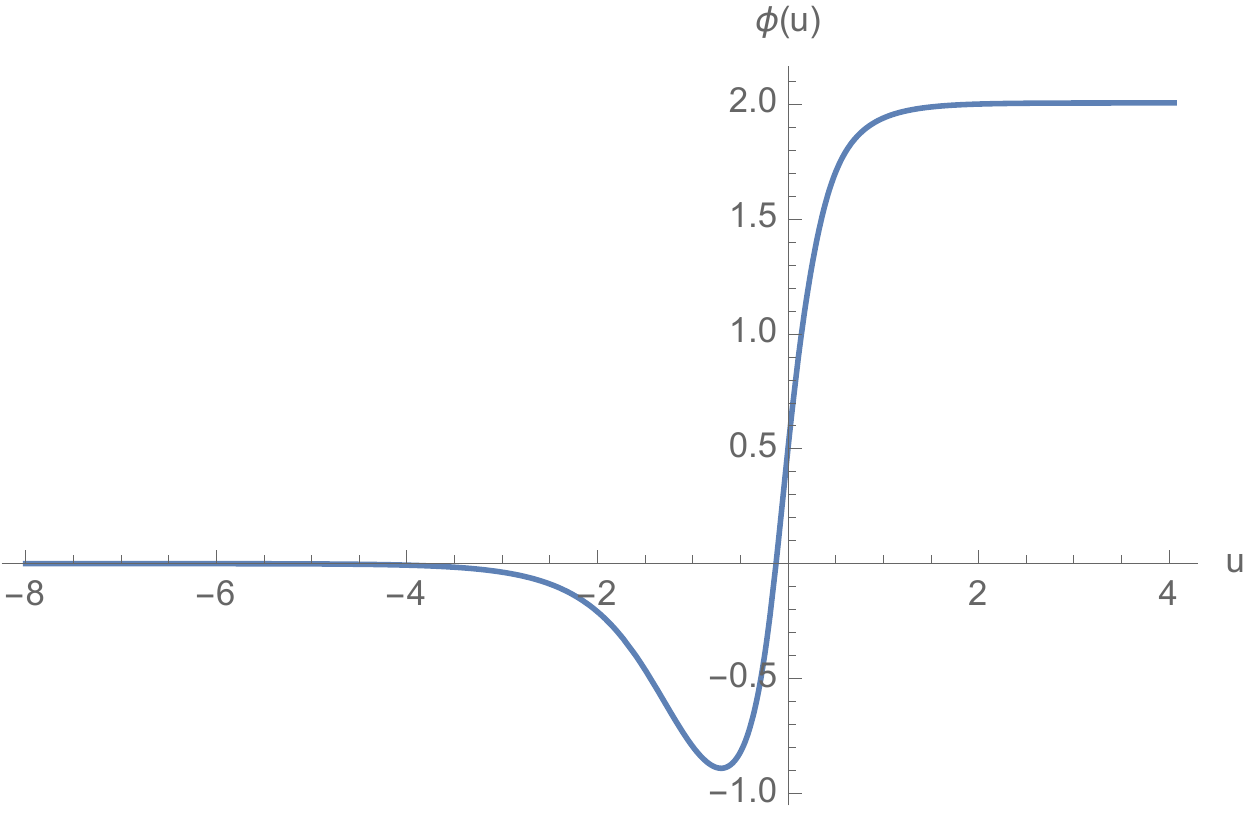}
\caption{Bouncing zero-temperature solution for the potential shown in figure \ref{FigPot}. }\label{FigPhi0}
\end{figure}

Bounces in the ground state solution occur when the dilaton potential
is sufficiently steep. For example, if the potential behaves as $V(\phi) \propto \exp(\Gamma \phi)$ for large $\phi$ then one finds a bouncing zero T solution as the exponent is larger than a critical value $\Gamma > \Gamma_c$. To be concrete, we consider the potential
\bea\lab{Vgauss}
V(\phi) &=& - \frac{d(d-1)}{\ell^2}\le(\cosh(\Gamma\phi)-\frac12 \phi^2 \Gamma^2 \ri)+\frac{1}{2\ell^2} \phi^2 \Delta_-(\Delta_- - d) \nonumber \\
{}&& - \frac{c}{\ell^2} \le(e^{-\frac{(\phi-\phi_0)^2}{2\sigma}}+e^{-\frac{(\phi+\phi_0)^2}{2\sigma}} + \phi^2 e^{-\frac{\phi_0^2}{2\sigma}}\le(\frac{1}{\sigma} - \frac{\phi_0^2}{\sigma^2}\ri)-2e^{-\frac{\phi_0^2}{2\sigma}}\ri)\, .
\eea
This corresponds to a potential with a cosh and a quadratic term modified with two gaussian peaks with width $\sigma$ and amplitude $c$ added at $\phi=\phi_0$ and $\phi=-\phi_0$. Without the gaussian modification, $\phi$ would run from the UV conformal fixed point at $\phi=0$, to the IR region at $\phi\to +\infty$ but the gaussians introduce an IR conformal fixed point at $\phi\approx \phi_0$ hence the IR behaviour is conformal. The quadratic terms in the potential above adjust the UV dimension of the perturbing operator to be  $\Delta_+ = d- \Delta_-$.
We plot this potential for a particular choice of parameters in figure
\ref{FigPot}, and we  show the bouncing zero-temperature (vacuum state) solution in figure \ref{FigPhi0}.
\begin{figure}[h!]
\centering
\includegraphics[width=12cm]{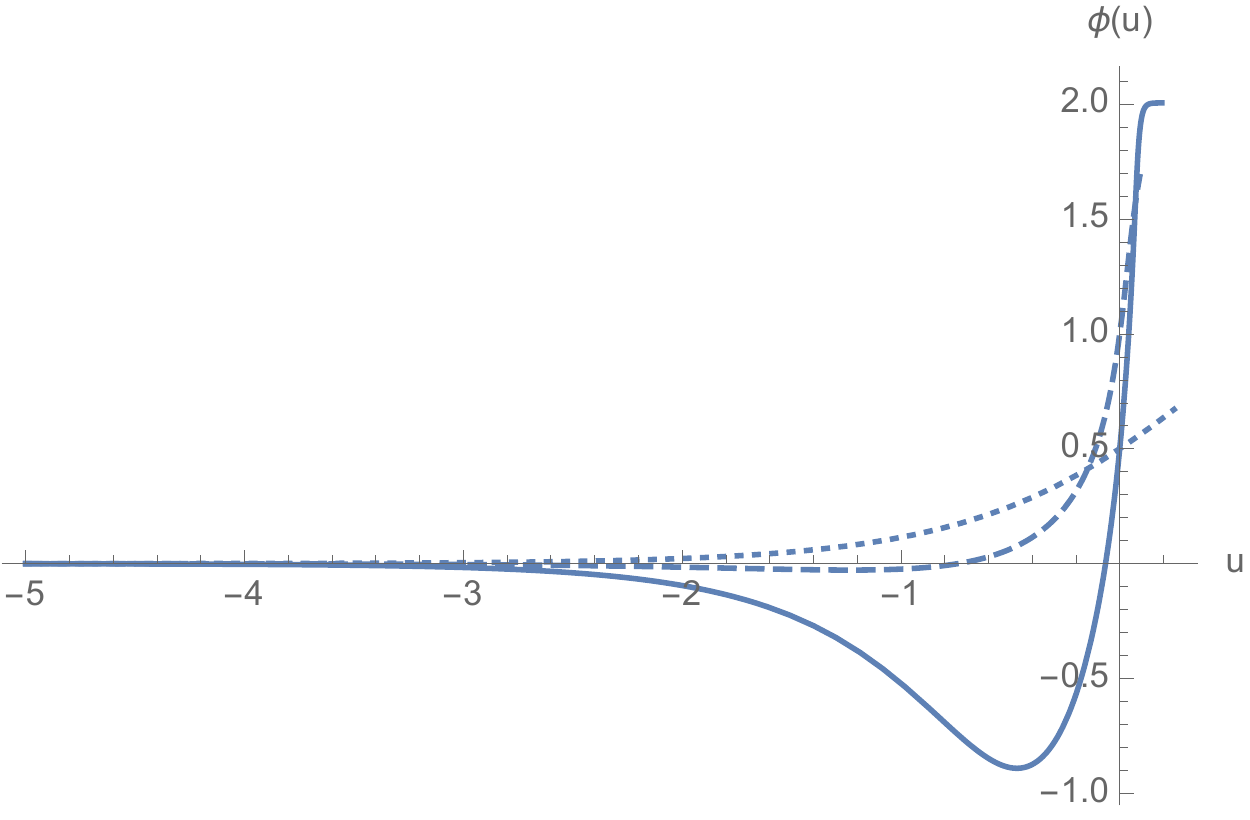}
\caption{Comparison of the vacuum solution (solid), a bouncing finite T solution with $\phi_h=1.7$ (dashed) and a non-bouncing finite T solution with $\phi_h=0.7$ (dotted) that arise from the potential shown in figure \ref{FigPot}. }\label{FigPhi}
\end{figure}

\begin{figure}[h!]
\centering
\includegraphics[width=10.5cm]{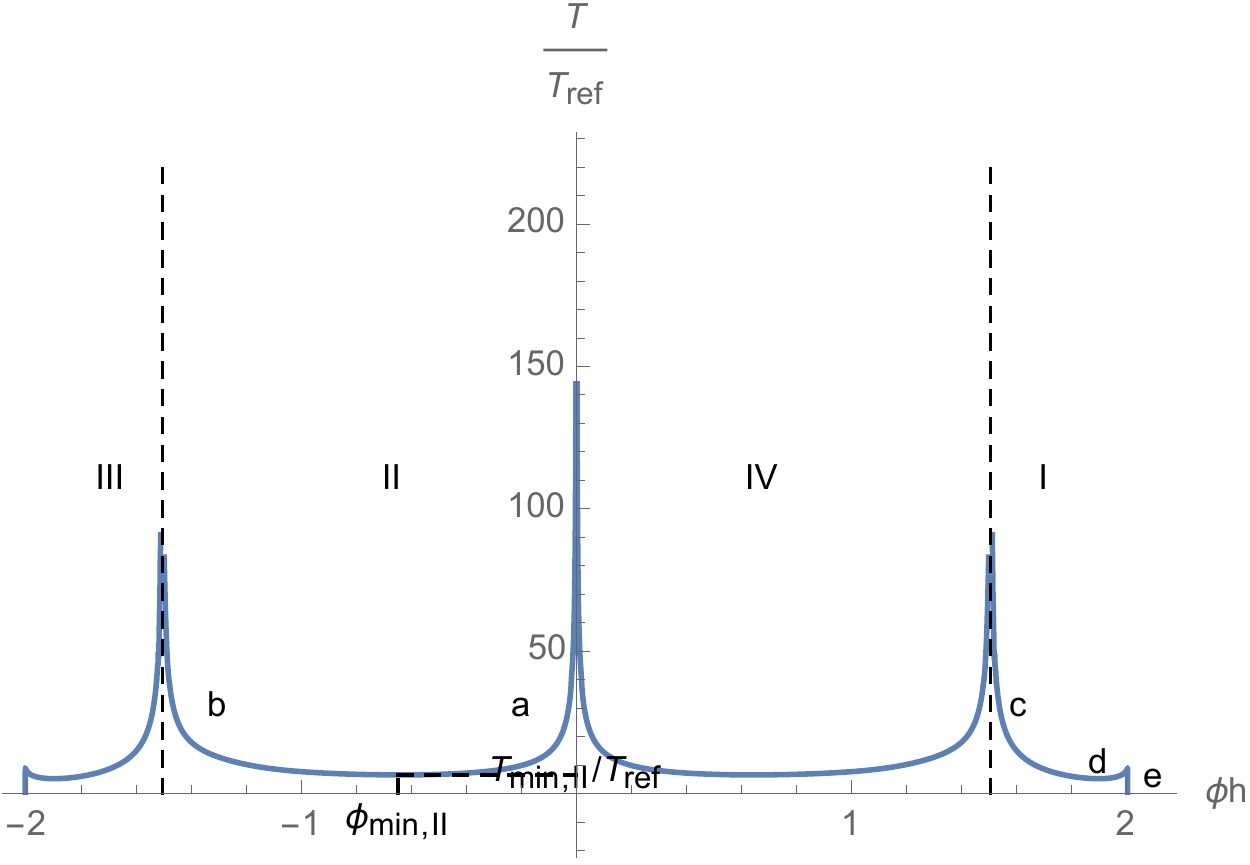}
\caption{T as a function of $\phi_h$ for a particular choice of the reference scale $T_{ref}$.  }\label{FigT}
\end{figure}

\begin{figure}[h!]
\centering
\hspace{0.2cm}\includegraphics[width=12.5cm]{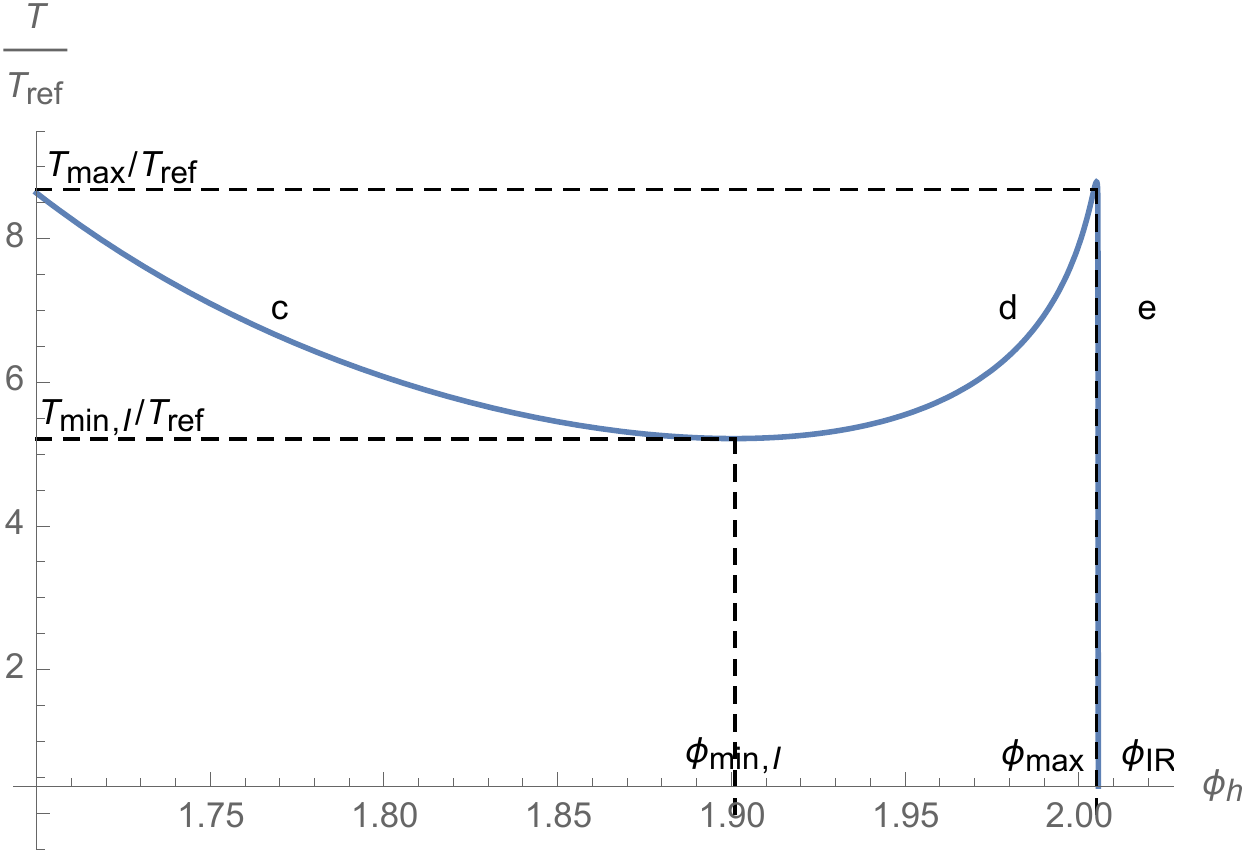}
\caption{  Blowing up the figure \ref{FigT} at very small temperatures reveals existence of there different blackhole solutions for $T$ fixed between a $T_{min,I}$ and a $T_{max}$.  }\label{FigT2}
\end{figure}

\subsection{Bounces at finite temperature and the phase diagram}

As in  section 3, one has to determine $T$ as a function of $\phi_h$
to describe all the black-hole solutions at a given $T$. Figure
\ref{FigPhi} shows a comparison of the bouncing vacuum solution, a
non-bouncing finite T solution with $\phi_h=0.7$ and a bouncing finite
T solution with $\phi_h=1.7$.

Because there can be more than one $\phi_h$ corresponding to the same
value of $T$,  one needs the free energy to distinguish between these multiple black-holes and to determine the dominant one with the lowest free energy.

Calculating $T$ as a function of $\phi_h$ reveals the different
 black-hole solutions at a given temperature. This can be obtained
 either by constructing the geometry and using equation (\ref{b1-1}) or more directly from the potential using the method of scalar variables  explained in appendix \ref{AppXY}, see equation (\ref{Teq1}). Following the latter method we obtain the function $T(\phi_h)$ shown in figure \ref{FigT}.

There are four separate intervals of $\phi_h$ that we label from I to
IV in this figure. First, we note the symmetry $T(\phi_h) =
T(-\phi_h)$. It is not hard to see that this is an immediate
consequence of the symmetry $V(\phi) = V(-\phi)$ of the dilaton
potential. Moreover, there exists a critical value of $\phi_h$, that
reads $\phi_c = 1.507$ for the choice of parameters in figure
\ref{FigPot}, above which the black-hole solution becomes
bouncing. This is demonstrated in figure \ref{FigPhi} where we show a
bouncing type black-hole for $\phi_h = 1.7$ and a non-bouncing type
for $\phi_h=0.7$. These two solutions belong to regions I and IV in
figure \ref{FigT}. It is important to realise that these two regions I
and IV {\em do not belong to the same theory}: as $\phi  = j\,
\ell^{\Delta_-}e^{\Delta_- u/\ell} + \cdots $ near the boundary, the solutions in region IV have $j>0$ whereas the solutions in region I have $j<0$ because of the bounce. Therefore the solutions in regions I and IV belong to different boundary theories.

Similarly, the solutions in region II have $j<0$ and region III have
$j>0$. In the following, we will choose $|j|=1$ with no loss of
generality (with this choice, the dimensionless temperature parameter
${\cal T}$ defined in equation (\ref{FE8}) coincides with the
temperature $T$). To simplify the presentation, we further define two reference scales $T_{ref}$ and $F_{ref}$ are defined in (\ref{TSref}).

The symmetry of $V(\phi)$ under parity implies that
we may only consider the solutions in regions I (bouncing) and II
(non-bouncing). The physics of regions III and IV are the same except
for the sign of $j$.

\subsubsection{Region I: bouncing black-holes}

First consider region I. As shown in figure \ref{FigT}, this region is further divided into three subregions, between $\phi_c \leq \phi_h < \phi_{min,I}$, $\phi_{min,I} \leq \phi_h < \phi_{max}$ and  $\phi_{max} \leq \phi_h < \phi_{IR}$,  labelled on the figure as ``c", ``d" and ``e" respectively,  where for our choices of parameters we have $\phi_{min,I} = 1.901$ and $\phi_{max} = 2.005$. The corresponding temperatures read $T_{min,I} = 5.209\, T_{ref}$ and $T_{max} = 8.685\, T_{ref}$.

As clear from figure \ref{FigT2}  there are three different black-hole solutions in this region at a given temperature between $T_{min,I}$ and $T_{max}$.

 Below $T_{min,I}$ and above $T_{max}$ there is a single type of black-hole solution. The free energy of these solutions can be calculated following the method described in appendix \ref{AppXY}. The result is shown in figure \ref{FplotR}. There is  a first order phase transition at $T_c\approx 6 T_{ref}$. This conclusion however will change once we consider the solutions in region II.

Another
point to note is the free energy of the branch between $T=0$ and $T=T_{max}$, as shown  figure \ref{FigT} (b). The free energy of this branch is almost constant (with almost vanishing entropy) as shown in figure \ref{FplotR}. In the limit $T\to 0$ this branch should smoothly turn into the vacuum solution. Indeed, the limiting value of the free energy as $T\to 0$ coincides precisely with the free energy of the vacuum solution.

\begin{figure}[h!]
\centering
\includegraphics[width=12cm]{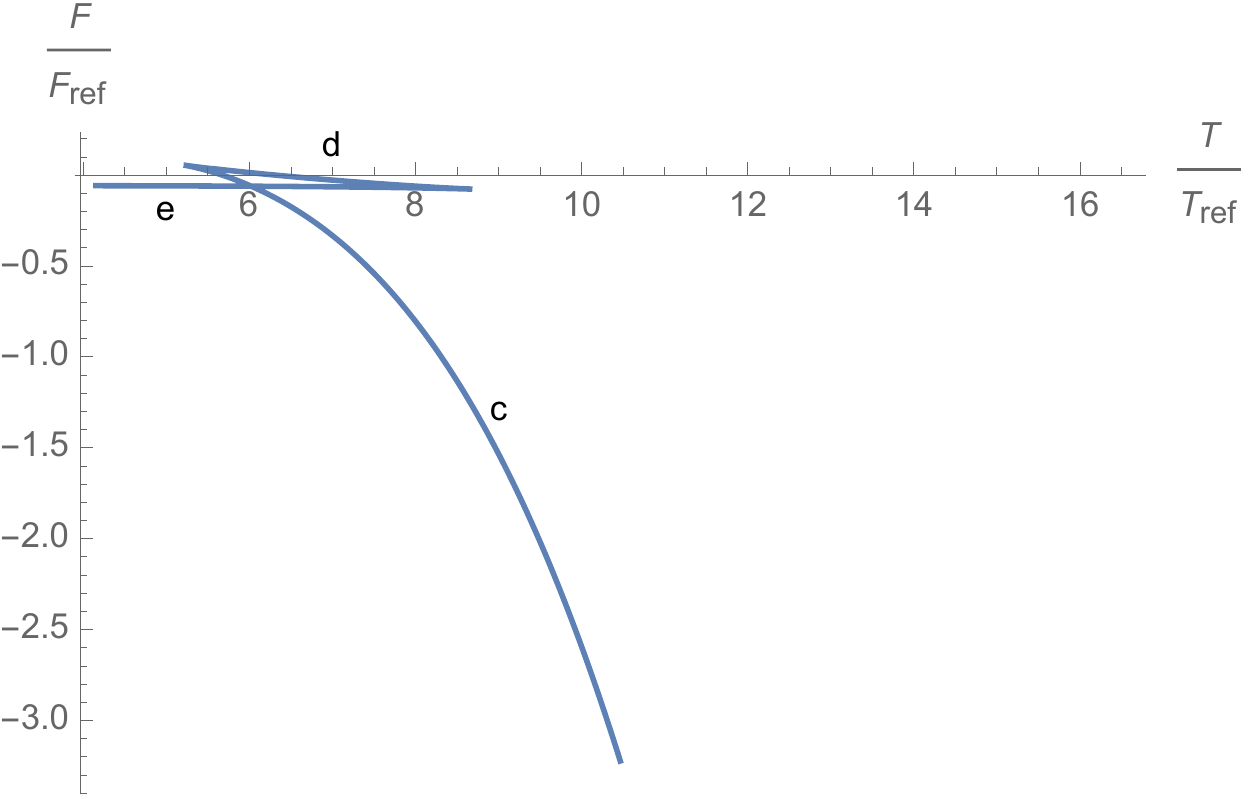}
\caption{Free energy of bouncing type solutions in region I as a function of T that follows from the potential in figure \ref{FigPot}. Latin letters label the different branches shown in figure \ref{FigT}.}\label{FplotR}
\end{figure}

\subsubsection{Region II: non-bouncing black-holes}

In this region there are three black-hole solutions. One observes at least two branches of solutions
in figure \ref{FigT}, for $0 \geq \phi_h > \phi_{min,II}$ and for $\phi_{min,II} \geq \phi_h > -\phi_c$ where
$\phi_{min,II} = -0.649$.  The corresponding temperature reads $T_{min,II} = 6.545\, T_{ref}$.

\begin{figure}[h!]
\centering
\includegraphics[width=8cm]{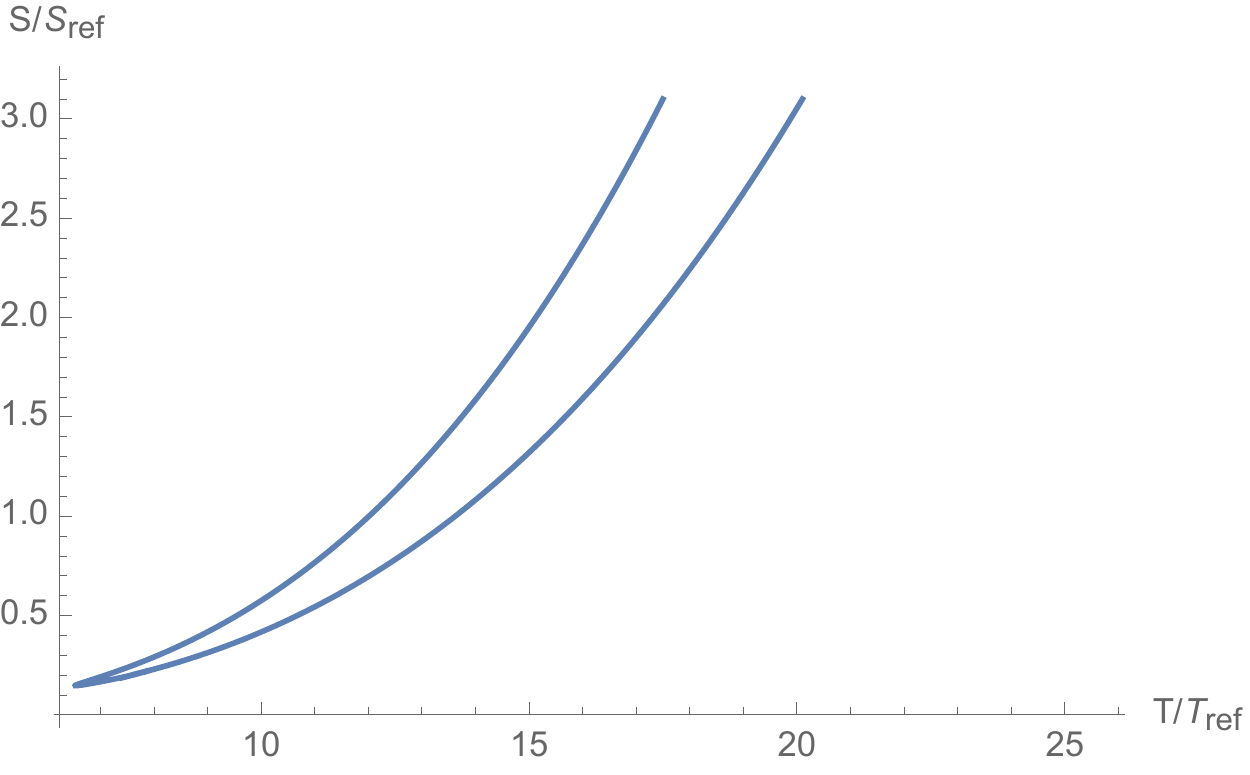}
\hspace{0.5cm}\includegraphics[width=6cm]{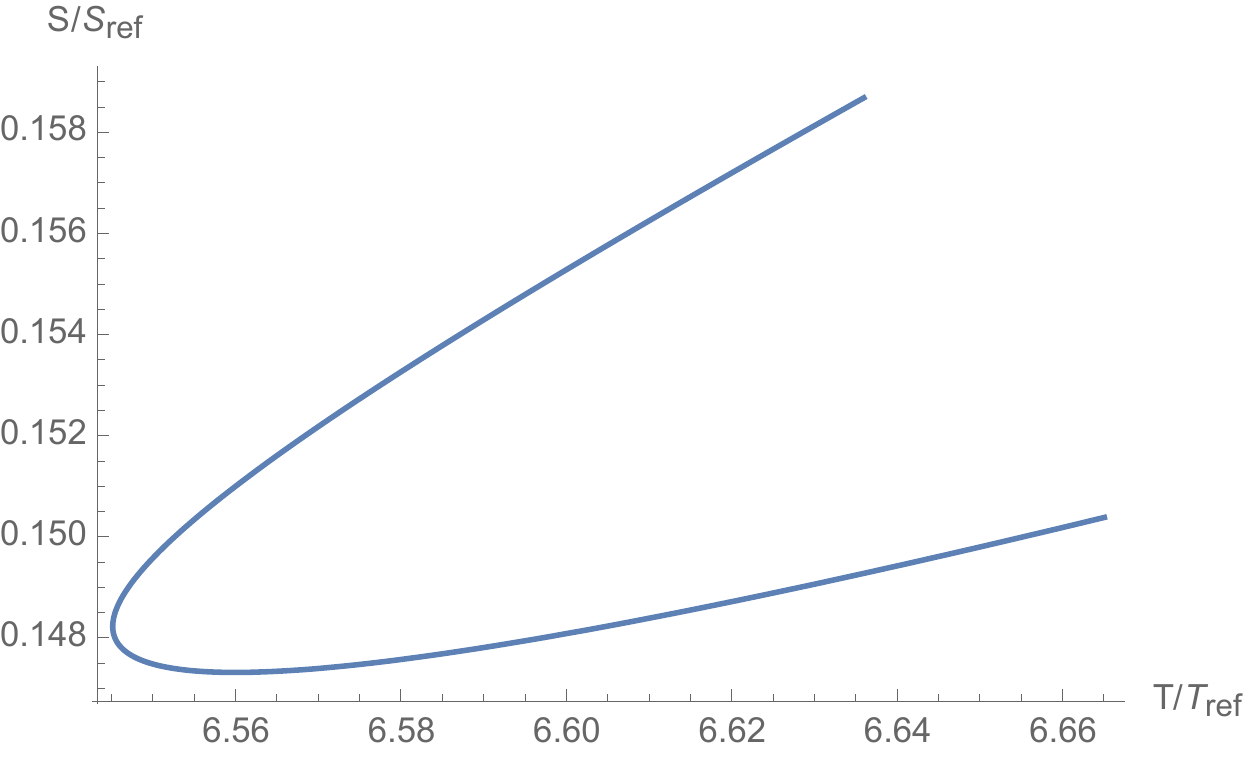}
\\ (a) \hspace{8cm} (b)
\caption{(a) S as a function of $T$ in region II. (b) Blow up of the
  figure around the minimum of $S$. }\label{STplotL}
\end{figure}

 However, there is a hidden third branch in this region, that is unveiled when one plots the entropy as a function of $T$. This is shown in figure \ref{STplotL} (a) and (b). Because of the shape of the function $s(T)$ near the minimum, the aforementioned second branch for $\phi_{min,II} \geq \phi_h > -\phi_c$ is further divided into two, as $T_{min,II} < T < T_s$ and $T_{s} < T < \infty$ where $T_s$ corresponds to the minimum of $s$ in figure \ref{STplotL} (b), with the value
$T_s =  6.560\, T_{ref}$. The corresponding value of $\phi_h$ is $\phi_s=-0.691$. Therefore, we need to divide the branch ``b'' in figure \ref{FigT} further into two: the branch for $\phi_{min,II} \geq \phi_h > \phi_s$ and for $\phi_s \geq \phi_h > -\phi_c$.

\begin{figure}[h!]
\centering
\includegraphics[width=12cm]{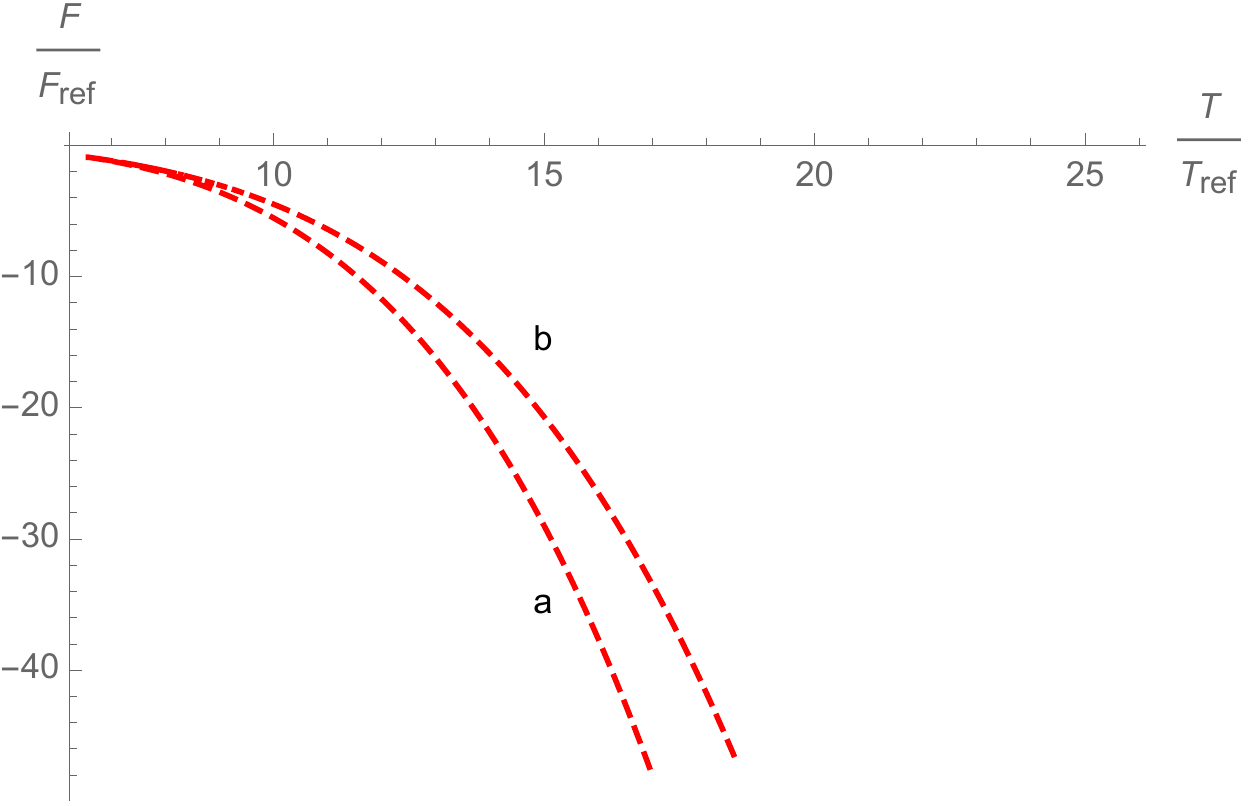}
\caption{Free energy of non-bouncing type solutions in region II as a function of T that follows from the potential in figure \ref{FigPot}. Latin letters label the different branches shown in figure \ref{FigT}.}\label{FplotL}
\end{figure}

The need for this further division of branches becomes more apparent upon consideration of the speed of sound in the plasma. The speed of sound can be expressed in terms of entropy and temperature as
\begin{equation}
c_s^2 = \frac{s}{T \frac{ds}{dT}}\, .
\end{equation}
As $s$ and $T$ are positive definite everywhere, and that $dS/dT$
vanishes (diverges) at $T=T_s$ ($T = T_{min,II}$) one finds that
$c_s^2$ ranges between  $0^-$ and $-\infty$ between
$T_{min,II}<T<T_s$. The reason that $c_s^2\to-\infty$ as $\phi_h\to \phi_s^+$ is because $ds/dT\to 0^-$ there. On the other hand $c_s^2\to+\infty$ as $\phi_h\to \phi_s^-$, thus $c_s^2$ of this solution jumps by an infinite amount at $T=T_s$. Therefore one has to characterise the solutions for $T<T_s$ and $T>T_s$ differently. That $ds/dT<0$ between
$T_{min,II}<T<T_s$ implies that this branch is thermodynamically unstable: it has negative
definite specific heat\footnote{Therefore there exists a very small thermodynamically unstable region on the (generically thermodynamically stable)
upper branch of solutions shown in figure \ref{FplotL} between the cusp at $T = T_{min,II}$ and $T_s$. At $T_s$ $C_v$ changes sign and becomes positive for $T>T_s$. This does not imply any non-analyticity for the free energy at $T_s$ however.} per unit volume $C_v = T ds/dT$.

To complete the discussion of the speed of sound, we find that $c_s^2$ ranges between\footnote{1/3 because the plasma becomes conformal as $\phi_h\to 0$ and $0^+$ as $\phi_h\to \phi_s$ because $dS/dT$ diverges there.} $1/3$ and $0^+$ in branch ``a" on figure \ref{FigT} with $0 \geq \phi_h > \phi_{min,II}$ and it ranges between $+\infty$ and some positive value (which should be determined by precise numerics near $-\phi_c$)  in the branch with $\phi_{s} \geq \phi_h > -\phi_c$. The fact that the speed of sound can exceed 1 in this branch most probably means that this branch is dynamically unstable, in analogy with the example studied in \cite{Gursoy:2016ggq}.

\begin{figure}[h!]
\centering
\includegraphics[width=12cm]{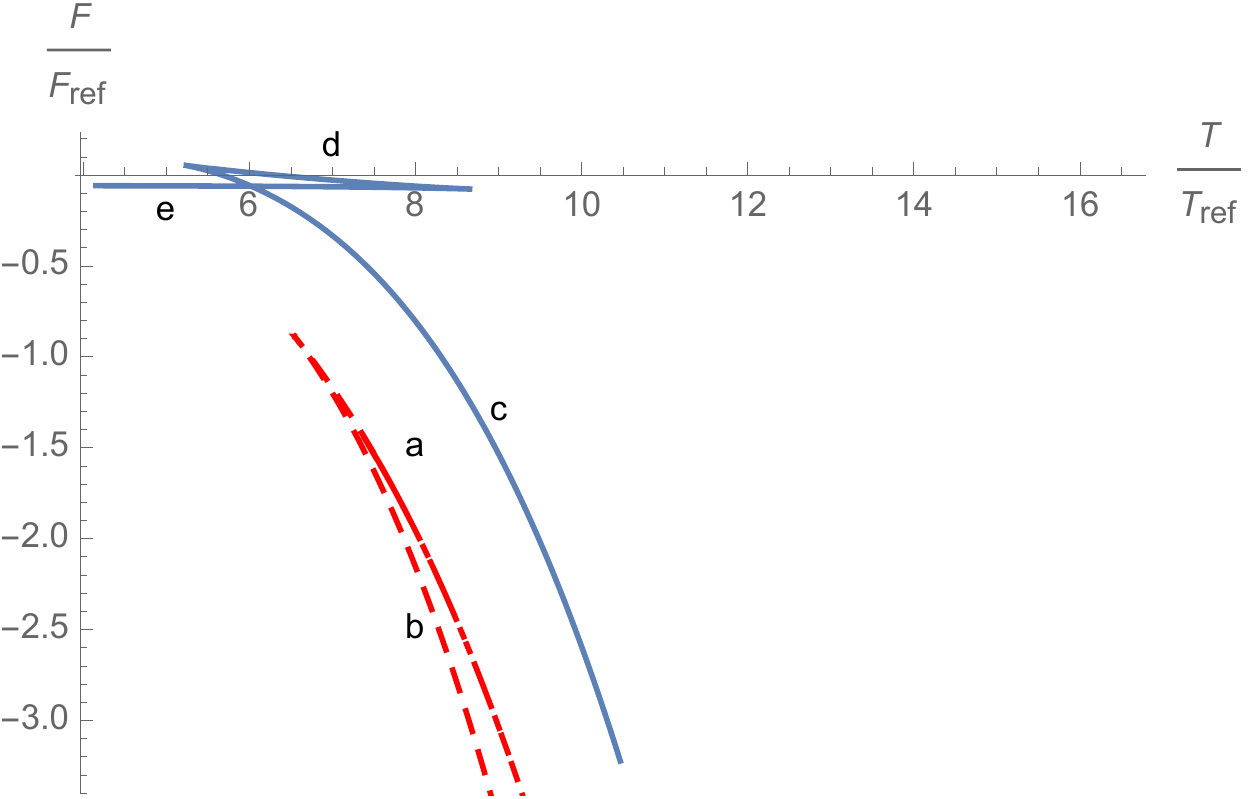}
\caption{Comparison of the free energies of the solutions that belong to region I (blue, solid) and II (red, dashed). }\label{Fcompare}
\end{figure}
Comparison of the free energies of the blackhole branches that belong to region I and II are shown in figure \ref{Fcompare}. One observes that the solutions in region II are always dominant. We observe that , as $T$ is increased first there is a first order transition between the blackhole branches ``e" and ``c" in region II at $T_c/T_{ref} =  6.1$ and then there seems to be a jump in the free energy at $T=T_{min,II}=6.545 T_{ref}$ from branch "c" to branch "b". This jump in the free energy may indicate that the finite temperature extension of bouncing holographic RG flows is ill defined.

\begin{figure}[h!]
\centering
\includegraphics[width=12cm]{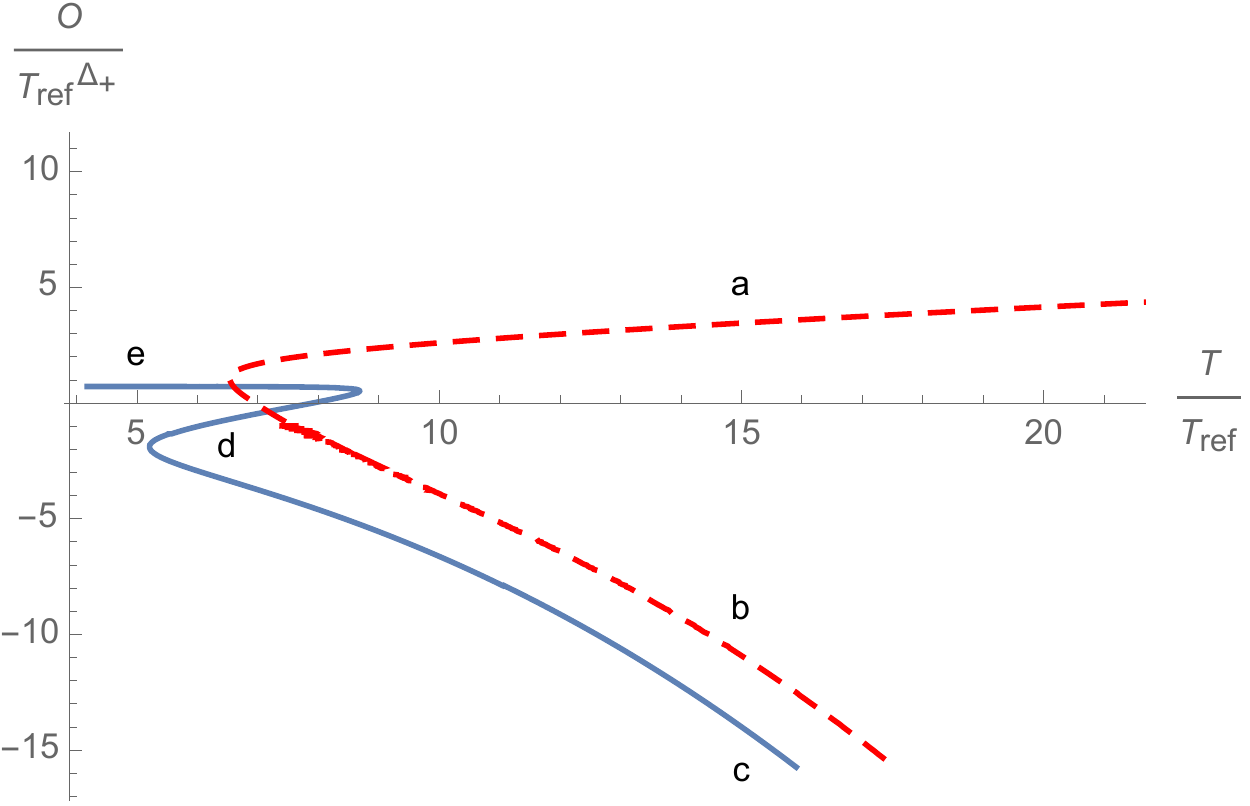}
\caption{Comparison of the VeVs of the scalar operator on the various black hole branches in region I (blue, solid) and region II (red, dashed). Latin letters label the separate branches shown in figure \ref{FigT}.}\label{FigVevs}
\end{figure}

Finally in figure \ref{FigVevs} we plot and compare the vacuum expectation values of the scalar operator $\langle {\cal O} \rangle$ on the blackhole branches in regions I and II.

\section{Sourceless black holes} \label{sec:vev}

As we have seen in the previous two sections, there are some special
values  of the horizon position in field space, $\phi_h$, which
correspond to singular limits:
\begin{itemize}
\item Approaching the values $\phi_c$ which separates between bouncing
  and non-bouncing solutions, the dimensionless thermodynamic quantities diverge;
\item In the case considered in section 2, there is a special point
  $\phi_*$ where (at fixed source) the temperature goes to zero,
  though this value does not obviously correspond to the endpoint of any vacuum RG flow.
\end{itemize}
In this section we clarify the meaning of these special points. As we
will see, they are new families of solutions,
which correspond to
flows driven by the vev of a relevant (in the case of $\phi_c$) or an
irrelevant (in the case of $\phi_*$) operator. Finally, we will
analyse the fate, at finite temperature, of solutions in special (fine-tuned)
theories which at $T=0$  admit regular sourceless flows from a minimum to a
minimum of the bulk potential, an example of which was presented in
\cite{multibranch}.

\subsection{Thermodynamics of vev-driven flows} \label{ss:vevthermo}

It is useful to analyse vev-driven flows in terms of the
superpotential formalism developed in Appendix \ref{app:first}.  In
this language, a sourceless flow is a solution of the first order
equations governed by the superpotential of the type $W^+$, of the
form given in equation (\ref{f3}),
\be \label{thvev1}
\dot{A} = -{1\over 2(d-1)} W^+ (\phi), \quad \dot{\phi} = {dW^+ \over
    d\phi}.
\ee

Unlike $W^-$, which contains the vev-related constant $C(\phi_h)$ as an integration constant
(see equation (\ref{f2})), the  $W^+$ solution is unique and does not admit continuous
deformations. Therefore the solution, when expressed in terms of
the scalar field as an independent variable in the superpotential
formalism, is completely fixed, including the horizon position. The integration constant of the first order equation,  $\phi_+$, is the only remaining integration
constant of the solution, (recall that
the integration constant in $A$ is fixed by the requirement that the
boundary metric is $\eta_{\mu\nu}$ without any scaling factor). Therefore,
vev-driven flows form a one-parameter family of solutions, parametrised
by either $T$ or $\phi_+$. This is to be contrasted with source-driven
flows, which are parametrised by the two independent data
$(T,\phi_-)$.

To make this more explicit, we can now repeat the scaling argument of section \ref{app:dim}, which
shows that solutions with different values of $\phi_+$ are generated
by the transformation
\be\label{thvev2}
\phi_+ \to e^{\Delta_+ v} \phi_+, \qquad T \to T e^{v},
\ee
which is the analog of the symmetry (\ref{dim4}-\ref{dim6}) in the
case of non-zero source. This transformation leaves
$\phi_+/T^{\Delta_+}$ invariant, therefore this quantity must have the
same value for all vev-driven flows\footnote{This applies
  for vev-driven flows which are connected to the same UV fixed
  point. If there are multiple UVs, there can be several vev-driven
  solutions, but there is at most one for each UV fixed point.}. We
conclude that, for vev-driven black holes attached to the same UV
fixed point,  the vev and the temperature are not independent
parameters but they must obey
\be\label{thvev3}
\langle O \rangle = \overline{{\cal O}} \,  T^{\Delta_+}
\ee
where $\overline{{\cal O}}$ is a fixed, temperature-independent, dimensionless constant.

Finally,  it is useful to repeat the
calculation of the free energy in the case of sourceless flows. The
calculation follows the same  steps detailed in Appendix
\ref{app:onshell}, except that instead of the solution $W^-$
(equation (\ref{f2}) we have
to use the superpotential of the type $W^+$ given in equation
(\ref{f3}). The crucial difference in this case is the absence of the $C
\phi^{d/\Delta_-}$, which contributed the second term in the free
energy (\ref{free}): in this case there is no finite term coming from
the sub-leading non-analytic part of the superpotential.

Therefore, for sourceless flows governed by the
superpotential $W^+$,  we simply have
\be \label{thvev4}
{\cal F}_{vev} = - {Ts \over d} \, V_{d-1}.
\ee
This implies that vev-driven flows have conformal thermodynamics:
integrating the relation $s = -(1/V_{d-1}) \de {\cal F} /\de T$ gives
\be\label{thvev5}
s = \overline{\sigma} \,T^{(d-1)}, \qquad {\cal F}_{vev} =
-{\overline{\sigma}\over d} \,  V_{d-1}\, T^d\, ,
\ee
where $\overline{\sigma}$ is a fixed constant which only depends on the details
of the bulk potential.

The fact that sourceless black holes display conformal thermodynamics
is expected since, for zero source, conformal invariance is always softly broken. An alternative derivation of this result, based on the effective
potential, will be presented in the next subsection.

\subsection{Relevant vev flows}

First, we  consider the black holes for which the dimensionless temperature (defined in (\ref{FE8}), ${\cal T} \to \infty$ as
the horizon approaches a critical value $\phi_c$, which is in between
extrema of the potential. As explained in the previous sections, the
value $\phi_c$ separates between bouncing and non-bouncing solutions,
which have opposite values of the source. This change of sign of the
source is perceived as a divergence in the dimensionless temperature
${\cal T}$, which can be interpreted in two different ways depending
how we approach $\phi_c$ in the $(T,\phi_-)$ space:
\begin{enumerate}
\item If we consider the theory with fixed, finite source, then none
  of the solutions can have a horizon at $\phi_h=\phi_c$. As we
  approach this limit, the temperature $T\to \infty$  and so do the
  entropy and the free energy. There is no regular solution in this
  limit.
\item
 On the other hand,  we can approach the $\phi_h=\phi_c$ by keeping $T$ {\em
   fixed} and sending $\phi_-\to 0$. Then we obtain {\em regular}
 black-hole solutions with finite free energy,  a horizon exactly at
 $\phi_c$, and zero source. For these black holes, the flow is driven
 instead by the vev of the (relevant) operator dual to $\phi$.
\end{enumerate}
In the rest of this subsection, we will examine further these
critical black-hole solutions with horizon at $\phi_c$.

We start by noting that  the limit in which the flow becomes vev-driven
corresponds to the following scaling limit  in the parameters entering
the solution (see equation
(\ref{FE6-2}))
\be \label{vev1}
\phi_- \to 0, \quad C \to +\infty, \quad {d\over
  \Delta_-}{ C |\phi_-|^{\Delta_+/\Delta_-} \over (d-2\Delta_-)} sign (\phi_-) =
\phi_+^c \;\; \text{fixed}
\ee
In this limit, the vev $\langle O \rangle = (d-2\Delta_-) \phi_+^c$ remains finite.  As we
will see shortly, the value $\phi_+^c$ is not free but it is determined
by the temperature.

The most transparent way to understand  the thermodynamics of the
vev-driven solutions ending at $\phi_c$ is to use the effective
potential derived in section \ref{ssec:veff}.
 From equations
(\ref{EP3}) and (\ref{EP5}), we see that  solutions with $\phi_- = 0$
correspond to  extrema  of the effective potential
(\ref{EP5}), for which
\be\label{vev2}
{\cal V}'({\cal O}) = 0 \qquad \Rightarrow \qquad {\cal O} = {\cal O}^c
\ee
where ${\cal O} = \<O \> /T^{\Delta_+}$ and ${\cal O}^c$ is a constant.   This equation therefore fixes the
vev in terms of the temperature,
\be\label{vev3}
\<O \>= {\cal O}^c \,  T^{\Delta_+}
\ee
Therefore, there is a one-parameter family of black holes,
parametrised by  $T$, all having
a horizon at $\phi=\phi_c$, zero source, and vev given by equation
(\ref{vev3}).

On-shell,  since the source  is zero, the effective potential is the
same as the free energy, and both are given by
\be\label{vev4}
{\cal F}  =  V_{eff}(T,{\cal O}^c)=  {\cal
  V}_c \,    T^d
\ee
where ${\cal V}_c = {\cal V}({\cal O}^c)$ is a temperature-independent constant. We have
recovered the general fact,  discussed in subsection \ref{ss:vevthermo}, that this family of  black holes
displays conformal thermodynamics, equation (\ref{thvev5}).  Comparing
equation (\ref{vev4}) with (\ref{thvev5}) we can
read-off
\be\label{vev4-i}
{\cal V}_c = -{\sigma^c\over d}V_{d-1}
\ee
where $\sigma^c$ is the (fixed) ratio $s/T^{d-1}$.

Notice  that, for fixed $T$ and $j=0$,  the theory has  another black
  hole solution with the same UV asymptotics:   it is the
  AdS-Schwarzschild black hole with constant scalar field
  $\phi=\phi_{UV}$, sitting at the maximum of the potential. This
  solution also has conformal thermodynamics, with
\be \label{sigmaconf}
{\cal F}_{AdSS} = -{\sigma_{conf}\over d}\, V_{d-1} T^d , \qquad \sigma_{conf}=
{(4\pi)^d \over d^{d-1}}
(M_p \ell_{UV})^{d-1}
\ee
where we have expressed $\sigma_{conf}$  using equation (\ref{Tsconf}). The
question then arises, which of these two solutions is
thermodynamically favored. Because of the simple scaling behaviour of
the free energy, the answer is the same at all $T\neq 0$, and it boils down to comparing the values of
$\sigma_{conf}$ and $\sigma^c$. A numerical computation shows that, in
the particular models we considered here,  $\sigma^c < \sigma_{conf}$,
meaning that the AdS-Schwarzshild solution is the dominant one. For relevant vev flows ending at $\phi_c$ in the model in section \ref{sec:skipping}, we find $\sigma^c =0.95 \sigma_{conf}$;  we come to the same conclusion for the corresponding solutions solutions  in the model considered in section \ref{sec:bouncing}, where we find  $\sigma_c = 0.635 \sigma_{conf}$.

Finally, we can go slightly off-shell and analyse the behaviour of the effective
potential around the critical value ${\cal O}^c$. Using the scaling
property (\ref{vev1}) it is easy to show that, as
${\cal T} \to \infty $, the quantities $\gamma({\cal T})$ and  $\sigma({\cal T})$ in equation
(\ref{FE11-b})  behave  as
\be
\sigma({\cal T}) \to \sigma^c , \qquad \gamma({\cal T}) \simeq {\cal
  T}^{-\Delta_-} \to 0.
\ee
Performing the Legendre transform  of equation (\ref{FE11}) explicitly
close to $\phi_-=0$ we find, as ${\cal O} \to {\cal O}^c$,
\be\label{vev6}
{\cal T} \simeq -{k \over {\cal O} - {\cal O}^c}, \quad  {\cal V}({\cal O}) \simeq {\cal V}_c + {|{\cal O} - {\cal
  O}^c|^{\Delta_-+1}   \over (\Delta_- +1) k^{\Delta_-}} ,
\ee
where $k$ is a constant.

\subsection{Fake zero-$T$ vacua and irrelevant flows}

We  now turn to another kind of special flows, which  correspond to
black holes ending at the  point $\phi_*$ found in section  3 and
represented in figure \ref{missing2}.   As we have explained at the
end of section 3, these black holes correspond to flows starting from
the point IR$_1$, located at $\phi=\phi_0$ in figure  \ref{missing2},
which in this case plays the role of a UV fixed point.

Since the UV is at a minimum of the potential, the conformal dimension
of the deforming operator is $\Delta_+ > d$, the operator is
irrelevant, and a source term is not allowed. Therefore, the solutions
with horizon at $\phi_*$ are driven by a vev, as those discussed in
the previous section. This is consistent with the fact that, if we
start from the UV at $\phi_0$,  there is a
{\em unique} value $\phi_*$  the scalar can take at the horizon: it is
fixed by the unique solution $W^+$ starting at $\phi=\phi_0$.

By the general discussion at the beginning of this section, there is a
one-parameter family of black holes, labeled by the temperature $T$,
for which the vev given by
\be \label{fake1}
\langle O \rangle = {\cal O}^* T^{\Delta_+},
\ee
where ${\cal O}^*$ is a constant.

The free energy is given by the conformal result
\be\label{fake2}
{\cal F} = -{\sigma^*\over d}  V_{d-1}\, T^d
\ee
where $\sigma^* = s/T^{(d-1)}$ is temperature-independent

Notice that the free energy is defined by renormalising with respect to IR$_2$,
i.e. the counter-term must be chosen to be
\be\label{fake3}
S_{ct} = M_p^{d-1} \int d^d x \, \sqrt{g}{2(d-1)\over \ell^2(\phi_0)}, \quad \ell(\phi_0)\equiv\sqrt{
-{d(d-1)\over V(\phi_0)}}.
\ee
Since they have a different UV boundary condition (and different
counter-terms) than those connecting to either UV$_1$ or UV$_2$, the
black holes ending at $\phi_*$ belong to a boundary theory different
from the ones considered in section 3 and in the previous
subsection. Therefore, unlike the case considered in the previous
section, we cannot describe the free energy in terms of an effective
potential at criticality, since there is no well-defined conjugate
variable to ${\cal O}$ which we can use to define a Legendre transform.

It is instructive to understand why  these solutions seem to arise as a
zero-temperature limit.
 As we have seen in section 3, when measuring the
  temperature as defined in  UV$_1$ or UV$_2$,
the  free energy receives a vanishing contribution from the part of
the solution which connects $\phi_0$ to $\phi_*$, and the limit
$\phi\to\phi_*$ looks like a zero-temperature limit.  This is due to the
fact that, from the point of view of, say,  UV$_1$, any solution
starting from IR$_1$ is seen as describing the far
infrared. Therefore, any finite temperature as measured in units of
IR$_1$ will be rescaled to zero in units of UV$_1$.

Notice that there is no sense in  ``glueing together''
the two solutions composed of
the flow from 0 to $\phi_0$ and the flow to $\phi_0$ to $\phi_*$ to
obtain a  new, exotic, zero-temperature solution: indeed, the
vev-driven solution from  $\phi_0$ to $\phi_*$ reaches  an asymptotically
AdS UV boundary as $\phi\to \phi_0$, where $e^A \to +\infty$. This
geometry is locally geodesically complete, and it cannot be glued across the
horizon of the flow reaching $\phi_0$ in the IR (where $e^{A} \to 0$).

Finally, a numerical computation of  $\sigma^*$ shows that, also
  in this case, the free energy (\ref{fake2}) is larger than the free
  energy of the AdS-Schwarzschild solution of the same temperature
  and  constant scalar field  $\phi=\phi_0$. The latter is therefore  the
  dominant solution at any temperature for $j=0$.


\subsection{Minimum-to-minimum irrelevant flows}

\begin{figure}[h!]
 \centering
\includegraphics[height=5cm]{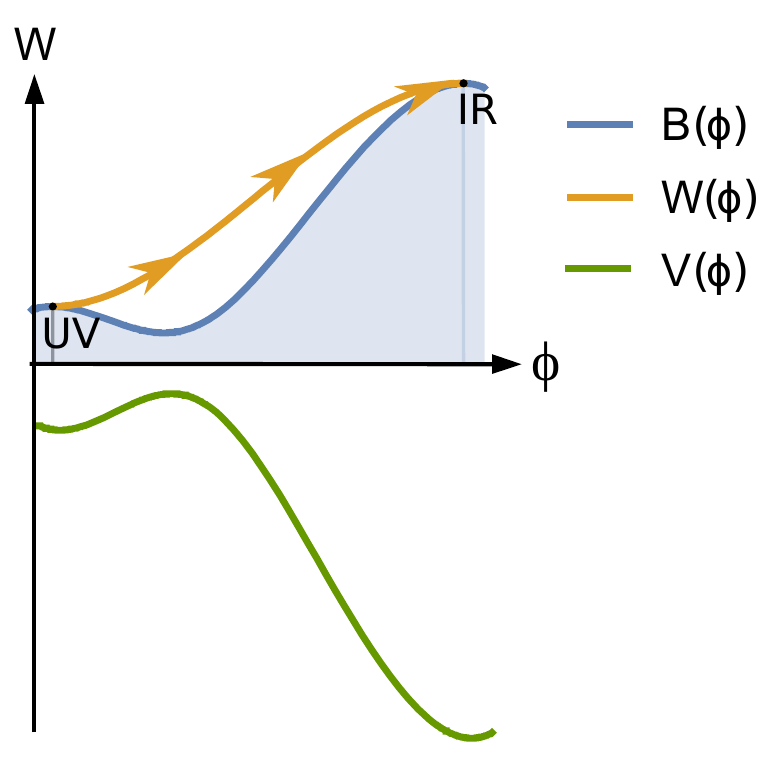}\hspace{1cm}
\includegraphics[height=5cm]{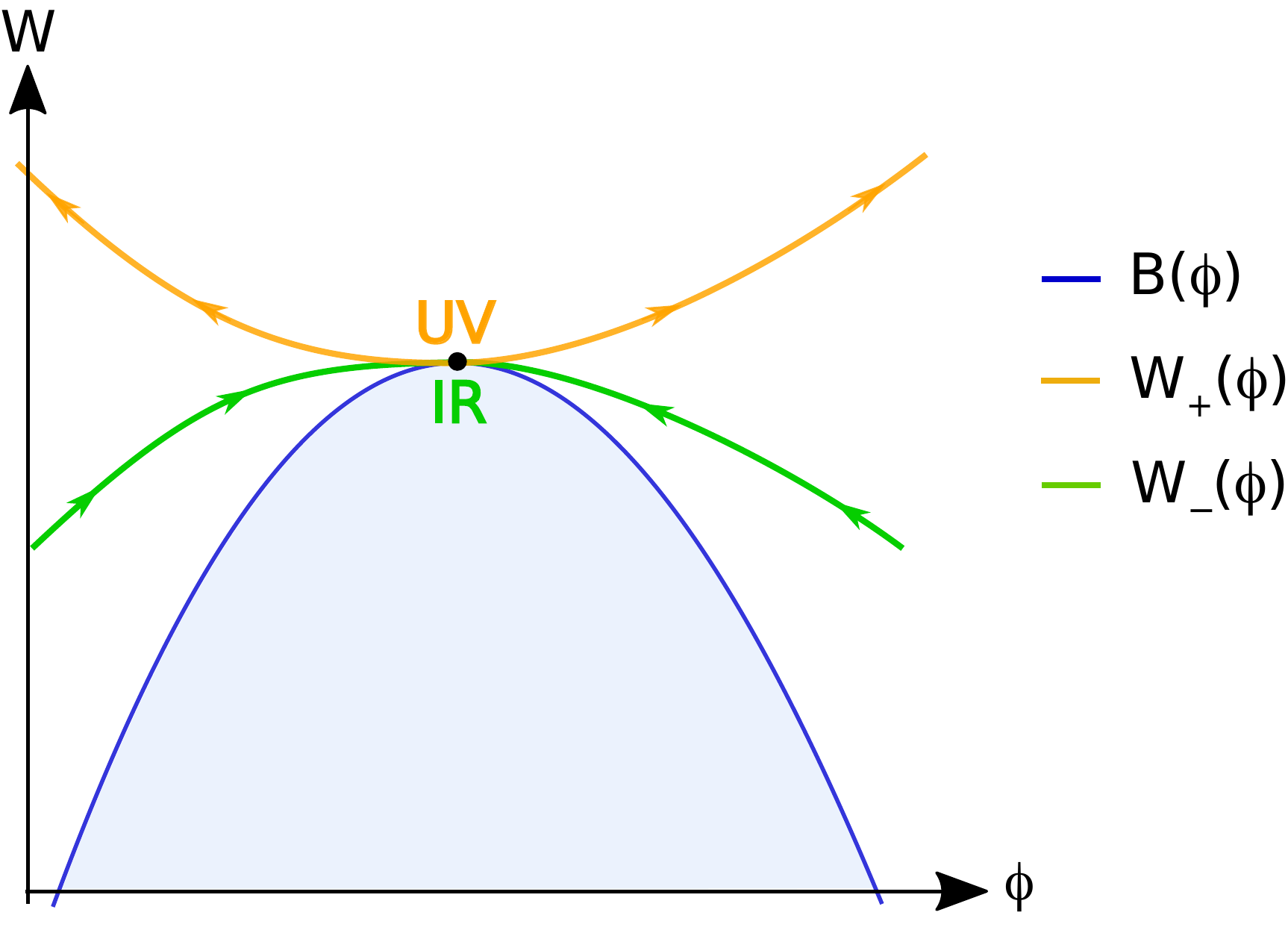}
\\ \hspace{-2cm}(a) \hspace{0.4\textwidth} (b)
\caption{A minimum to minimum flow is displayed. The left figure (a)
  shows a schematic picture of the
  superpotential (yellow curve) and the bulk potential (green
  curve). The curve $B(\phi)$ is  defined via equation
  \eqref{BoundW} and is the boundary of the forbidden region below
  which the zero-temperature superpotential cannot go. The right
  figure (b) shows a sketch of solutions of the superpotential
  equation with critical point at a local minimum of the potential
  (local maximum of $B(\phi)$). The $W_+(\phi)$ and $W_-(\phi)$
  solutions correspond each to  two asymptotically AdS geometries in
  the UV and IR, respectively. The corresponding  geometries are not
  connected, as each flow stops (or starts) at the fixed point. The arrows on $W_-(\phi)$ and  $W_+(\phi)$ indicate the direction of the holographic RG flow. }
\label{fig:Min2Min0T}
\end{figure}

In this section, we consider the finite-temperature generalisation of
the flows connecting two {\em minima} of the potential (one serving as a  UV fixed point, the other as an  IR fixed point). These flows, discussed in \cite{multibranch},  are
driven by the vev of an {\em irrelevant} operator, in contrast
to the zero-temperature solutions  discussed in the sections \ref{sec:skipping} and \ref{sec:bouncing}, for which
the operator was always relevant. They are
shown schematically in terms of the associated superpotential in figure
\ref{fig:Min2Min0T}.

As we discuss below,  at zero temperature, these theories display a
moduli space of vacua\footnote{A field theory example of such a flow is the flow driven by baryon vevs in the baryonic branch of N=1 sQCD.}. We will see that this is lifted  when
temperature is turned on, and that the only point in the moduli space that is left is the
AdS black hole with constant scalar field profile.

Minimum-to-minimum flows are interesting because they provide the only
regular zero-temperature flows in this setup which are driven by the vacuum expectation
value of an irrelevant operator. Such flows arise from spontaneous
breaking of scale invariance of the boundary theory: the source  $J$ of the UV theory operator $O$ is zero as it is non-renormalisable and, yet, this operator acquires a non-zero VEV which is related to the asymptotic behaviour of $\phi(u)$ through
\eql{mod}{
\phi(u) = \phi_+ \, \ell^{\Delta_+} e^{\Delta_+ u } + \ldots, \qquad \langle O \rangle =
\left(M_p \ell\right)^{d-1}
(d-2\Delta_-) \phi_+\,.
}
Equation (\ref{mod}) is valid for both zero and finite temperature. For generic potentials and at zero temperature, the bulk geometries in
the class \eqref{mod} are  singular. However, for special potentials, it is possible to make the flow reach a second minimum of
$V$ in the IR, providing a regular solution with IR AdS asymptotics
\cite{multibranch}.

The first-order or superpotential formalism of appendix
\ref{app:first} provides a single description of {\it all}
zero-temperature flows of the form \eqref{mod} as well as an easy
implementation of the regularity condition. The flows of the form
\eqref{mod} with non-zero $\le<O\ri>$ which start from a given minimum of
$V$ (seen as a UV fixed point) are associated with a {\it unique}
superpotential of the type $W^+(\phi)$, with an asymptotic expansion
of the form (\ref{f3}). This is the yellow curve in
figure \ref{fig:Min2Min0T} (b)), where, for comparison, we also displayed the
$W_-$-type solution arriving at the same minimum for which this point
is seen as an IR fixed point.The blue curve in figure \ref{fig:Min2Min0T}, $B(\phi)$, is
defined though equation \eqref{BoundW} and it is the lower bound on
$W$ at zero temperature: the shaded blue region is not allowed, as a
consequence of the superpotential equation \eqref{a8} with $f\equiv1$
(see appendix \ref{app:bound} for more details).

At $T=0$, these models display a moduli space of vacua, parametrised by
$\phi_+$. The reason is that the superpotential which describes these
flows is of the type $W^+$ which, as we explained
in subsection \ref{ss:vevthermo},  does not contribute a finite part $\sim C
\phi^{d/\Delta_-}$  to the renormalised
on-shell action. Equation (\ref{freeT0}) then  implies  that the
zero-temperature free energy ${\cal F} = 0$ for any value
$\le<O\ri>$.

This one-parameter degeneracy of vacua is continuously
connected\footnote{The limit $\phi_+ \to 0$ is not uniform however:
  for any non-zero $\phi_+$, eventually the scalar field reaches its
  fixed IR value, $\phi(u=+\infty)$,  which is different from the UV
  value $\phi(u=-\infty)$.  As $\phi_+ \to
  0$, a significant  departure from the UV value $\phi_{UV}$ happens
  for larger and larger $u$: the domain wall ``moves to infinity''
  leaving the UV-AdS solution at any finite $u$.} to the AdS vacuum of the unbroken theory, which corresponds
to $\le<O\ri> =0$ and to a constant scalar field profile and also has
vanishing free energy\footnote{More precisely, the free energy is zero
  in our renormalisation scheme, in which we have chosen $C_{ct}=0$ in
  the counter-term action, see
  equation (\ref{ct3}).  In a more general  scheme ${\cal F}$ will be
  a non-zero, $\langle O \rangle$-independent constant.
 }.

At finite temperature, as we have seen in subsection
(\ref{ss:vevthermo}),  we  generically expect for vev-driven black holes a
relation between $\langle O \rangle$ and $T$ of the form
\eqref{vev3}. This means that at any fixed temperature at most one
solution is expected, and the moduli space is lifted.  Moreover, taking
the $T\to 0$ limit we only obtain an AdS solution with
$\langle O \rangle=0$ i.e. these black holes, if they exist, are
connected only to the constant-$\phi$ solution and to no other solution
in the moduli space.

The previous considerations  suggest that the solutions with non-zero
$\langle O \rangle$ and non-trivial scalar field profile do not have
any finite-temperature generalisation: to obtain a finite $\langle O
\rangle$  at $T=0$ one would need the constant $\overline{\cal O}$ in equation (\ref{thvev3})
to be  infinite, in which case the finite temperature VEV would
diverge.
Since the $W^+$-type potential admits
no continuous deformations close to the UV
\cite{wenliang,multibranch},  there cannot be regular black holes
with $\langle O \rangle= 0$ starting from the same UV minimum as the
zero-temperature regular flow.
The reason is that there exists a $W^+$ regular
solution for $f=1$ and therefore  any other regular solution with
non-trivial $f$ will ``miss'' the UV fixed point and flow somewhere
else.

We will now move to a concrete example in which we find that these
expectations are correct: in the theory that admits a regular  minimum-to-minimum flow at $T=0$, there are no
black-hole solutions which start from the same UV.

To be concrete, we consider  the model presented in reference
\cite{multibranch}, which we will subsequently study at finite
temperature. In $d$ boundary dimensions, the following family of superpotentials parametrised by $k$, $v$ and $W_*$,
\eql{ir2}{
W(\phi)=kv^2\le(\phi\over v\ri)\le[
						1-{1\over 3}\le(\phi\over v\ri)^2
						\ri]+W_*
}
allows for very simple kink scalar field profiles:
\eql{ir3}{
	\phi(u)=v \tanh(k~(u-u_*))~.
}
Each superpotential in \eqref{ir2} solves \eqref{a8} with $f(\phi)=1$ and with the potential
\eql{ir4}{
V(\phi)={(kv)^2\over 2} \le[1-\le(\phi\over v\ri)^2\ri]^2
			-\frac{d}{4(d-1)}\le\{
							kv^2\le(\phi\over v\ri)\le[
								1-{1\over 3}\le(\phi\over v\ri)^2
							\ri]+W_*
						\ri\}^2~.
}
The flows of the form \eqref{ir3} interpolate between the extrema of $V(\phi)$ at $\phi=-v$ and $\phi=v$.
A region of parameter space which includes the point
\eql{ir5}{
v=k=1,~~ d=4~~ \nd~~W_*=1.8
}
is such that the extrema of $V(\phi)$ at $\phi=\pm v$ are both local
minima. The potential \eqref{ir4} is not everywhere negative, however
it is negative between $-v$ and $v$ for the values \eqref{ir5} of the
parameters. As we argued in \cite{multibranch}, the sign of $V(\phi)$
outside the range of definition of a superpotential $W(\phi)$ does not
affect the existence of this kind of flow because the equations of
motion are local in field space. In this particular model, the
 maximum of $V$ in between the two minima violates the BF bound, so it
 does not provide a consistent UV for the theory. This is
 inconsequential as far as the minimum-to-minimum flow is concerned.

At finite temperature, we need to solve the system (\ref{a8}- \ref{a9-ii}).
Contrary to the zero-temperature case where the bound \eqref{BoundW}
holds as long as the potential is strictly negative,
at finite temperature even strictly negative potentials allow a
positive $W(\phi)$ to cross into the forbidden region and to be arbitrarily small, as follows for example from equation \eqref{WT4}.

\begin{figure}[t]
 \centering
\begin{overpic}
[width=0.7\textwidth]{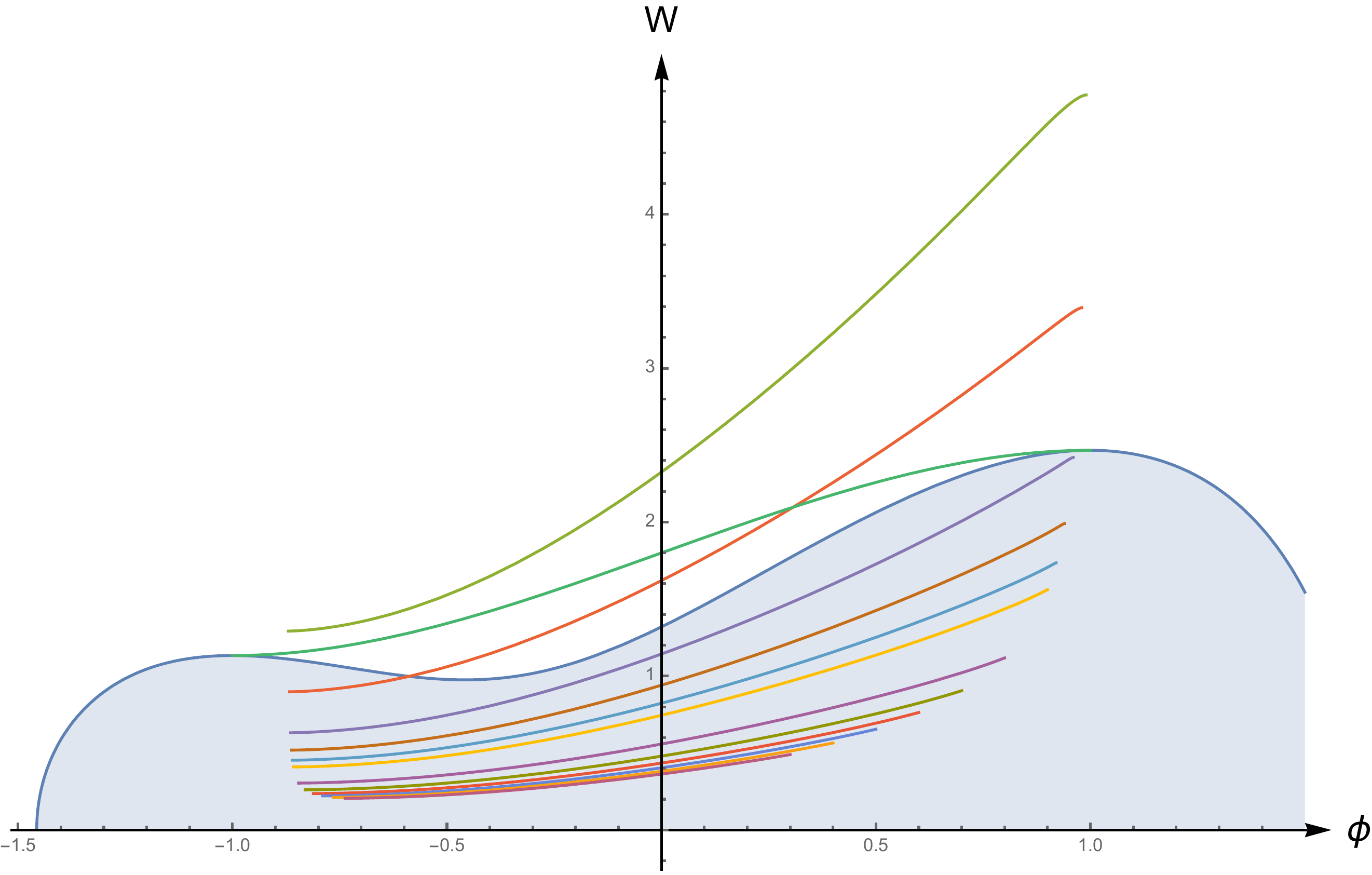}
\end{overpic}
\caption{The blue curve bounding the shaded region represents the curve $B(\phi)$ which bounds superpotentials at zero temperature (see \protect\eqref{BoundW}). It is presented here to locate the extrema of the potential between which the zero temperature solution flows. The green curve which starts at -1 with value $B(-1)$ and extends until $\phi=1$ where its value is $B(1)$ corresponds to the zero-temperature solution. The remaining curves correspond to finite-temperature superpotentials solving \protect\eqref{a8} and \protect\eqref{a9-ii} with boundary conditions (\protect\ref{bc1}-\protect\ref{bc2}) for the potential (\protect\ref{ir4}-\protect\ref{ir5}). The values of $W$ increase as the horizon position takes values closer and closer to the IR extremum at $\phi=v$. None of the black-hole solutions reach the local minimum of $V$ at $\phi=-1$, they all bounce for $\phi>-1$. The position of the bounces and their progression as $\phi_h$ approaches $v$ is clearer when looking at $W'(\phi)$, as is done in figures \protect\ref{fig:dWT} and \protect\ref{fig:dWT2}.}
\label{fig:WT}
\end{figure}

\begin{figure}[t]
 \centering
\begin{overpic}
[width=0.7\textwidth]{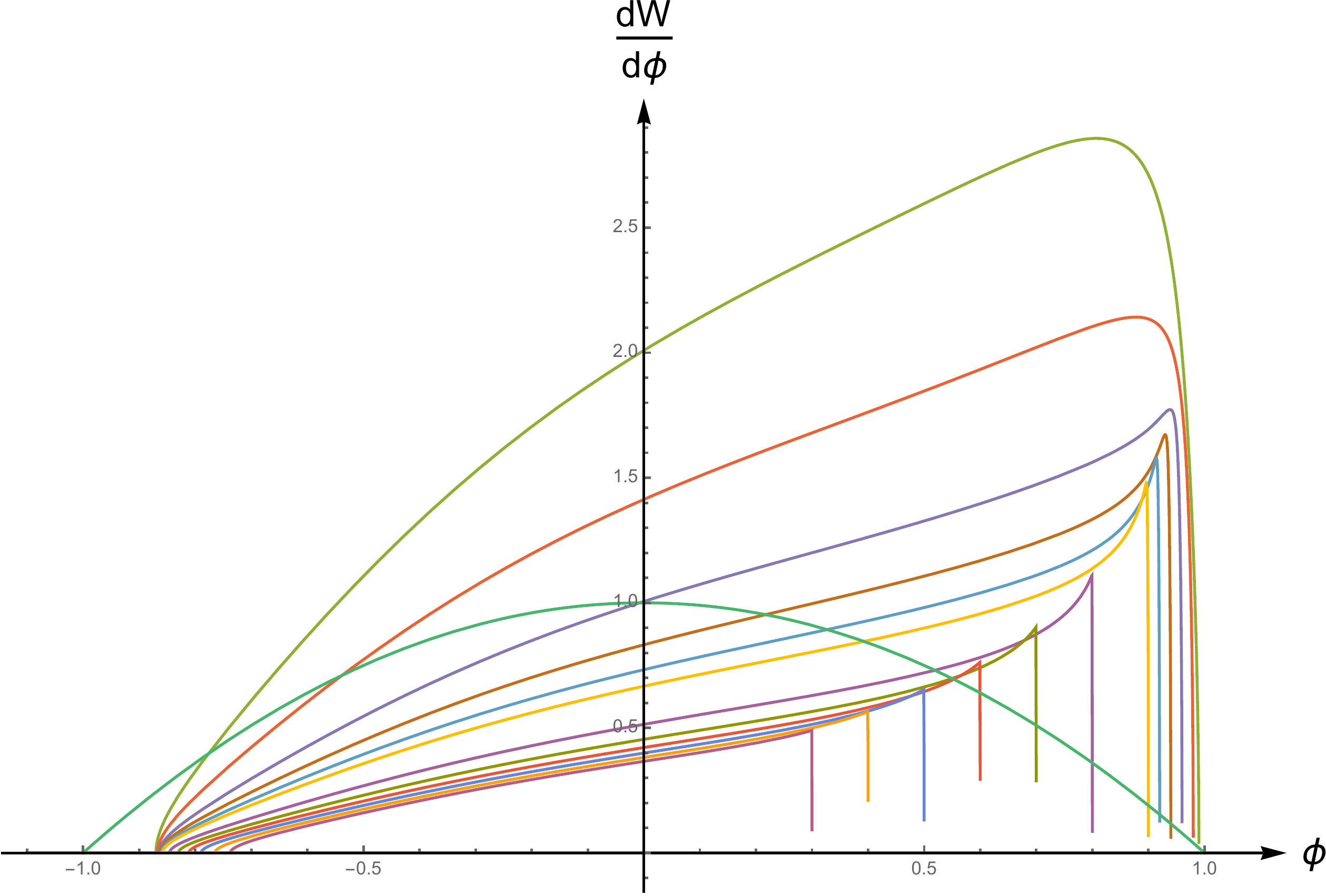}
\end{overpic}
\caption{This figure shows the first derivative of the functions displayed in figure \protect\ref{fig:dWT}.  The curve which extends from $-1$ to $1$ corresponds to the zero-temperature solution. The remaining curves are the derivatives with respect to $\phi$ of the finite-temperature superpotentials displayed in figure \protect\ref{fig:WT}. The rightmost zero (or the tendance of $W'$ to vanish) corresponds to the horizon position which approaches $\phi=1$ without reaching it and the leftmost zero corresponds to the bounce position. The closer the horizon is to $\phi=1$ the closer the bounce gets to $\phi=-1$ but the superpotential never reaches $-1$. This means that the zero-temperature minimum-to-minimum flow is not associated with a black hole. The bouncing points are better visualised in figure \protect\ref{fig:dWT2}.}
\label{fig:dWT}
\end{figure}

\begin{figure}[t]
 \centering
\begin{overpic}
[width=0.7\textwidth]{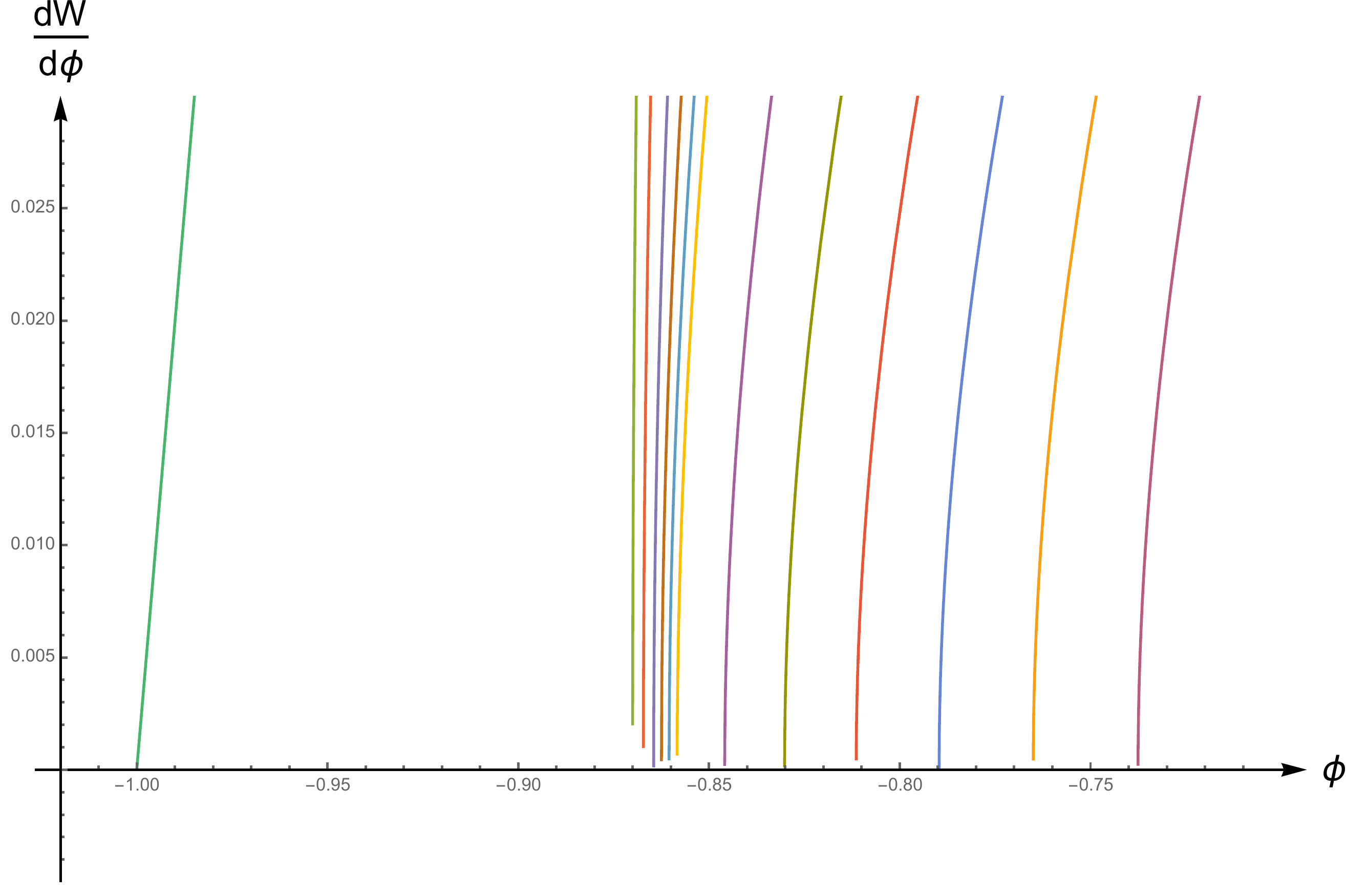}
\end{overpic}
\caption{A closer look at the near-bounce region of figure \protect\ref{fig:dWT}. The leftmost curve corresponds to the zero temperature solution and the remaining curves are superpotentials at finite temperature. The derivative of each finite-temperature superpotential vanishes at a different bouncing point, where the each solution is glued to the next branch. The closer to $-1$ the zero of $dW/d\phi$ is, closer the horizon is to $\phi=1$ and lower is the black-hole temperature. No finite-temperature solution reaches $\phi=-1$, thus showing that the minimum-to-minimum zero temperature flow from \protect\cite{multibranch} does not have an associated black-hole solution with a running scalar.}
\label{fig:dWT2}
\end{figure}

To scan for black-hole solution numerically, we choose a set of values
of horizon position $\phi_h$ in the interval $(-v,v)$ and  and solve numerically the system
(\ref{a8}- \ref{a9-ii}) providing  boundary conditions at
$\phi_h+\delta \phi$  using the relations \eqref{bc1}. What we find is
that no solution reaches the would be UV point $\phi = -v$, but all of
them bounce at some larger $\phi$. The resulting
superpotentials are displayed in figure \ref{fig:WT}, where only the
branch above the bound is displayed and we set $v=1$. The closer the
horizon position $\phi_h$ is to $\phi=v$,  the larger   $W$  and  the closer the bounce is to  $\phi=-1$.
No finite-temperature solution reaches $\phi=-1$, no matter how close
to $v=1$ we set the horizon.   This is further emphasised in figures \ref{fig:dWT} and \ref{fig:dWT2} which display $W'(\phi)$ for different values of $\phi_h$.

The conclusion is that all the black holes in this model miss the UV
minimum at $v=-1$ and flow towards the intermediate maximum. The
latter however violates the BF bound, so these states  are not
part of healthy theory. A part from this,  conceptually they are in
the same class as the black holes with UV asymptotics at a maximum of
$V$, which were discussed in the previous sections.

\section*{Acknowledgements}\label{ACKNOWL}
\addcontentsline{toc}{section}{Acknowledgements}

We thank Dio Anninos, Jerome Gauntlet, Jewel Kumar Ghosh, Lukas Witkowski, Aron
Jansen, David Mateos, Kyriakos Papadodimas, Christopher Rosen, Sergey Sibiryakov for
discussion and comments.

\noindent This work was supported in part  by the Advanced ERC grant SM-grav, No 669288.

\appendix
\renewcommand{\theequation}{\thesection.\arabic{equation}}
\addcontentsline{toc}{section}{Appendices\label{app}}
\section*{Appendices}

\section{First order formalism} \label{app:first}
\label{1st}
As for the zero-temperature solutions, we can introduce a function $W(\phi)$ such that, on-shell
\be
\label{m8}
\dot{A}(u) = -{1 \over 2(d-1)}W(\phi(u)), \quad  \dot \phi(u)={d W(\phi(u)) \over d\phi}
\ee
This can always be achieved piecewise in any region where $\phi(u)$ is
monotonic, by inverting the relation $\phi(u) \to u(\phi)$.
With the definition (\ref{m8})  equation \eqref{b3d} is identically
satisfied, and the remaining equations (\ref{b3a}-\ref{b3c}) become a
system of scalar equations for the functions $W(\phi)$, $f(\phi)
\equiv f(u(\phi))$ \cite{KN},
\be
\label{a8}
\le(\ha \le({dW\over
  d\phi}\ri)^2-\frac{d}{4(d-1)}W^2\ri)f+{W\over 2}\frac{df}{d\phi}{dW\over d\phi}= V
~,
\ee
\be\label{a9-ii}
{d^2 f \over d\phi^2}  {dW \over d\phi} + {d f \over d\phi}  {d^2W
  \over d\phi^2} - {d\over 2(d-1)} W  {d f \over d\phi} =0 .
\ee
For $f(\phi)=1$ , equation (\ref{a8})  becomes the usual superpotential
equation.

Among the transformations \ref{inv} which leave the original equations of motion invariant, only \eqref{inv3} affects \eqref{a8} and \eqref{a9-ii}. The corresponding transformations for $W(\phi)$ and $f(\phi)$ are
\eql{resc}{
\le(\tilde f(\phi), \tilde W(\phi)\ri)\to \Big(f(\phi),W(\phi)\Big)= \Big(
\l^2\tilde  f\le(\phi\ri),\l^{-1}\tilde  W\le(\phi\ri)\Big)
}
and $\phi$ is left invariant.


\subsection{The integration constants for the superpotential
  equations} \label{ssec:int}

The system (\ref{a8}-\ref{a9-ii}) is third order\footnote{Since both $\de_\phi^2 W$ and
  $\de_\phi^2 f$ appear, at fist sight it  may look as
  the  system is fourth order, but this is not the case, as one can
  easily realise by algebraically solving equation (\ref{a8}) for
  $\de_\phi W$ and inserting the result into equation (\ref{a9}).} and can be solved in terms of  three
integration constants which can be
chosen in different ways. It is most convenient to view it as a
boundary value problem for $f(\phi)$ and an initial value problem (at
the black-hole horizon) for $W(\phi)$, and fix the integration
constants by
\be\label{a9-iii}
f(0)=1, \qquad f(\phi_h) = 0, \qquad W(\phi_h) = W_h.
\ee
The first equation follows from the requirement that the  leading asymptotic
boundary metric is the same for all solutions. This leaves  the two
arbitrary parameters parameters $\phi_h$ and $W_h$. As we will
show below however, $W_h$ is also  fixed uniquely by the choice the
horizon position $\phi_h$, resulting in  a one-parameter
family of solutions parametrised by the value of $\phi_h$.

A very simple way to see that the system (\ref{a8}- \ref{a9-ii}) is third order and which is helpful to give boundary conditions for solving the system numerically goes as follows. By differentiating equation \eqref{a8} with respect to $\phi$, a term with $f''(\phi)$ is generated and can be used to eliminate $f''(\phi)$ from \eqref{a9-ii}. The resulting equation is:
\eql{nh1}{
W'(\phi ) \left[f(\phi ) \left(\frac{d W(\phi )}{2-2 d}+W''(\phi )\right)+f'(\phi ) W'(\phi )\right]=V'(\phi ),
}
Equations \eqref{a8} and \eqref{nh1} demand three integration constants which completely specify $W(\phi)$ and $f(\phi)$. Evaluating \eqref{a8} and \eqref{nh1} at the horizon we obtain
\begin{subequations}\label{nh2}
\begin{align}
&\le[{W}{f'}{W'}\ri]_{\phi_h}= 2V(\phi_h), \label{nh2a}
\\
&\le[W'f' W'\ri]_{\phi_h}=V'(\phi_h) ~.\label{nh2b}
\end{align}
\end{subequations}
For any choice of the integration constant $f'(\phi_h)$ and imposing that $W$ is positive, the system \eqref{nh2} completely specifies $W(\phi_h)$ and $W'(\phi_h)$. Therefore one can chose boundary conditions close to the horizon as follows:
\begin{subequations}\label{bc1}
\begin{align}
&f(\phi_h+\d\phi)= \d\phi~f'(\phi_h) +\cO(\d\phi^2)\qquad \text{with} \quad f'(\phi_h)\d\phi>0\label{bc1a},
\\
&W'(\phi_h+\d\phi)=\sqrt{\frac{V'(\phi_h)}{f'(\phi_h)}} +\cO(\d\phi)~,\label{bc1b}
\\
&W(\phi_h+\d\phi)=-\frac{2 V(\phi_h)}{\sqrt{f'(\phi_h) V'(\phi_h)}}+\d\phi  \sqrt{\frac{V'(\phi_h)}{f'(\phi_h)}} +\cO(\d\phi^2)~.\label{bc1c}
\end{align}
\end{subequations}

The boundary conditions \eqref{bc1} will typically lead to $f(\phi_{UV})\neq1$, however this apparent inconvenience is immediately circumvented with a rescaling of the form \eqref{resc} with parameter
\eql{bc2}{
\l=(f(\phi_{UV}))^{-1/2}~.
}

Another way to specify the integration constants is to explicitly identify the boundary theory source in $f(\phi)$ and the product $T S$ in $W(\phi)$, making a more explicit contact with the thermodynamics. To proceed, we chose a value $\phi_h$. First, notice that equation (\ref{a9-ii}) can also be written
equivalently as
\be\label{a9}
{d\over d\phi} \left[{df\over d\phi } {dW \over d\phi} e^{d{\cal A}(\phi)}\ri]
  = 0
\ee
where
\be \label{a10-A}
{\cal A}(\phi) = -{1\over
  2(d-1)}\int^\phi d\hat \phi {W \over W'}
\ee
The function ${\cal A}(\phi)$ depends on an arbitrary additive
constant but clearly this does not affect equation (\ref{a9}), and in
fact it can be reabsorbed in one of the other  integration constants.
Integrating equation  (\ref{a9}) gives
\be \label{a10-1}
{df \over d\phi}= -{\tilde{D} \over W'} e^{-d{\cal A}(\phi)},
\ee
where   $\tilde{D}$ is a constant. As we mentioned above,  an additive
constant in ${\cal A}(\phi)$ can be absorbed in a redefinition of
$\tilde{D}$ and we can  fix this redundancy (which does {\em not}
correspond to one of the integration  constants of the system
(\ref{a8}-\ref{a9-ii}))  so that ${\cal A}(\phi_h) = 0$:
\be \label{a10-4}
 \qquad {\cal A}(\phi) = -{1\over
  2(d-1)}\int_{\phi_h}^\phi d\hat \phi {W \over W'}
\ee
Equation (\ref{a10-1}) can be further
integrated to give
\be\label{a10-2}
f(\phi)  = \tilde{F} - \tilde{D}\int_0^\phi  {dy \over W'(y)} e^{-d{\cal A}(y)}.
\ee
As we will see in a moment, the second term vanishes as $\phi\to
0$. Therefore,  imposing the boundary conditions for $f(\phi)$ from equation (\ref{a9-iii})
fixes
\be \label{a10-3}
\tilde{F} = 1, \quad \tilde{D} =  \left( \int_0^{\phi_h}  {dy \over W'(y)} e^{-d{\cal A}(y)}
\right)^{-1}.
\ee
Finally, evaluating equation (\ref{a8}) at $\phi_h$ and using
(\ref{a10-1})  we find
\be
W(\phi_h)  = -{2 V(\phi_h) \over \tilde{D}}.
\ee
This condition is equivalent to the requirement that the horizon is
regular, and it fixes the remaining free parameter $W_h$ in the
initial conditions (\ref{a9-iii}).\\

To summarise we have found that, for regular asymptotically $AdS$
black holes, solutions to the system (\ref{a8}-\ref{a9}) depend on a single
continuous free parameter, which we can be taken to be the horizon position $\phi_h$.  In the next subsection we
will clarify the physical meaning of the quantity
$\tilde{D}(\phi_h)$.

\subsection{Lower bounds on the superpotential} \label{app:bound}

At zero temperature, an important property of $W(\phi)$ is that it is bounded from below. This can be shown by considering equation \eqref{a8} with $f(\phi)\equiv 1$ and algebraically solving for $W$:
\eql{BoundW0}{|W(\phi)|= \sqrt{\frac{4(d-1)}{d}\le(\ha W'^2-V(\phi)\ri)}\geqslant\sqrt{-\frac{4(d-1)}{d}V(\phi)}\equiv B(\phi)>0.}
When $V(\phi)$ is negative, $W$ can never approach zero, meaning that the branches with $W>0$ and $W<0$ which solve \eqref{BoundW0} are disconnected. Furthermore, they are physically equivalent as one can pass from one to the other by the transformation $(u,W)\to(-u,-W)$. Therefore, without loss of generality we can choose a positive $W$, meaning that $\dot A(u)$ is negative, and see \eqref{BoundW0} as a lower bound on $W$ and not just on its absolute value:
\eql{BoundW}{W(\phi)\geqslant B(\phi)\equiv \sqrt{-\frac{4(d-1)}{d}V(\phi)}>0.}
Another useful property to understand the numerical results shown in figures \ref{fig:WT}, \ref{fig:dWT} and \ref{fig:dWT2} which is related to the normalisation of $f(\phi)$ is that $W$ can lie below $B(\phi)$ defined in \eqref{BoundW}. It is instructive to algebraically solve equation \eqref{a8} for $W(\phi)$ keeping into account the positivity of $W$ which is necessary for $\dot A$ to be negative
\eql{WT4}{
W(\phi)={(d-1) \over {d }}{f'W'\over f}+\sqrt{{\le({(d-1)\over d} {f' W'\over {f}}\ri)^2+{2 (d-1) \over d }\le(W'^2-{2 V\over f}\ri)}}
}
As long as $V(\phi)$ is negative, it is guaranteed that $W$ is real at stationary points, i.e. when $W'$ vanishes. Negativity of $V(\phi)$ also guarantees that giving $W$ a positive value at any fixed $\phi=\phi_0$ implies the positivity of $W$ for all $\phi$. In contrast to the zero temperature case, the function $f(\phi)$ permits the existence of solutions such that the limit
\eql{WT5}{
\le({f' W'\over {f}}\ri)^2\gg{2d\over d-1}\le(W'^2-{2 V\over f}\ri)
}
holds, thus including arbitrarily small superpotentials in the space of solutions. On the other hand, from \eqref{WT4} it follows that the value of $W(\phi)$ at critical points, i.e. those where $W'(\phi)$ vanishes, is set by $f(\phi)$ and $V(\phi)$,
\eql{WT6}{
W'=0 \quad \nd \quad {f' W'\over {f}}=0 \qquad \implies \qquad W=\sqrt{{-{4 (d-1) \over d }{V\over f}}}
}
one of the few remaining properties from the zero temperature setup.

\subsection{Near-boundary solution: universal part}
To obtain the solution close to the boundary,  we solve the system of equations (\ref{a8}-\ref{a9})
perturbatively close to $\phi=0$, where $f\simeq 1$.  To lowest order we
obtain the analytic part $W_0$   of the superpotential,
\be\label{a11}
W_0^{\pm} (\phi) = {2(d-1)\over \ell} + {{\Delta_\pm} \over 2\ell}
\phi^2 + O(\phi^4),
\ee
which coincide with the perturbative expansion in the vacuum solution,
see e.g. \cite{Papadimitriou:2007sj,wenliang,multibranch}. The two
choices, $\Delta_-$ or $\Delta_+$, in the quadratic term, correspond
respectively to a flow driven by a source or by a vev of the operator dual to
$\phi$. Generically we will be describing flows with non-zero source,
therefore from now on we will choose $W_0 = W_0^{-}$. Vev-driven flows
will play an important role however, and will be discussed separately
later on.

Up to higher order corrections,  we can replace $W$ by $W_0$ in ${\cal A}$ in
(\ref{a10-1}) and close to the boundary we find
\be\label{a12}
{\cal A}(\phi) = {\cal A}_0 -  {1\over \Delta_-} \log |\phi|  + O(\phi)
\ee
where ${\cal A}_0$ is a  constant which is also completely determined by
equation (\ref{a10-4}) and depends only on $\phi_h$.

Using equations (\ref{a11}) and (\ref{a12})  we can  approximate  equation (\ref{a10-2}) for
 close to $\phi=0$,
\be \label{a15}
f(\phi) = 1 - {\tilde{D}e^{-d{\cal A}_0} \ell\over d}  |\phi|^{d\over \Delta_-}.
\ee
As announced, the $\phi-$dependent part vanishes as $\phi \to 0$, and
$f$ satisfies the correct boundary condition.

Finally, integrating equations (\ref{m8}) give the expected leading
order scaling,
\be \label{a16}
A(u) = -{u \over \ell} + O(e^{u}) , \qquad \phi(u) = j\, \ell^{\Delta_-} e^{\Delta_-
  u/\ell} + \ldots, \qquad u \to -\infty,
\ee
and introduces as additional free parameter  $j$, the constant
corresponding to the source of the dual operator, whose choice is part
of the boundary data one has to give to fix the holographic
theory. The other piece of boundary data is the leading  asymptotic
behaviour of the boundary metric: in writing equation  (\ref{a16}) we have chosen the  constant term in $A$ to vanish, so that in
the holographic dictionary the dual field theory metric is
$\eta_{\mu\nu}$ with unit coefficient. \\

Thus, the full solution $(A(u), \phi(u), f(u))$ depends on only two
parameters: the value of the source $j$ and the horizon position
$\phi_h$. The black-hole temperature and entropy density are therefore
determined by these two quantities. In the next section we will make
this dependence  more explicit.

\subsection{Dimensionless temperature and entropy} \label{app:dim}

Equations (\ref{a15}-\ref{a16}) reproduce  the expected  scaling
(\ref{max0}) near the boundary. Comparing  equations  (\ref{a15}) and
(\ref{max0c}), and using the relation (\ref{a10-1})  we can relate
$\tilde{D}$ to the  black-hole temperature $T$  and entropy density $s$:
\be \label{dim1}
\tilde{D}e^{-d{\cal A}_0} = {1 \over  M_p^{d-1}}\, \,
{T s \over  |j|^{d/\Delta_-}}.
\ee
As the right hand side depends only on $\phi_h$, the same must be true
for the combination on the right hand side. In fact we will now  show that
the two  quantities
\be\label{dim2}
{\cal T} \equiv {T\over |j|^{1/\Delta_-}} , \qquad {\cal S} \equiv
  {s\over  |j|^{(d-1)/\Delta_-}}
\ee
 are  function of $\phi_h$ only. They represent the temperature and
 entropy density in units of the UV source $j$\footnote{They are
   the finite-temperature analog of the dimensionless curvature  parameter ${\cal R}$
   used in \cite{curvedRG}.}.

To see that the ${\cal T}$ and ${\cal S}$ defined above depend on
$\phi_h$ only, and not on both $\phi_h$ and $j$, we use the invariance property
(\ref{inv1}-\ref{inv2}),
\be \label{dim3}
u \to u'= u+ v, \qquad A(u) \to A'(u') = A(u') + \bar{A}.
\ee
Starting from a solution, these
transformation produces a new solution of the same form (\ref{b1}) but
with different UV data, temperature and entropy: the source term
transforms as
\be\label{dim4}
j \to e^{\Delta_- v} j
\ee
and the horizon data as
\be\label{dim5}
u_h \to u_h +v, \quad e^{A(u_h)} \to e^{A(u_h+v) + \bar{A}}, \quad
\dot{f}(u_h) \to \dot{f}(u_h + v)
\ee
However if we want to keep the boundary metric fixed to be
$\eta_{\mu\nu}$ we have to choose $\bar{A} = v/\ell$, as we have to
cancel a constant term in the asymptotic expansion of $A(u)$, equation
(\ref{a16}). Taking this into account, we find that under (\ref{dim3})
\be \label{dim6}
T \to e^{v} T, \qquad s \to e^{(d-1)v}s
\ee
On the other hand, the horizon value $\phi_h$ is unaffected by this
transformation, so all solutions related by the transformation
(\ref{dim3}) have the same $\phi_h$ {\em and} the same values of the
combinations (\ref{dim2}). We   conclude that both  ${\cal T}$ and
${\cal S}$ are determined solely by the horizon field value
$\phi_h$. Since ${\cal T}$ (as opposed  to $\phi_h$) is directly
related to the horizon data $(T,j$, it is useful to invert the
relation ${\cal T}(\phi_h)$ (at least piecewise) and use ${\cal T}$ to
parametrise black-hole solutions with the same $\phi_h$.

\subsection{Near-boundary solution: sub-leading  term} \label{ssec:sub}
The sub-leading, non-analytic  contributions to $W(\phi)$  beyond the
leading behaviour
(\ref{a11}) contains the
information about the vevs, see \cite{wenliang, multibranch} and
gives finite contributions to the free energy. It can be obtained
by setting $W = W_0 + \delta W$,  linearizing
equation (\ref{a8}) around the solution (\ref{a11})  and keeping in
mind that $f = 1+ O(\phi^{d\Delta_-})$.
The resulting equation is
\be \label{a13}
W_0' \delta W' - {d\over 2(d-1)} W_0 \delta W_0 =  (f_0-1) V + f_0'
    W_0' W_0
\ee
Using the expression for $f_0$ in equation (\ref{a15}) and evaluating
$V$ at the origin (recall $V(0) = -d(d-1)\ell^{-2}$)  we find that the
right hand side vanishes at the lowest order $\phi^{d/\Delta_-}$, and
it starts at order  $\phi^{d/\Delta_-+2}$. To lowest order the
solution for $\delta W$ is
\bea\label{a14}
\delta W = \ell^{-1}C(\phi_h) |\phi|^{d/\Delta_-} \left(1 + O(\phi^2)\right)
\eea
where $C(\phi_h)$ is an integration constant which, following the
discussion in section \ref{ssec:int}, is determined by the horizon
value $\phi_h$. Equation (\ref{a14})  has exactly the same
form as the generic non-analytic part of the superpotential at zero
temperature, the only difference being that  the integration constant
is fixed by the horizon position in field space.

Rather than $\phi_h$, we will find it convenient to  consider ${\cal T}(\phi_h)$ as the independent parameter, since it
directly relates to the boundary quantities $T,j$. Therefore we will
often write $C({\cal T})$ instead of $C(\phi_h$).

In terms of the domain-wall coordinate $u$, using the leading order
behaviour $\phi \simeq je^{\Delta_- u}$ , we find that $\delta W$
scales as $e^{du}$. More precisely,
\be\label{a17}
 \delta W\simeq \ell^{-1}C({\cal
  T})|j|^{\Delta_+/\Delta_-}e^{du/\ell}, \quad u\to -\infty.
\ee
Finally, we can integrate the first order equation $\dot{\phi}(u) =
W'(\phi(u))$ to sub-leading order as $u \to -\infty$, to find
\be\label{a18}
\phi(u) = j \, \ell^{\Delta_-}e^{\Delta_- u} + \ldots + \phi_+\, \ell^{\Delta_+} e^{\Delta_+ u},
\quad \phi_+ = {d\over \Delta_-} {C({\cal T}) \over (d-2\Delta_-)
}|j|^{\Delta_+/\Delta_-} \, sign(j).
\ee
Therefore, $C({\cal T})$ plays the role as a dimensionless vev
parameter, $\langle O\rangle/ |j|^{\Delta_+/\Delta_-}$, which depends on ${\cal
  T}$ (or equivalently, on $\phi_h$) only.

\subsection{Superpotentials for vev-driven flows}

To conclude this appendix we discuss  solutions with $j=0$ which
nevertheless describe a non-trivial flow. For these solutions, the
appropriate (analytic part of the) superpotential corresponds to the
choice $\Delta_+$ in equation (\ref{a11}),
\eql{W+}{
W^+ = {2(d-1)\over \ell }  + {\Delta_+ \over 2\ell} \phi^2 + O(\phi^4)
}
 In this case, a similar
analysis to the one in section \ref{ssec:sub} shows that a
sub-leading, non-analytic part is {\em not} allowed (see
\cite{wenliang,multibranch} for a more extended discussion). Therefore
the solution of the kind $W^+$  is unique, it is completely fixed by
its analytic expansion,  and does not admit a continuous parameter family of deformations.

Since they have, $j = 0$ these solutions correspond to taking  the limit
${\cal T}\to +\infty$, which occurs (at most) at isolated special
values of $\phi_h$.

After integrating the first order flow equations for $A(u)$ and
$\phi(u)$, the full solution  is parametrised by the vev
$\phi_+$ (which now is a free integration constant), and the
near-boundary expansion of $\phi(u)$, solving $\dot{\phi} = W'$, is
\be
\phi(u) = \phi_+ \, \ell^{\Delta_+} e^{\Delta_+ u } + \ldots, \qquad \langle O \rangle =
\left(M_p \ell\right)^{d-1}
(d-2\Delta_-) \phi_+.
\ee
Fixing $\phi_+$
also fixes the temperature of the solution since, by a similar
argument as the one presented in section \ref{ssec:int}, it is easy to
show that for all such solutions\footnote{That is, all solutions with
  the same value of $\phi_h$ corresponding to one of the special points in parameter space})  the ratio $T/\phi_+^{1/\Delta_+}$ is
fixed.

\section{The on-shell action} \label{app:onshell}

The on-shell Euclidean action  is finite if we include the appropriate counter-term
action in equation  (\ref{b0}). It was systematically analysed in \cite{KN}.
 Here we discuss the three terms in (\ref{b0})
separately. In doing so, each term is finite only if we  introduce  a
UV cut-off $u=u_{UV}$ in the asymptotic boundary region. In the final step
 one  removes the cut-off by taking $u_{UV} \to -\infty$.

\paragraph{Bulk Action.}
This can be written in terms of the UV asymptotic data, as follows.  The trace of the Einstein's equations of motion following from \eqref{b0} lead to the following expression for the Ricci scalar:
\eql{OS1}{
	R=
		{\dot \phi^2\over 2f}+{d+1\over d-1}V,
}
which can be used to eliminate $R$ from the bulk part of the  action $S^{bulk}$,
i.e. the first term in equation
\eqref{b0}, leaving
\eql{OS2}{
S^{bulk}=-{2M_P^{d-1}\over d-1}\int_{\MM} \dd^{d+1}x ~e^{d A(u)}V(\phi(u))
}
Using equation \eqref{b3b} to eliminate V and equation  \eqref{b3d} to
eliminate $\dot \phi^2$ we find
\begin{align}
	S^{bulk}
	=
	&
	{2M_P^{d-1}}\int_{u_{UV}}^{u_h}  \int d^{d}x
	~e^{d A}
		\le[
			\dot f\dot A
			+\le(
				d\dot A^2
				+\ddot A
			\ri)f
		\ri]
	\nonumber
	\\
	=
	&
	{2M_P^{d-1}}\beta V _{d-1}
		~\le[
			e^{d A}\dot A f
		\ri]^{u_h}_{u_{UV}}
		\label{OS3}
\end{align}
where $V_{d-1}$ is the $d-1$-dimensional spatial  volume (which we
take to be finite, e.g. by considering the system in a spatial square
box of size $L$). Because $f(u_h)=0$,  the
horizon does not contribute, and the contribution from the bulk part
of the action is:
\be\label{OS3-1}
S^{bulk} =  -{2M_P^{d-1}}\beta V_{d-1}
		~\le[
			e^{d A}\dot A f
		\ri]_{u=u_{UV}}
\ee

\paragraph{Gibbons-Hawking-York term.}
To compute the Gibbons-Hawking-York boundary term $S_{GHY}$ from
\eqref{b0} we define the outward-oriented unit normal vectors to the
boundary at $u=u_{UV}$. Notice that we should {\em not} include a GHY
term at the horizon, since this is a regular point in the interior of
the Euclidean geometry, rather than a boundary.

The unit normal to the boundary is given by
\eql{OS6}{
n^a=-\sqrt{f}\d^c_u\quad \text{at}\quad u=u_{UV}
}
and the extrinsic curvature is
\eql{OS7}{
K=\nabla_a n^a
=-{1\over \sqrt{f}}\le(
{\dot f\over 2}
+d \dot A f
\ri)
}
The on-shell expression for $S_{GHY}$ from \eqref{b0} for solutions of
the form \eqref{b1} is then
\eql{OS8}{
S_{GHY}^{on-shell}=
M_P^{d-1}\beta V_{d-1}
\le[
	e^{d A}\le(
		{\dot f}
		+2d \dot A f
	\ri)
\ri]_{u=u_{UV}}~.
}

Combining the two terms  \eqref{OS3-1} and \eqref{OS8} we obtain the
regularised on-shell action,
\eql{OS9}{
S_{on-shell}^{reg}
=M_P^{d-1}\beta V_{d-1}
\le[
	e^{d A}\le(
		{\dot f}
		+2(d-1) \dot A f
	\ri)
\ri]_{u=u_{UV}}~.
}
This expression is divergent as we let $u_{UV}\to -\infty$. Before we
remove the cut-off we must add appropriate counter-terms, which we discuss
below.

\paragraph{counter-term action.} The counter-terms are universal, and are
written in the general form \cite{papa,wenliang}
\be\label{ct1}
S_{ct} = M_p^{d-1}\int_{u=u_{UV}} d^dx \sqrt{\gamma} \left[ W_{ct}(\phi) +
  U_{ct}(\phi)   R^{(\gamma)} \right]
\ee
where $u_{UV}$ is a UV cut-off, $\gamma$ is the induced metric on
$u=u_{UV}$, $R^{(\gamma)}$  its intrinsic curvature,  and   $W$ and
$U$ are appropriate functions of the scalar field.

In the case at hand, the $u=const.$
hyper-surfaces are flat, so $R^{(\gamma)} = 0$ and the curvature
counter-term vanishes identically. The appropriate function
$W_{ct}(\phi)$ must solve the {\em superpotential equation,}
\be\label{ct2}
{1\over 2}\le(dW_{ct} \over d\phi \ri)^2  - {d \over 4(d-1)} W_{ct} = V
\ee
In the case $d/2> \Delta_->d/4$, to which we restrict here ,
the solution for $W$ close to the boundary has the universal form:
\be\label{ct3}
W_{ct}(\phi) = W_0(\phi)  + {C_{ct} \over \ell}|\phi|^{d\over
  \Delta_-}  + \ldots,
\ee
up to term which vanish faster than $e^{-dA}$ as $\phi \to
0$. In the expression above, $W_0$ is the analytic solution
whose expression can be found in equation (\ref{a11}), taken up to
quadratic order;    $C_{ct}$ is an arbitrary constant which encodes the scheme
dependence of the subtraction. We choose to work in a minimal scheme,
in which we set $C_{ct}=0$. The resulting counter-term is quadratic
in $\phi$ and we will denote it by

Evaluating the counter-term on-shell gives:
\be\label{ct4}
S_{ct} = M_p^{d-1} \beta V_{d-1} \left[e^{dA }\sqrt{f} W_{0}\ri]_{u=u_{UV}}.
\ee

\subsection{Calculation of the Free energy}

The renormalised Free energy is found by adding the counter-term
(\ref{ct4}) to the regularised on-shell action (\ref{OS9}),
\be\label{FE1}
\beta {\cal F} =  M_P^{d-1}\beta V_{d-1}
\le[
	e^{d A}\le({\dot f} +
		2(d-1) \dot A f +  W_{0} \sqrt{f} \ri)\ri]_{u=u_{UV}}
\ee
Let us look at the two terms in the square brackets separately.
\begin{enumerate}
\item Using equations (\ref{a7}) and (\ref{a10})  we can rewrite the first term simply as follows:
\be\label{FE2}
   M_P^{d-1}\beta V_{d-1}
\le[
	e^{d A}
		{\dot f} \right]_{u_{UV}} = - \beta Ts \, V_{d-1}.
\ee
\item
The second term  can be recast in a simpler form if we
revert to the first order formalism developed in Appendix A, and we
write $\dot A$ in terms of the finite temperature superpotential
$W$ using equation (\ref{m8}):
\be\label{FE3}
\le(2(d-1)\dot{A} f  +
\sqrt{f}W_0\ri) = - \sqrt{f}\left( W \sqrt{f} -  W_0\ri)e^{dA}.
\ee

As shown in Appendix \ref{1st}, close to the boundary the
superpotential takes the form
\be \label{Wexp}
W = W_0(\phi) + {C({\cal T}) \over \ell}  |\phi|^{d/\Delta_-}  + \ldots
\ee
up to terms which  vanish faster than $e^{du/\ell}$ as  $u\to
-\infty$. In the equation above,  $W_0(\phi)$ is the same universal power-series  which solves the
zero-temperature superpotential equation (\ref{ct2}), and which enters
the counter-term (\ref{ct3}).

We now expand $\sqrt{f}$ and $e^{dA}$  using
(\ref{max0}),
\be\label{FE4}
\sqrt{f} \simeq  1 - {\ell D \over 2d} e^{du/\ell} + O\le(e^{2du
  /\ell}\ri), \qquad e^{dA} \simeq e^{-du} (1 + O(e^{2\Delta_- u/\ell })).
\ee
Inserting the expansions (\ref{Wexp}) and (\ref{FE4}),  the  right
hand side of equation (\ref{FE3}) takes the form,
\be\label{FE5}
\sqrt{f}\left( W \sqrt{f} -  W_0\ri)e^{dA} \simeq -W_0 {\ell D \over 2d} -
  {C({\cal T})\over \ell}  |\phi|^{d/\Delta_-} e^{dA} \to   {(d-1) \over d} {Ts
    \over M_p^{d-1}} - C({\cal T}) \ell^{d-1} |j|^{d/\Delta_-}
\ee
where in the last step we have taken the limit $u_{UV}\to -\infty$ and
we have expressed $D$ in terms of $Ts$ by equation (\ref{a10}).
\end{enumerate}
Putting everything together we finally arrive at the expression for
the free energy,
\be\label{free-app}
{\cal F} = - {Ts \over d} \, V_{d-1} - (M_p \ell)^{d-1}C({\cal
  T})|j|^{d/\Delta_-}  V_{d-1} .
\ee
\subsection{Thermal Vev} \label{app:vev}
In this appendix we show the validity  of the relation
\be  \label{vev1a}
{\de {\cal F}(T,j)  \over \de j} = - \langle O \rangle V_{d-1}
\ee
where the dual operator vev $\langle O\rangle$ is related to $C({\cal
  T})$ and $j$ by equation (\ref{a18}),
\be \label{vev1-ia}
\langle O \rangle = (M_p \ell)^{d-1}{d\over \Delta_-} C({\cal T})|j|^{\Delta_+/\Delta_-} \, sign(j).
\ee
For simplicity of notation we suppose $j > 0$, but the final
result holds for either sign of $j$.

We start from the expression (\ref{FE11}),
\be\label{vev2a}
{\cal F} = T^{d}\left(-{\sigma ({\cal T})\over d} + \gamma({\cal T})
\right), \qquad \gamma({\cal T}) \equiv (M_p \ell)^{d-1}V_{d-1} {C({\cal
    T})\over {\cal T}^{d}}, \qquad {\cal T} ={T\over  j^{\Dm}}.
\ee
First, we differentiating ${\cal F}$ with respect to $T$, and use the identities
\be\label{vev3a}
{\de {\cal T} \over \de T} = {1\over j^{\Dm}}, \quad {\de {\cal
    T} \over \de j}  = -{1\over \Dm j}{\cal T},
\ee
to  obtain
\be\label{vev4a}
{{\de {\cal F}} \over \de T} = -T^{d-1}\left[\left({\sigma' \over d} +
  \gamma'\right){\cal T} + d \left({\sigma  \over d} +
  \gamma\right)\right],
\ee
where a prime denotes a derivative with respect to the argument,
${\cal T}$.
On the other hand, we make use of  the thermodynamic relation
\be\label{vev5a}
s = - {1\over V_{d-1}}{{\de {\cal F}} \over \de T},
\ee
to replace the left hand side of equation (\ref{vev4a}) by
$-T^{d-1}\sigma$. This leads to the relation
\be\label{vev6a}
\left({\sigma' \over d} + \gamma'\right) = - d\gamma .
\ee
Next, we differentiate equation (\ref{vev2a}) with respect to $j$,
\be\label{vev7a}
{\de {\cal F}(T,j)  \over \de j} =  T^d\left({\sigma' \over
    d} + \gamma'\right){{\cal T} \over \Dm j} = -{d \over \Dm}
{T^d \over j} \gamma,
\ee
where we have used the result (\ref{vev6a}) in the second
equality. Finally, using the definition of $\gamma$ from equation
(\ref{vev2a}),  we obtain,
\be
{\de {\cal F}(T,j)  \over \de j} =  -  (M_p\ell)^{d-1}V_{d-1}
{d\over \Dm} C({\cal T})
j^{\Dp/\Dm}
\ee
Comparing the right hand side with the equation above with the
relation (\ref{vev1-ia}), we obtain the desired result
(\ref{vev1a}).
\section{Determination of the free energy using scalar variables}
\label{AppXY}
\subsection{The phase variables}

We define the two phase variables $X$ and $Y$ as:
\begin{equation}\label{XY1}
  X(\phi)\equiv \frac{\g}{d}\frac{\phi'}{A'}, \qquad  Y(\phi)\equiv
  \frac{1}{d}\frac{g'}{ A'}.
\end{equation}
where the function $g$ is defined as $g = \log{f}$ and the
constant $\g$ is given by,
\begin{equation}\label{gam1}
  \g = \sqrt{\frac{d}{2(d-1)}}.
\end{equation}
These functions satisfy:
\bea\label{Xeq}
\frac{dX}{d\phi} &=& -\g~(1-X^2+Y)\le(1+\frac{1}{2\g}\frac{1}{X}\frac{d\log V}{d\phi}\ri),\\
\frac{dY}{d\phi} &=& -\g~(1-X^2+Y)\frac{Y}{X }. \label{Yeq}
\eea
This second order system is sufficient to determine all of the
thermodynamic properties (and dissipation) of the gravitational
theory \cite{gkmn}. This is a reduction of the fifth order
Einstein-scalar system to an equivalent second order system.

It is straightforward to show that these equations combined with
the following three,
\bea\label{Ap}
\frac{dA}{du} &=& -\frac{1}{\ell} e^{-\g\int^{\phi}_{0} X(t)dt},\\
\label{fp} \frac{d\phi}{du} &=&  -\frac{1}{\ell} \frac{d}{\g} X(\phi) e^{-\g\int^{\phi}_{0} X(t)dt},\\
\label{gp} \frac{dg}{du} &=& -\frac{1}{\ell}~d~Y(\phi)
e^{-\g\int^{\phi}_{0} X(t)dt}, \eea
solve the original Einstein equations in the domain-wall variables defined by the ansatz \eqref{b1}.
The solution in the conformal coordinates is found by the change of variables $du = \exp(A) dr$,
\begin{equation}\label{BH}
  ds^2 = e^{2A(r)}\le(f^{-1}(r)dr^2 + dx_{d-1}^2
 + dt^2 f(r) \ri), \qquad \Phi= \Phi(u).
\end{equation}

 One can also express $g$ and $A$ in terms of
the phase variables directly from the definitions (\ref{XY1}):
\bea
A(\phi) &=& A(\phi_c) + \frac{\g}{d}\int_{\phi_c}^{\phi} \frac{d\tilde{\phi}}{X},\label{Aeq1}\\
f(\phi) &=&  \exp\le(\g \int_{0}^{\phi} \frac{Y}{X}d\tilde{\phi}\ri).\label{feq1}
\eea
Here $\Phi_c$ denotes a surface
near the boundary where we will apply the UV matching conditions of the TG and the BH solution in the following.
Another useful equation relates the scalar potential to the phase variables, that follows from (\ref{Xeq}):
\be\lab{V1} V(\phi) = \frac{d(d-1)}{\ell^2} \le(1+Y-X^2\ri)
e^{-2\g\int_{0}^{\phi} (X(t)-\frac{Y(t)}{2X(t)})dt}. \ee
The precise form of the overall coefficient follows from inserting (\ref{Ap}), (\ref{fp}) and (\ref{gp}) in the Einstein's equations.

The temperature $T$ and the entropy density $s$ of the black-hole are  given by
\bea \lab{Teq1} T(\phi_h) &=&
\frac{\ell}{4\pi(d-1)}~e^{A(\phi_h)}~V(\phi_h)~e^{\gamma
\int_{0}^{\phi_h} X(\phi)~d\phi},\\
\label{ent31}
s &=& 4\pi M_p^{d-1}e^{(d-1)A(\phi_h)}.
\eea
In the first equation we used
\be\lab{Tcomp1}
-4\pi T = f'(r_h) = \frac{df}{d\phi}\frac{d\phi}{du}\frac{du}{dr}\bigg|_{\phi_h} = \frac{d}{\ell} Y(\phi_h) e^{A(\phi_h) + \gamma \int_0^{\phi_h} \le( \frac{Y}{X} - X\ri) d\tilde{\phi}}\, ,
\ee
and the fact that $Y(\phi)$ diverges as $\phi\to \phi_h$  to express it in terms of the same limit of the potential (\ref{V1}).  As $A(\phi)$ from (\ref{Aeq1}) diverges as the cut-off $\phi_c$ is removed, equations that explicitly involve $A(\phi_h)$ are not very efficient practically. Instead, one can derive the following equations with no reference to $A(\f)$ after a little bit of algebra:
\be
T = T_{ref} \le(V(\phi_h)\ell^2 e^{\g \int_0^{\phi_h} X}\ri)^{1-\frac{1}{d}}Y_0(\phi_h)^\frac{1}{d}, \quad
s = s_{ref}\le(Y_0(\phi_h)^{-1} V(\phi_h)\ell^2 e^{\g \int_0^{\phi_h} X}\ri)^{\frac{1}{d}-1} \lab{TSphase}\, ,
\ee
where $T_{ref}$ and $s_{ref}$ are constants given by :
\be
\lab{TSref}
T_{ref} =\frac{d^{\frac{1}{d}}(d-1)^{\frac{d}{d-1}}}{4\pi} |j|^{\frac{1}{\Delta_-}} \, , \qquad s_{ref} = 4\pi (d(d-1))^{1-\frac{1}{d}} (M_P\ell)^{d-1} |j|^{\frac{d-1}{\Delta_-}}\, .
\ee
Finally we note that the variable $X$ is related to the superpotential defined in section \ref{app:first} as
\be\lab{spotX}
W(\phi) = \frac{2(d-1)}{\ell} e^{-\gamma \int_0^\phi X(t) dt}\, .
\ee

\subsection{UV and IR asymptotics}

We first discuss the UV asymptotics in the zero T solution. Equation (\ref{Xeq}) near $\phi=0$ yields
\be\lab{X0uv}
X_0(\phi) = - X_\pm \phi + \cO(\phi^2)
\ee
where
\be\lab{Xpm}
X_\pm = \frac{\g}{2}\le( 1\pm \sqrt{1+\frac{4m^2\ell^2}{d^2}}\ri)\, .
\ee
Here the integration constant of  equation (\ref{Xeq}) is not visible in the Taylor expansion near the UV. In fact it is given by a non-analytic term. As discussed in \cite{gkmn} this integration constant is completely fixed by the choice of asymptotics in the IR. In passing we note the relationship:
\be\lab{relDeltaX}
\frac{\gamma}{X_\pm} = \frac{d}{\Delta_\pm}\, .
\ee
As mentioned above the relevant deformations correspond to $m^2<0$. The BF bound can be read off from (\ref{Xpm}) as,
\be\lab{BFbound}
\frac{m^2\ell^2}{d^2\xi} \geq -\frac12.
\ee
By computing the expansion of $\phi$ near the boundary one learns that the choice $X\to -X_-\phi$ corresponds to a deformation of the UV conformal theory by a source,
and $X\to -X_+\phi$ corresponds to a VeV, hence spontaneous breaking of conformal symmetry. We will assume deformation by a source below.

At finite temperature, solution of (\ref{Xeq}) and (\ref{Yeq}) near $\phi$ yields,
\bea
\lab{Yuv}
Y(\phi) &=& Y_0(\phi_h)~\phi^{\frac{d}{\Delta_-}} + \cdots\\
X(\phi) &=& X_0(\phi) + \delta X_0(\phi_h)~ \phi^{\frac{d}{\Delta_-}-1} + \cdots
\lab{Xuv}
\eea
Here, $\delta X$ is defined as the deformation due to the presence of the BH, i.e. $\delta X = 0$ for the TG solution. We indicated the dependence of the integration constants in the UV on the location of the
horizon $\phi_h$.

Using these asymptotics in (\ref{Ap}) and (\ref{fp}) we find
\bea\lab{phiuv}
\phi(u) &=& \phi_- e^{\Delta_- \frac{u}{\ell}} + \cdots\\
A(u) &=& \tilde{A}_0 - \frac{u}{\ell} +
\cdots\lab{Auv}
\eea
where $u\to -\infty$ corresponds to the boundary and
$\phi_0$ and $\tilde{A}_0$ are the integration constants.
Because the equations of motion are invariant under the change of variables $u\to u+ const$, {\em the integration constant $\tilde{A}_0$ can be set to zero. This is what we will do in the following.} Changing to the conformal frame near the boundary where $r/\ell = \exp(u/\ell)$, we have
\bea\lab{phiuvr}
\phi(r) &=& j r^{\Delta_-} + \phi_+ r^{\Delta_+}\cdots\\
A(r) &=& - \log \frac{r}{\ell} +
\cdots\lab{Auvr}
\eea
We further note the constant limit
\be\lab{A0}
\lim_{\phi\to 0} e^{A(\phi)} \phi^{\frac{\g}{dX_-}} = \ell |j|^{\frac{1}{\Delta_-}} \, .
\ee
UV asymptotics of $X$ and $Y$ cannot depend on any other integration constant by the following simple argument.
One solves (\ref{Xeq}) and (\ref{Yeq}) starting from the horizon. A priori one expects two integration constants. Location of the horizon can be viewed as one, therefore there remains one. However, demanding a regular horizon of the form
\be\lab{reghor}
f(\phi) = const. \times (\phi_h-\phi), \qquad near\,\,\, \phi_h,
\ee
from the equations (\ref{feq1}) and (\ref{Yeq}) means that this remaining integration constant is completely fixed, as
one should require,
\be\lab{XYh} X(\phi) = - \frac{1}{2\g} \frac{V'(\phi_h)}{V(\phi_h)} + \cO(\phi_h-\phi); \qquad Y(\phi) = - \frac{X(\phi_h)}{\g(\phi_h-\phi)} +  \cO(1),
\ee
near $\phi\approx \phi_h$.
In general, this argument shows that one does not have to worry about the other integration constant which can be thought of as the source $j$---that we discuss below---if one
derives the thermodynamics  directly from the $X,\,Y$ system. Below, we show this in more detail.

\subsection{The free energy} \label{AppXY:free}

One can calculate the free energy  directly from the on-shell value of the GR action by using the solution expressed in terms of the phase variables above. The blackhole metric in the $\phi$ frame reads,
\be
ds^2_{BH} =  B^2(\phi)\le( dt^2 F(\phi) + d{\vec{x}}^2 + \frac{d\phi^2}{F(\phi)D(\phi)^2}\ri).\lab{lambdaBH}
\ee
Here the various metric functions are defined as follows:
\bea\lab{B0B}
B(\phi) &=& B(\phi_0) e^{\frac{\g}{d} \int_{\phi_c}^{\phi} \frac{d\tilde{\phi}}{X}},\qquad
D(\phi) =  -\frac{d}{\gamma \ell}X(\phi)B(\phi)e^{-\gamma \int_0^{\phi} d\tilde{\phi} X},
\lab{D0D}\\
F(\phi) &=& e^{\gamma \int_{0}^{\phi} d\tilde{\phi} \frac{Y}{X}}\, ,\lab{FL}
\eea
where $\phi_c$ is a UV cut-off that we will remove at the end of the calculation.
They are obtained directly from the the expressions for the metric functions defined in the text
 in terms of the radial variable $r$, viz. (\ref{Aeq1}),(\ref{feq1}) and (\ref{Ap}-\ref{gp}).
We call the metric functions in $\l$ with the capital letters to distinguish them from the
analogous functions of $r$. The relations are explicitly given by
$B(\phi) = b\le(r(\phi)\ri), \qquad F(\phi) = f\le(r(\phi)\ri) $
where $r$ is determined by
$$ r(\phi) = \int_0^{\phi} \frac{d\tilde{\phi}}{D(\tilde{\phi})}\, .$$
The expressions above completely determine the map between the r-frame and the $\phi$-frame.

\paragraph{Einstein contribution:\\}
\vspace{.5cm}
We first compute the  Einstein (bulk) contribution to the free energy,
i.e. the first term in equation (\ref{b0}).  The bulk on-shell action $S^{bulk}$ (after using the Einstein's equations) is generally given by
the frame-independent expression,
\be\lab{E1}
S^{bulk} = \frac{2}{d-1} M^{d-1} \int_{\cal M} \sqrt{g} V.
\ee
${\cal M}$ is the manifold with a boundary. We regulate the integral in the
$\phi$-frame by placing a cut-off at $\phi_c$. Thus, using the metric functions
defined above, one obtains the following expression in the $\phi$ variable
 \be\lab{E6}
S^{bulk} = \frac{2}{d-1} M^{d-1} \b V_{d-1} \int_{\phi_c}^{\phi_h} B(\phi)^{d+1} V(\phi) D(\phi)^{-1}.
\ee
We now substitute the expression for $D(\phi)$, $B(\phi)$  and $V(\phi)$
from (\ref{D0D}), (\ref{B0B}) and (\ref{V1}), and obtain,
\be\lab{E7}
S^{bulk} = -\frac{2\g}{\ell} M^{d-1} \b V_{d-1} B(\phi_c)^d e^{-\g\int_0^{\phi_c} (X-\frac{Y}{X}) d\tilde{\phi} }
\int_{\phi_c}^{\phi_h} d\phi \frac{1-X^2+Y}{X} e^{\g \int_{\phi_c}^{\phi} d\tilde{\phi} \frac{1-X^2+Y}{X}}.
\ee
Integrand is a total derivative, thus
\be\lab{E8}
S^{bulk}= -\frac{2}{\ell} M^{d-1} \b V_{d-1} B(\phi_c)^d e^{-\g\int_0^{\phi_c} \le(X-\frac{Y}{X}\ri) d\tilde{\phi} }
e^{\g \int_{\phi_c}^{\phi} d\tilde{\phi} \frac{1-X^2+Y}{X}}\bigg|_{\phi_c}^{\phi_h}.
\ee
This can be simplified further: using (\ref{Yeq}), one realises that the integrand in the exponent is a total derivative of
$\log Y(\phi)$. Thus, one has,
\be\lab{E9}
S^{bulk} = -\frac{2}{\ell} M^{d-1} \b V_{d-1} B(\phi_c)^d e^{-\g\int_0^{\phi_c} \le(X-\frac{Y}{X}\ri) d\tilde{\phi} } \le( \frac{Y(\phi_c)}{Y(\phi_h)}-1\ri).
\ee
But $Y(\phi_h)=\infty$ by regularity condition at the horizon (see  section (\ref{reghor})), hence we have the final expression
for the Einstein contribution on the BH geometry:
\be\lab{E10}
S^{bulk} = \frac{2}{\ell} M^{d-1} \b V_{d-1} B(\phi_c)^d e^{-\g\int_0^{\phi_c} \le(X-\frac{Y}{X}\ri) d\tilde{\phi} } .
\ee

\paragraph{Gibbons-Hawking contribution:\\}
\vspace{.5cm}

We move on to the Gibbons-Hawking term that is given by the
frame-independent expression,  the second term in (\ref{b0}):

\begin{equation}
   S_{GH} = -2M^{d-1} \int_{\partial M}d^dx \sqrt{h}~K
    \label{app1}\end{equation}
    with
    \be
K_{\m\n}\equiv  \nabla_\mu n_\nu = {1\over 2}n^{\rho}\partial_{\rho}h_{\m\n}\sp K=h^{ab}K_{ab}
\label{app2}\ee
where $h_{ab}$ is the induced metric on the boundary and $n_{\m}$ is the (outward directed) unit
normal to the boundary. In the $\phi$-frame (\ref{lambdaBH}), it is given by
\be
n^{\mu}=-{1\over \sqrt{ g_{\phi\phi}}}\left({\partial\over
\partial \phi}\right)^{\mu}={{\delta^{\mu}}_{\phi}\over \sqrt{ g_{\phi\phi}}}.
\label{app3}\ee
The determinant of the induced metric on the boundary and the extrinsic curvature now are
\be
\sqrt{h} = B(\phi_c)^d\sqrt{F(\phi_c)},
\label{hBH}\ee
and
\be
K = \frac{\g D(\phi_c)\sqrt{F(\phi_c)}}{X(\phi_c)B_0(\phi_c)}\le( 1+\frac{Y(\phi_c)}{2}\ri).
\label{KBH}\ee
Therefore one finds,
\begin{equation}
   S_{GH}  =-\frac{2d}{\ell} M^{d-1} \b V_{d-1} B(\phi_c)^d F(\phi_c) e^{-\g\int_0^{\phi_c} X}\le( 1+\frac{Y(\phi_c)}{2}\ri)\, .
    \label{GHBH}\end{equation}

\paragraph{counter-term:\\}
\vspace{.5cm}
The counter-term action is expressed in terms of the superpotential in equation (\ref{ct4}). Using equation (\ref{spotX}) we obtain
\be\lab{ctac}
S_{ct} = \frac{2(d-1)}{\ell} M^{d-1} \b V_{d-1} B(\phi_c)^d  e^{-\g\int_0^{\phi_c} X_{ct} - \frac{Y}{2X}}\, ,
\ee
where $X_{ct}$ is the analytic solution to (\ref{Xeq}) obtained in terms of odd powers of $\phi$. We will only need the first term $X_{ct} = -X_- \phi +\cdots$ in what follows.
\paragraph{The total free energy:\\}
\vspace{.5cm}

The total free energy now can be obtained as the sum of (\ref{E10}), (\ref{GHBH}) and (\ref{ctac}). Using the relation between the free energy and the on-shell gravity action $F = S/\b$, equations (\ref{Xuv}), (\ref{Yuv}) and (\ref{A0}) we find in the limit $\phi_c\to 0$,
 \be\lab{Ftot}
 F = (M_p\ell)^{d-1}  V_{d-1}  |j|^{\frac{d}{\Delta_-}} \le(2(d-1) \, \textrm{sgn}(j) X_-\delta X_0(\phi_h) - Y_0(\phi_h)\ri) \, .
 \ee
 The physical meaning of the constants $\delta X_0$ and $Y_0$ are as
 follows. Solving (\ref{fp}) near the boundary and matching onto
 (\ref{max0a}) and using (\ref{vev}) one finds that the VeV is given in terms of $\delta X_0$ as,
 \be\lab{VeVX}
 \langle {\cal O} \rangle = \frac{d}{\gamma} |j|^{\frac{d}{\Delta_-}-1} \delta X_0(\phi_h)\, .
 \ee
On the other hand solving $Y$ from (\ref{XY1}) near the boundary using the near boundary expansions, and the definitions of $T$ and $s$ from (\ref{b1-1}) we find $Y_0$ in terms of enthalpy as
\be\lab{EnthalpyY}
Y_0(\phi_h) = \frac{Ts}{d} (M_p\ell)^{1-d}  |j|^{\frac{d}{\Delta_-}} \, .
 \ee
 Then the free energy (\ref{Ftot}) can directly be expressed in terms of the enthalpy and the VeV as in (\ref{free}).


\addcontentsline{toc}{section}{References}
\begin{thebibliography}{99}

\bibitem{maldacena}
J.~M.~Maldacena,
  {\em The Large N limit of superconformal field theories and supergravity,}
  \href{https://doi.org/10.1023/A:1026654312961}{Int.\ J.\ Theor.\ Phys.\  {\bf 38}, 1113 (1999)},
  [Adv.\ Theor.\ Math.\ Phys.\  {\bf 2}, 231 (1998)],
  \hre{hep-th}{9711200}.

\bibitem{witten}
E.~Witten,
  {\em Anti-de Sitter space and holography,}
  \href{https://doi.org/10.1023/A:1026654312961}{Adv.\ Theor.\ Math.\ Phys.\  {\bf 2}, 253 (1998)},
  \hre{hep-th}{9802150}.


\bibitem{GKP}
S.~S.~Gubser, I.~R.~Klebanov and A.~M.~Polyakov,
  {\em Gauge theory correlators from noncritical string theory,}
   \href{https://doi.org/10.1016/S0370-2693(98)00377-3}{Phys.\ Lett.\ B {\bf 428}, 105 (1998)},
  \hre{hep-th}{9802109}.


\bibitem{Boonstra:1998mp}
  H.~J.~Boonstra, K.~Skenderis and P.~K.~Townsend,
  {\em The domain wall / QFT correspondence,}
  \href{https://doi.org/10.1088/1126-6708/1999/01/003}{JHEP {\bf 9901}, 003 (1999)},
  \hre{hep-th}{9807137}.


\bibitem{Girardello:1998pd}
  L.~Girardello, M.~Petrini, M.~Porrati and A.~Zaffaroni,
  {\em Novel local CFT and exact results on perturbations of N=4 superYang Mills from AdS dynamics,}
   \href{https://doi.org/10.1088/1126-6708/1998/12/022}{JHEP {\bf 9812}, 022 (1998)},
  \hre{hep-th}{9810126}.
  

\bibitem{Balasubramanian:1999jd}
  V.~Balasubramanian and P.~Kraus,
  {\em Space-time and the holographic renormalisation group,}
  \href{https://doi.org/:10.1103/PhysRevLett.83.3605}{Phys.\ Rev.\ Lett.\  {\bf 83}, 3605 (1999)},
  \hre{hep-th}{9903190}.
  


\bibitem{Freedman}
  D.~Z.~Freedman, S.~S.~Gubser, K.~Pilch and N.~P.~Warner,
    {\em renormalisation group flows from holography supersymmetry and a c theorem,}
  \href{https://doi.org/10.4310/ATMP.1999.v3.n2.a7}{  Adv.\ Theor.\ Math.\ Phys.\  {\bf 3}, 363 (1999)},
  \hre{hep-th}{9904017}.

  \bibitem{deboer}
  J.~de Boer, E.~P.~Verlinde and H.~L.~Verlinde,
  {\em  On the holographic renormalisation group, }
  \href{https://doi.org/10.1088/1126-6708/2000/08/003}{JHEP {\bf 0008} (2000) 003},
 \hre{hep-th}{9912012}.
  

\bibitem{Bianchi:2001de}
  M.~Bianchi, D.~Z.~Freedman and K.~Skenderis,
  {\em  How to go with an RG flow, }
    \href{https://doi.org/10.1088/1126-6708/2001/08/041}{JHEP {\bf 0108}, 041 (2001)},
  \hre{hep-th}{0105276}.
  

\bibitem{deHaro:2000vlm}
  S.~de Haro, S.~N.~Solodukhin and K.~Skenderis,
  {\em Holographic reconstruction of space-time and renormalisation in the AdS / CFT correspondence,}
  \href{https://doi.org/10.1007/s002200100381}{Commun.\ Math.\ Phys.\  {\bf 217}, 595 (2001)},
  \hre{hep-th}{0002230}.


\bibitem{papaskenderis}
 I.~Papadimitriou and K.~Skenderis,
    {\em Correlation functions in holographic RG flows, }
  \href{https://doi.org/10.1088/1126-6708/2004/10/075}{JHEP\ {\bf 0410}, 075  (2004)},
  \hre{hep-th}{0407071}.
  


\bibitem{multibranch}
  E.~Kiritsis, F.~Nitti and L.~Silva Pimenta,
  {\em Exotic RG Flows from Holography,}
  \href{http://dx.doi.org/10.1002/prop.201600120}{Fortsch.\ Phys.\  {\bf 65} (2017) no.2,  1600120},
     \hri{1611.05493}{[hep-th]}.

  \bibitem{multifield}
  F.~Nitti, L.~Silva Pimenta and D.~A.~Steer,
  {\em  On multi-field flows in gravity and holography, }
  \hri{1711.10969}{[hep-th]}.
  


  \bibitem{dilaton1}
  C.~Hoyos, U.~Kol, J.~Sonnenschein and S.~Yankielowicz,
  {\em  The a-theorem and conformal symmetry breaking in holographic RG flows, }
      \href{http://dx.doi.org/10.1007/JHEP03(2013)063}{ JHEP {\bf 1303} (2013) 063},
  \hri{1207.0006}{[hep-th]};\\
  C.~Hoyos, U.~Kol, J.~Sonnenschein and S.~Yankielowicz,
  {\em  The holographic dilaton, }
       \href{http://dx.doi.org/10.1007/JHEP10(2013)181}{JHEP {\bf 1310} (2013) 181},
  \hri{1307.2572}{[hep-th]}.



\bibitem{curvedRG}
  J.~K.~Ghosh, E.~Kiritsis, F.~Nitti and L.~T.~Witkowski,
  {\em  Holographic RG flows on curved manifolds and quantum phase transitions, }
\hri{1711.08462}{[hep-th]}.
  


  \bibitem{Gursoy:2016ggq}
U.~Gürsoy, A.~Jansen and W.~van der Schee,
    {\em New dynamical instability in asymptotically anti-de Sitter spacetime,}
    \href{http://dx.doi.org/10.1103/PhysRevD.94.061901}{Phys.\ Rev.\ D {\bf 94}, no. 6, 061901 (2016)},
\hri{1603.07724}{[hep-th]}.
  
\bibitem{Janik:2016btb}
  R.~A.~Janik, J.~Jankowski and H.~Soltanpanahi,
  {\em Quasinormal modes and the phase structure of strongly coupled matter}
        \href{http://dx.doi.org/10.1007/JHEP06(2016)047}{JHEP {\bf 1606} (2016) 047},
\hri{1603.05950}{[hep-th]}.

\bibitem{super}
  S.~S.~Gubser,
  {\em  Breaking an Abelian gauge symmetry near a black hole horizon, }
  \href{https://doi.org/10.1103/PhysRevD.78.065034}{Phys.\ Rev.\ D {\bf 78} (2008) 065034},
\hri{0801.2977}{[hep-th]};\\
  S.~A.~Hartnoll, C.~P.~Herzog and G.~T.~Horowitz,
  {\em  Building a Holographic Superconductor, }
  \href{https://doi.org/10.1103/PhysRevLett.101.031601}{Phys.\ Rev.\ Lett.\  {\bf 101} (2008) 031601},
  \hri{0803.3295}{[hep-th]}.


\bibitem{David}
Y.~Bea and D.~Mateos, {\em Heating up exotic RG flows with holography,} to appear.

\bibitem{gkmn}
  U.~Gursoy, E.~Kiritsis, L.~Mazzanti and F.~Nitti,
  {\em  Holography and Thermodynamics of 5D Dilaton-gravity, }
    \href{https://doi.org/10.1088/1126-6708/2009/05/033}{JHEP {\bf 0905} (2009) 033},
  \hri{0812.0792}{[hep-th]}.



  \bibitem{hami}
  I.~Papadimitriou and K.~Skenderis,
  {\em  AdS / CFT correspondence and geometry, }
       \href{http://dx.doi.org/}{IRMA Lect.\ Math.\ Theor.\ Phys.\  {\bf 8} (2005) 73},
  \hre{hep-th}{0404176}.


\bibitem{Kiritsis:2011zq}
  E.~Kiritsis, V.~Niarchos,
 {\em   Josephson Junctions and AdS/CFT Networks, }
   \href{https://doi.org/10.1007/JHEP07(2011)112}{ JHEP {\bf 1107}, 112 (2011)},
  \hri{1105.6100}{[hep-th]}.


  \bibitem{KN}
  E.~Kiritsis and V.~Niarchos,
  {\em  The holographic quantum effective potential at finite temperature and density, }
        \href{http://dx.doi.org/}{JHEP {\bf 1208} (2012) 164},
\hri{1205.6205}{[hep-th]}.


\bibitem{Bourdier}
  J.~Bourdier and E.~Kiritsis,
    {\em Holographic RG flows and nearly-marginal operators,}
  \href{http://dx.doi.org/10.1088/0264-9381/31/3/035011}{Class.\ Quant.\ Grav.\  {\bf 31}, 035011 (2014)},
  \hri{1310.0858}{[hep-th]}.


  \bibitem{Papadimitriou:2007sj}
  I.~Papadimitriou,
  {\em  Multi-Trace Deformations in AdS/CFT: Exploring the Vacuum Structure of the Deformed CFT, }
    \href{https://doi.org/10.1088/1126-6708/2007/05/075}{JHEP {\bf 0705}, 075 (2007)},
  \hre{hep-th}{0703152}.
  

\bibitem{wenliang}
E.~Kiritsis, W.~Li and F.~Nitti,
  {\em  Holographic RG flow and the Quantum Effective Action, }
  \href{http://dx.doi.org/10.1002/prop.201400007}{Fortsch.\ Phys.\  {\bf 62}, 389 (2014)},
 \hri{1401.0888}{[hep-th]}.


    \bibitem{papa}
  I.~Papadimitriou,
  {\em  Holographic renormalisation of general dilaton-axion gravity, }
    \href{https://doi.org/10.1007/JHEP08(2011)119}{JHEP {\bf 1108} (2011) 119},
\hri{1106.4826}{[hep-th]}.







\end{thebibliography}
\end{document}